\documentclass[12pt]{article}
\usepackage{epsfig}
\usepackage{axodraw}
\usepackage{epsfig}                            
\usepackage{graphicx}
\usepackage{rotate}
\usepackage{latexsym}
\newcommand\pubnumber{}
\newcommand\pubdate{\today}
\newcommand\hepnumber{hep-ph/0112004}

\def\csumb{Dipartimento di Fisica Teorica, Universit\`a di Torino, Italy\\
INFN, Sezione di Torino, Italy}
\def\support{\footnote{Work supported by the
European Union under contract HPRN-CT-2000-00149.}} 

\def\Title#1{\begin{center} {\Large\bf #1 } \end{center}}

\newcommand{\Authors}[2]{\begin{center}{ \sc #1 \hspace{0.1cm} {\rm and} 
\hspace{0.1cm} #2} 
\end{center}}
\def\Address#1{\begin{center}{ \it #1} \end{center}}

\newcommand\pubblock{\rightline{\begin{tabular}{l} \pubnumber\\
         \pubdate\\ \hepnumber \end{tabular}}}
\newenvironment{Abstract}{\begin{quotation}  }{\end{quotation}}

\def\Acknowledgments{\bigskip  \bigskip \begin{center}
          \large\bf Acknowledgments\end{center}}

\makeatletter
\def\section{\@startsection{section}{0}{\z@}{5.5ex plus .5ex minus
 1.5ex}{2.3ex plus .2ex}{\large\bf}}
\def\subsection{\@startsection{subsection}{1}{\z@}{3.5ex plus .5ex minus
 1.5ex}{1.3ex plus .2ex}{\normalsize\bf}}
\def\subsubsection{\@startsection{subsubsection}{2}{\z@}{-3.5ex plus
-1ex minus  -.2ex}{2.3ex plus .2ex}{\normalsize\sl}}

\renewcommand{\@makecaption}[2]{%
   \vskip 10pt
   \setbox\@tempboxa\hbox{\small #1: #2}
   \ifdim \wd\@tempboxa >\hsize     % IF longer than one line:
       \small #1: #2\par          %   THEN set as ordinary paragraph.
     \else                        %   ELSE  center.
       \hbox to\hsize{\hfil\box\@tempboxa\hfil}
   \fi}

 \def\citenum#1{{\def\@cite##1##2{##1}\cite{#1}}}
\def\citea#1{\@cite{#1}{}}
 
\newcount\@tempcntc
\def\@citex[#1]#2{\if@filesw\immediate\write\@auxout{\string\citation{#2}}\fi
  \@tempcnta\z@\@tempcntb\m@ne\def\@citea{}\@cite{\@for\@citeb:=#2\do
    {\@ifundefined
       {b@\@citeb}{\@citeo\@tempcntb\m@ne\@citea\def\@citea{,}{\bf }\@warning
       {Citation `\@citeb' on page \thepage \space undefined}}%
    {\setbox\z@\hbox{\global\@tempcntc0\csname b@\@citeb\endcsname\relax}%
     \ifnum\@tempcntc=\z@ \@citeo\@tempcntb\m@ne
       \@citea\def\@citea{,}\hbox{\csname b@\@citeb\endcsname}%
     \else
      \advance\@tempcntb\@ne
      \ifnum\@tempcntb=\@tempcntc
      \else\advance\@tempcntb\m@ne\@citeo
      \@tempcnta\@tempcntc\@tempcntb\@tempcntc\fi\fi}}\@citeo}{#1}}
\def\@citeo{\ifnum\@tempcnta>\@tempcntb\else\@citea\def\@citea{,}%
  \ifnum\@tempcnta=\@tempcntb\the\@tempcnta\else
  {\advance\@tempcnta\@ne\ifnum\@tempcnta=\@tempcntb \else\def\@citea{--}\fi
    \advance\@tempcnta\m@ne\the\@tempcnta\@citea\the\@tempcntb}\fi\fi}
\makeatother

\newcommand{\nl}{\nonumber\\}

\newcommand{\ds}{\displaystyle}

\newcommand{\lpar}{\left(}                            % bracketing
\newcommand{\rpar}{\right)}

\newcommand{\bq}{\begin{equation}}                    % equationing
\newcommand{\eq}{\end{equation}}
\newcommand{\bqa}{\arraycolsep 0.14em\begin{eqnarray}}
\newcommand{\eqa}{\end{eqnarray}}
\newcommand{\ba}[1]{\begin{array}{#1}}
\newcommand{\ea}{\end{array}}
\newcommand{\ben}{\begin{enumerate}}
\newcommand{\een}{\end{enumerate}}
\newcommand{\bei}{\begin{itemize}}
\newcommand{\eei}{\end{itemize}}
\newcommand{\eqn}[1]{Eq.(\ref{#1})}
\newcommand{\eqns}[2]{Eqs.(\ref{#1})--(\ref{#2})}

\newcommand{\tabn}[1]{Tab.~\ref{#1}}

\newcommand{\tabns}[2]{Tabs.~\ref{#1}--\ref{#2}}

\newcommand{\fig}[1]{Fig.~\ref{#1}}

\newcommand{\sect}[1]{Section~\ref{#1}}

\newcommand{\subsect}[1]{Subsection~\ref{#1}}
\newcommand{\subsectm}[2]{Subsection~\ref{#1} -- \ref{#2}}

\newcommand{\GeV}{\mathrm{GeV}}

\def\Re{\mathop{\operator@font Re}\nolimits}
\def\Im{\mathop{\operator@font Im}\nolimits}
\newcommand{\ord}[1]{{\cal O}\lpar#1\rpar}

\newcommand{\ib}{i}
\newcommand{\asums}[1]{\sum_{#1}}

\newcommand{\ph}{\gamma}

\newcommand{\wb}{W}

\newcommand{\zb}{Z}

\newcommand{\baru}{\overline u}

\newcommand{\mone}{m_1}
\newcommand{\mtwo}{m_2}

\newcommand{\mones}{m^2_1}
\newcommand{\mtwos}{m^2_2}

\newcommand{\seffsf}[1]{\sin^2\theta^{#1}_{\rm{eff}}}

\newcommand{\spro}[2]{{#1}\cdot{#2}}
\newcommand{\gam}{\lpar 1-\gamma_5\rpar}

\newcommand{\li}[2]{\mathrm{Li}_{#1}\lpar\displaystyle{#2}\rpar} % polylog
\newcommand{\lkall}[3]{\lambda\lpar#1,#2,#3\rpar}       % Kallen's lambda
\newcommand{\egam}[1]{\Gamma\lpar#1\rpar}               % Euler's Gamma
\newcommand{\drii}[2]{\delta_{#1#2}}                    % Kronecker's delta
\newcommand{\intmomii}[3]{\int\,d^{#1}#2\,\int\,d^{#1}#3}
\newcommand{\intfx}[1]{\int_{\scriptstyle 0}^{\scriptstyle 1}\,d#1}
\newcommand{\intfxy}[2]{\int_{\scriptstyle 0}^{\scriptstyle 1}\,d#1\,
                        \int_{\scriptstyle 0}^{\scriptstyle #1}\,d#2}

\newcommand{\MSB}{\overline{MS}}

\newcommand{\bff}[4]{B_{#1}\lpar #2;#3,#4\rpar}

\newcommand{\Ddr}{{\ds\frac{1}{{\bar{\varepsilon}}}}}

\newcommand{\ep}{\epsilon}
\newcommand{\tHs}{\mu}

\newcommand{\tHss}{\mu^2}
\newcommand{\Reb}{{\rm{Re}}}
\newcommand{\Imb}{{\rm{Im}}}
\newcommand{\xvar}{x}

\newcommand{\lnx}{\ln\xvar}

\newcommand{\pmoms}{p^2}

\newcommand{\upar}[1]{u}

\newcommand{\ssI}{{\scriptscriptstyle{I}}}

\newcommand{\ssL}{{\scriptscriptstyle{L}}}
\newcommand{\ssM}{{\scriptscriptstyle{M}}}
\newcommand{\ssN}{{\scriptscriptstyle{N}}}

\newcommand{\ssR}{{\scriptscriptstyle{R}}}
\newcommand{\ssS}{{\scriptscriptstyle{S}}}

\newcommand{\ssU}{{\scriptscriptstyle{U}}}
\newcommand{\ssV}{{\scriptscriptstyle{V}}}
\newcommand{\ssW}{{\scriptscriptstyle{W}}}

\newcommand{\bqas}{\begin{eqnarray*}}
\newcommand{\eqas}{\end{eqnarray*}}
\newcommand{\pc}{\,\%}
\def\app#1#2 {{\it Acta. Phys. Pol.} {\bf#1},#2}
\def\cpc#1#2 {{\it Computer Phys. Comm.} {\bf#1},#2}
\def\np#1#2 {{\it Nucl. Phys.} {\bf#1},#2}
\def\pl#1#2 {{\it Phys. Lett.} {\bf#1},#2}
\def\prep#1#2 {{\it Phys. Rep.} {\bf#1},#2}
\def\prev#1#2 {{\it Phys. Rev.} {\bf#1},#2}
\def\prl#1#2 {{\it Phys. Rev. Lett.} {\bf#1},#2}
\def\zp#1#2 {{\it Zeit. Phys.} {\bf#1},#2}
\def\sptp#1#2 {{\it Suppl. Prog. Theor. Phys.} {\bf#1},#2}
\def\mpl#1#2 {{\it Modern Phys. Lett.} {\bf#1},#2}
\def\jetp#1#2 {{\it Sov. Phys. JETP} {\bf#1},#2}
\def\fpj#1#2 {{\it Fortschr. Phys.} {\bf#1},#2}
\def\afp#1#2 {{\it Acta.Phys. Polon.} {\bf#1},#2}
\def\err#1#2 {{\it Erratum} {\bf#1},#2}
\def\ijmp#1#2 {{\it Int. J. Mod. Phys} {\bf#1},#2}
\def\nc#1#2 {{\it Nuovo Cimento} {\bf#1},#2}
\def\ap#1#2 {{\it Ann. Phys.} {\bf#1},#2}
\def\cmp#1#2 {{\it Comm. Math. Phys.} {\bf#1},#2}
\def\el#1#2 {{\it Europhys. Lett.} {\bf#1},#2}
\def\hpa#1#2 {{\it Helv. Phys. Acta} {\bf#1},#2}
\def\yf#1#2 {{\it Yad. Fiz.} {\bf#1},#2}
\def\nim#1#2 {{\it Nucl. Instrum. Meth.} {\bf#1},#2}
\def\spz#1#2 {{\it Sov. Pisma Zhetf} {\bf#1},#2}
\def\jetpl#1#2 {{\it JETP Lett.} {\bf#1},#2}
\def\sjnp#1#2 {{\it Sov. J. Nucl. Phys.} {\bf#1},#2}
\def\ptp#1#2 {{\it Progr. Theor. Phys. (Kyoto)} {\bf#1},#2}
\def\rmp#1#2  {{\it Rev. Mod. Phys.} {\bf#1},#2}
\def\zhetf#1#2 {{\it ZhETF} {\bf#1},#2}
\def\prs#1#2 {{\it Proc. Roy. Soc.} {\bf#1},#2}
\def\phys#1#2 {{\it Physica} {\bf#1},#2}

\newcommand{\egams}[1]{\Gamma^2\lpar#1\rpar}               % Euler's Gamma

\newcommand{\intfxx}[2]{\int_{\scriptstyle 0}^{\scriptstyle 1}\,d#1\,
                        \int_{\scriptstyle 0}^{\scriptstyle 1}\,d#2}
\def\bfi{\begin{figure}}
\def\efi{\end{figure}}
\newcommand{\intmomsii}[3]{\int\,d^{#1}#2\,d^{#1}#3}
\newcommand{\intfz}{\int_{\scriptstyle 0}^y\,dz}
\newcommand{\intfzp}{\int_{\scriptstyle 0}^{1-y}\,dz}
\newcommand{\intfy}{\int_z^{\scriptstyle 1}\,dy}
\newcommand{\intfzs}{\int_y^{\scriptstyle 1}\,dz}
\newcommand{\hyper}[4]{{}_2F_1(#1\,,\,#2\,;\,#3\,;\,#4)}
\newcommand{\aba}{{_{\scriptstyle{121}}}}
\newcommand{\aca}{{_{\scriptstyle{131}}}}
\newcommand{\acan}[1]{{_{\scriptstyle{#1;131}}}}

\newcommand{\bbab}{{_{\scriptstyle{221}|b}}}
\newcommand{\bbabn}[1]{{_{\scriptstyle{221}|b|#1}}}

\newcommand{\bban}[1]{{_{\scriptstyle{#1;221}}}}
\newcommand{\aban}[1]{{_{\scriptstyle{#1;121}}}}

\newcommand{\bba}{{_{\scriptstyle{221}}}}

\newcommand{\sst}{\scriptstyle}

\begin{document}
\begin{titlepage}
\pubblock

\vfill
\def\thefootnote{\fnsymbol{footnote}}
\Title{Algebraic-Numerical Evaluation of Feynman Diagrams:\\
Two-Loop Self-Energies}
\vfill
\Authors{Giampiero Passarino\support}{Sandro Uccirati}
\Address{\csumb}
\vfill
\begin{Abstract}
A recently proposed scheme for numerical evaluation of Feynman diagrams is
extended to cover all two-loop two-point functions with arbitrary internal and
external masses. The adopted algorithm is a modification of the one proposed
by F.~V.~Tkachov and it is based on the so-called generalized Bernstein 
functional relation. On-shell derivatives of self-energies are also considered
and their infrared properties analyzed to prove that the method which is 
aimed to a numerical evaluation of massive diagrams can handle the infrared 
problem within the scheme of dimensional regularization. Particular care is 
devoted to study the general massive diagrams around their leading and 
non-leading Landau singularities.
\end{Abstract}
\vfill
\vfill
\begin{center}
\end{center}
\end{titlepage}
\def\thefootnote{\arabic{footnote}}
\setcounter{footnote}{0}
%--
\section{Introduction}
%--
The evaluation of multi-loop Feynman diagrams has a long
history~\cite{Pais:1986nu} and, roughly speaking, we can say that there are
theories more simple than others, noticeably QED (see for instance
ref.~\cite{Mastrolia:2000va}) and also to some extent QCD (see for
instance ref.~\cite{Chetyrkin:2000dq}),
where we have few masses and the analytical approach can be pushed very far.
Conversely the full electroweak Lagrangian shows several masses, ranging
over a wide interval of values, therefore making the analytical evaluation of
Feynman diagrams a complicated task. Equivalently complicated is the
situation in any extension of the minimal standard model.

Soon or later the analytical approach will collapse and one can easily
foresee that this failure will show up at the level of a complete two-loop
calculation in the standard model or beyond. This calculation is, for
instance, required to produce quantities as $\seffsf{l}$~\cite{Bardin:1999ak}
with a theoretical precision of $1\,\times\,10^{-6}$. For this reason one is 
lead to consider an alternative approach to the whole problem, namely to 
abandon the analytical way in favor of a fully automatic, numerical evaluation
of multi-loop diagrams. The Holy Graal (General Recursive Applicative and 
Algorithmic Language) requires fast and accurate procedures to deal with the
singularities of an arbitrary diagram.

For a fast and accurate numerical evaluation of arbitrary multi-loop
Feynman diagrams there is some interesting proposal in the
literature~\cite{Tkachov:1997wh} that has been recently applied to the
two-loop two-point sunset topology~\cite{Passarino:2001wv}.
Older approaches can be found in ref.~\cite{Ghinculov:2001uh}.

The Bernstein-Tkachov theorem~\cite{Tkachov:1997wh} (hereafter BT) tells us
that for any finite set of polynomials $V_i(x)$, where $x = \,\lpar
x_1,\dots, x_{\ssN}\rpar$ is a vector of Feynman parameters, there exists an 
identity of the following form:
%--
\bq
{\cal P}\,\lpar x,\partial\rpar \prod_i\,V_i^{\mu_i+1}(x) = B\,
\prod_i\,V_i^{\mu_i}(x).
\label{functr}
\eq
%--
where ${\cal P}$ is a polynomial of $x$ and $\partial_{i} =
\partial/\partial x_i$; $B$ and all coefficients of ${\cal P}$ are
polynomials
of $\mu_i$ and of the coefficients of $V_i(x)$.
Any multi-loop Feynman diagram can be cast into the form of \eqn{functr}.
Iterative applications of the BT functional relations, followed by
integration
by parts, allows us to express the integrand in parametric space as
a combination of (polynomials)${}^N\,\times\,$ logarithms of the same
polynomials, therefore achieving a result that is well suited for numerical
integration. The $B$ coefficients of \eqn{functr} will contain all Landau
singularities~\cite{Landau:1959fi} of the corresponding diagram.

For general one-loop diagrams we have an explicit solution for the
polynomial
${\cal P}$ which is due to F.~V.~Tkachov~\cite{Tkachov:1997wh} (see also
ref.~\cite{Bardin:2000cf} and ref.~\cite{Passarino:2001sq}).
Any one-loop Feynman diagram $G$, irrespectively from the number of external
legs, is expressible as
%--
\bq
G = \int_{\ssS}\,dx\,V^{-\mu}(x),
\eq
%--
where the integration region is $x_i \ge 0, \,\asums{i}\,x_i \le 1$ and
where $V(x)$ is a quadratic form of $x$,
%--
\bq
V(x) = x^t\,H\,x + 2\,K^t\,x + L.
\label{defHKL}
\eq
%--
The solution to the problem of determining the polynomial ${\cal P}$ is as
follows:
%--
\bq
{\cal P} = 1 - {{\,\lpar x+X\rpar\,\partial_x}\over {2\,\lpar\mu+1\rpar}},
\qquad
X = K^t\,H^{-1}, \qquad B = L - K^t\,H^{-1}\,K.
\label{rol}
\eq
%--
For $N$-loop diagrams with $N \ge 2$ a minimal BT approach has
been recently proposed in ref.~\cite{Passarino:2001wv} where
we have adopted a different strategy aimed to deal with arbitrary diagrams.
It represents a compromise based on the simple observation that we know how
to
apply the BT-iterative procedure for arbitrary one-loop diagrams. Therefore,
given any two-loop diagram $G$ we apply the BT functional relation to
$G_{\ssL}$, the one-loop sub-diagram of $G$ which has the largest number of
internal lines. In this way the integrand for $G$ in $x$-space can be made
smooth, a part from the factor $B$ of \eqn{functr} which is now a polynomial
in $x_{\ssS}$, the Feynman parameters needed for the complementary
sub-diagram of $G$ with the smallest number of internal lines, $G_{\ssS}$.
The sub-diagram $G_{\ssS}$, after integration over its momentum,
becomes an additional -- $x_{\ssS}$-dependent and with non-canonical
power -- propagator for $G_{\ssL}$.
This procedure can be immediately generalized to any number of loops.
Furthermore, one should realize that the BT procedure, in short {\em
raising}
powers, does not introduce singularities through $B(x_{\ssS})$, a part from
the singularities in the external parameters of the original diagram.
Therefore, before performing the $x_{\ssS}$-integration we move the
integration contour into the complex hyper-plane, thus avoiding the
crossing of apparent singularities, see also the work of
ref.~\cite{Ghinculov:2001cz}.

As we have anticipated, in this paper we extend the minimal BT-approach to
cover all two-loop, two-points functions.
In dealing with the sunset topology we have discovered a remarkable property
that will be generalized in this paper to all two-loop two-point
functions but that remains true for all two-loop three-point planar
topologies. For all these diagrams, essentially, we do not have enough
external momenta to make the matrix $H$ of \eqn{defHKL} non-singular. As
a consequence, a change of variables is always possible in the quadratic
form $V$ that makes the $B$-coefficient of \eqn{functr} independent on
$x_{\ssS}$, at least if we use only one iteration of the algorithm. 
Therefore, in these cases, no distortion of the integration path is needed.
In this way we are able to be closer to the original idea of the BT-method.
However, as already stressed in the original paper~\cite{Tkachov:1997wh},
this $B$ will vanish at some threshold. We are able to show that this occurs
at some non-leading Landau singularity of the diagram where additional
analytical work is needed before starting the numerical evaluation.

At the same time we have considered the problem of an efficient evaluation
of infrared divergent multi-loop diagrams. Although the massless world is
efficiently treated in QED and QCD within the analytical approach, any
multi-loop calculation in the standard model or beyond is plagued by
infrared divergences and we must contrive a scheme for dealing with them in a 
purely numerical way. For some of the two-loop two-point functions, the 
infrared divergent configurations are simple enough to allow for BT treatment 
with a consequent and straightforward analytical evaluation. The remaining 
topologies, in particular the one containing overlapping infrared 
configurations, require a novel approach.

Our solution is derived by adapting the general algorithm of
ref.~\cite{Binoth:2000ps}: the residues of the infrared poles and the finite
part of a multi-loop diagram can be cast into a form which allows for a
reliable numerical integration.
The algorithm is perfectly defined for all integrands that may vanish only
at the boundaries of the parametric space. Here the strategy differs somehow
from the BT approach. In the latter the main attempt is toward {\em raising}
powers in the integrand while here the powers remain untouched and the whole
idea is about factorizing the singular behavior into simple factors that can
be integrated analytically, leaving non-singular terms to be treated
numerically.

The outline of the paper will be as follows: in \sect{atld} we introduce the
definitions and the basic properties for arbitrary two-loop diagrams.
In \subsect{rtld} we discuss the so-called reducible two-loop diagrams
while, in \sect{sbtr} we present special cases of the general BT functional
relation. In \sect{sfourf} we start the evaluation of two-loop two-point
functions with four internal lines.
In \sect{gonot} a general class of two-loop diagrams, those that contain a
self-energy insertion, is discussed.
Explicit methods for evaluating diagrams with four internal lines are given
in \subsectm{sabameto}{spfpsaba}, while their Landau singularities are
computed in \subsect{lesaba}.
Further refinements for diagrams with four internal lines are discussed in
\subsectm{oisaba}{esabamtf}.
The on-shell derivative of these diagrams and their infrared poles are
analyzed in \subsect{dsabaip}. In \subsect{tisabaf} the tensor integrals
are studied.
The two-loop two-point diagrams with five internal lines are presented and
discussed in \sect{sfivet}. In particular we discuss in \subsect{sacat}
the simplest case in this class, in \subsect{lesacasc} its Landau
singularities and in \subsect{esacasp} its explicit evaluation.
The corresponding on-shell derivative and its infrared poles are shown in
\subsect{dsacaip}.
The most difficult topology with five internal lines is presented in
\subsect{sbabt}, its Landau singularities in \subsect{lesbab}, its
evaluation in \subsectm{esbabmo}{esbabmtwo}. The corresponding on-shell
derivative and its infrared poles are discussed in \subsect{dsbabip} while
in \subsect{itsbabf} we analyze the tensor integrals of this class with a
specific and realistic example shown in \subsect{are}.
Finally, numerical results and comparisons for those few cases where
analytical work has been done are shown and discussed in \sect{numres}.
Technical details are discussed in various Appendices.
%--
\section{Arbitrary two-loop diagrams\label{atld}}
%--
In this section we present basic definitions and properties of diagrams.
An arbitrary two-loop diagram has the following expression:
%--
\bqa
{}&{}&
(\alpha;m_1,\dots,m_{\alpha};\eta_1|\gamma;m_{\alpha+1},\dots,m_{\alpha+\gam
ma};
\eta_{12}|\beta;m_{\alpha+\gamma+1},\dots,m_{\alpha+\gamma+\beta};\eta_2) =
\nl
{}&{}&
\frac{\mu^{2\ep}}{\pi^4}\,
\intmomsii{n}{q_1}{q_2}\,\prod_{i=1}^{\alpha}\,(k^2_i+m^2_i)^{-1}\,
\prod_{j=\alpha+1}^{\alpha+\gamma}\,(k^2_j+m^2_j)^{-1}\,
\prod_{l=\alpha + \gamma+1}^{\alpha+\gamma+\beta}\,(k^2_l+m^2_l)^{-1},
\label{Gdiag}
\eqa
%--
where $n = 4 - \ep$, with $n$ being the space-time dimension, and where
$\alpha, \beta$ and $\gamma$ give the number of lines in the $q_1, q_2$ and
$q_1-q_2$ loops respectively. Furthermore we have
%--
\[
\ba{ll}
k_i = q_1+\sum_{j=1}^{\ssN}\,\eta^1_{ij}\,p_j, \;&\; i=1,\dots,\alpha  \\
k_i = q_1-q_2+\sum_{j=1}^{\ssN}\,\eta^{12}_{ij}\,p_j, \;&\;
i=\alpha+1,\dots,
\alpha+\gamma  \\
k_i = q_2+\sum_{j=1}^{\ssN}\,\eta^2_{ij}\,p_j, \;&\;
i=\alpha+\gamma+1,\dots,
\alpha+\gamma+\beta,
\ea
\]
%--
$N$ being the number of external legs, $\eta^a = \pm 1,$ or $0$ and
$\{p\}$ the set of external momenta. Furthermore, $\tHs$ is the arbitrary
unit of mass. Diagrams that can always be reduced to combinations of other
diagrams with less internal lines will never receive a particular name. 
Otherwise we will denote a two-loop diagram with
%--
\bq
G^{\alpha\beta\gamma},
\eq
%--
where $G = S, V, B\,$ etc. stands for two-, three-, four-point etc,
and $\alpha, \beta$ and $\gamma$ give the number of lines in the $q_1, q_2$
and $q_1-q_2$ loops respectively. Family of diagrams with the same number
$N$ of internal lines will be denoted collectively by $S_{\ssN}$, etc.

Next we recall few basic properties of two-loop
diagrams~\cite{'tHooft:1972fi}. In any two-loop diagram there are three
one-loop sub-diagrams, called $\alpha\gamma, \beta\gamma$ and $\alpha\beta$
sub-diagrams respectively.
\begin{description}

\item[Definition 1:] the $\alpha\beta\gamma$ diagram is overall ultraviolet
(UV)convergent if $\alpha + \beta + \gamma > 4$, logarithmically divergent
if
$\alpha + \beta + \gamma = 4$, linearly etc. divergent if $\alpha + \beta +
\gamma = 3,\dots$.

\item[Definition 2:] the $\alpha\beta$ sub-diagram is convergent if $\alpha
+
\beta > 2$, logarithmically divergent if $\alpha + \beta = 2$, linearly etc.
divergent if $\alpha + \beta = \frac{3}{2},\dots$.

\end{description}

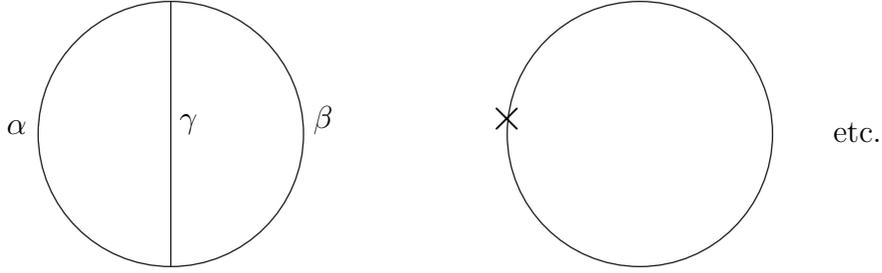
\begin{figure}[th]
\vspace{0.5cm}
\[
  \vcenter{\hbox{
  \begin{picture}(150,0)(0,0)
  \CArc(100,0)(50,0,360)
  \Line(100,50)(100,-50)
  \Text(42,0)[cb]{$\alpha$}
  \Text(158,0)[cb]{$\beta$}
  \Text(107,0)[cb]{$\gamma$}
  \end{picture}}}
\qquad
  \vcenter{\hbox{
  \begin{picture}(150,0)(0,0)
  \CArc(100,0)(50,0,360)
  \Text(50,0)[cb]{{\Large$\times$}}
  \end{picture}}}
\qquad
\mbox{etc.}
\]
\vspace{0.5cm}
\caption[]{The arbitrary two-loop diagram $G_{\ssL}^{\alpha\beta\gamma}$
of \eqn{Gdiag} and one of the associated subtraction sub-diagrams.}
\label{subdia}
\end{figure}

If any of the one-loop sub-diagrams diverge we have counter-terms associated
with them. Therefore, in addition to the $\alpha\beta\gamma$ diagram we will
also consider the subtraction diagrams of \fig{subdia}.
%--
\bq
G^{{\underline\alpha}\beta{\underline\gamma}}, \quad
\mbox{etc.},
\eq
%--
The arbitrary $G^{\alpha\beta\gamma}$ diagram has the following representation
in parametric space:
%--
\bqa
G^{\alpha\beta\gamma}
&=& -\,\lpar\frac{\mu^2}{\pi}\rpar^{\ep}\,\egam{\alpha+\beta+\gamma-4+\ep}\,
\int\,d[x]\,\int\,d[y]\,\Bigl[X(1-X)\Bigr]^{2-\alpha-\gamma-\ep/2}\,
y^{\alpha+\gamma-3+\ep/2}_1
\nl
{}&\times& \Bigl\{ M^2_x\,y_1 +
\sum_{l=\alpha+\gamma+1}^{\alpha+\beta+\gamma}\,
(P^2_{2l}+m^2_l)\,y_{l-\alpha-\gamma+1} -
(P_x\,y_1 + \sum_{l=\alpha+\gamma+1}^{\alpha+\beta+\gamma}\,P_{2l}\,
y_{l-\alpha-\gamma+1})^2\Bigr\}^{4-\alpha-\beta-\gamma-\ep},
\eqa
%--
where the integration measures are
%--
\bqa
d[x] &=& \int_0^1\,\prod_{i=1}^{\alpha+\gamma}\,dx_i\,\delta(
\sum_{i=1}^{\alpha+\gamma}\,x_i-1),
\quad
d[y] = \int_0^1\,\prod_{i=1}^{\beta+1}\,dy_i\,\delta(
\sum_{i=1}^{\beta+1}\,y_i-1),
\eqa
%--
and parameter dependent masses and momenta are defined by
%--
\bqa
M^2_x &=& \frac{R^2 - K^2}{X(1-X)}, \quad P_x = \frac{P_1}{1-X} -
\frac{P_{12}}{X},
\nl
K &=& P_1 + P_{12},
\quad
P_1 = \sum_{i=1}^{\alpha}\,x_i\,P_{1i},
\quad
P_{12} = \sum_{i=\alpha+1}^{\alpha+\gamma}\,x_i\,P_{12i},
\nl
R^2 &=& \sum_{i=1}^{\alpha}\,x_i\,(P^2_{1i}+m^2_i) +
\sum_{i=\alpha+1}^{\alpha+\gamma}\,x_i\,(P^2_{12i}+m^2_i),
\nl
P_{ai} &=& \sum_{k=1}^{\ssN}\,\eta^a_{ik}\,p_k,
\quad
X = \sum_{i=\alpha+1}^{\alpha+\gamma}\,x_i,
\eqa
%--
%--
\subsection{Reducible two-loop diagrams\label{rtld}}
%--
There are two-loop diagrams that can be easily reduced to products of
one-loop diagrams, some of them occurring in the reduction of tensor integrals.
A typical example is represented by
%--
\bq
AB(p^2;m_1,m_2,m_3) = \frac{\mu^{2\ep}}{\pi^4}\,\int\,\frac{d^nq_1\,d^nq_2}
{(q^2_1+m^2_1)(q^2_2+m^2_2)((q_2+p)^2+m^2_3)}
= A_0(m_1)\,\bff{0}{p^2}{m_2}{m_3},
\eq
%--
where $A_0$($B_0$) is the one-loop, one-(two-)point scalar
function~\cite{'tHooft:1979xw}. This function satisfies 
%--
\bq
AB(p^2;m_1,m_2,m_3) = AB(p^2;m_1,m_3,m_2)
\eq
%--
\section{Minimal BT algorithm\label{mbta}}
%--
For an arbitrary diagram the application of minimal BT algorithm requires
distortion of the integration hyper-contour. First of all let us discuss how
this can be done: we rely on the fact that applying BT functional relations 
to $G_{\ssL}$ does not introduce new singularities.
Consider any diagram: the integration hyper-contour can be distorted away from
its original real location, a simplex described by $x_i \ge 0$ and $
\sum x_i \le 1$, and when the possibility of this distortion ceases, we 
encounter a singularity of the function~\cite{elop}. The difficulty in 
locating the singularities lies in imagining what happens in the 
multi-dimensional (complex) space of integration. Consider now a simple example
%--
\bqa
G &=& \intfxy{x}{y}\,V^{-1-\ep}(x,y) = \intfx{y}\,f(y),
\eqa
%--
\bqa
V(x,y) &=& H\,x^2 + 2\,K(y)\,x + L(y).
\eqa
%--
Minimal BT algorithm transforms $G$ into
%--
\bqa
G &=& \intfxy{x}{y}\,\frac{1}{b(y)}\,\Bigl[ 1 + \frac{1}{2} \ln V(x,y)\Bigr]
\nl
{}&-& \frac{1}{2}\,\intfx{y}\, \frac{1}{b(y)}\,\Bigl\{
\Bigl[ 1 + X(y)\Bigr]\,\ln V(1,y) - \Bigl[ y + X(y)\Bigr]\,\ln
V(y,y)\Bigr\},
\label{expex}
\eqa
%--
with $b = L-K^2/H$ and $X = K/H$.
Let $y_0$ be a root of $b$, i.e.\ $b(y_0) = 0$. What happens to $f(y_0)$?
Since $b(y_0) = 0$ we can use another identity
%--
\bq
\Bigl[ 1 + \frac{1}{2\,(1+\ep)}\,(x + X)\,\partial_x\Bigr]\,
V^{-1-\ep}(x,y_0) = 0,
\eq
%--
to derive
%--
\bq
f(y_0) = - \frac{1}{2\,(1+\ep)}\,\int_{y_0}^1\,dx\,(x+X)\,\partial_x\,
V(x,y_0).
\eq
%--
If there is a pinch singularity at $x = x_0$ with $y_0 < x_0 < 1$
then $\{x_0, y_0\}$ is already a singularity of $G$.
%--
If not we integrate by parts and obtain
%--
\bq
(1+2\,\ep)\,f(y_0) = (y_0+X)\,V^{-1-\ep}(y_0,y_0) -
(1+X)\,V^{-1-\ep}(1,y_0).
\eq
%--
If $V(y_0,y_0) = 0$ or $V(1,y_0) = 0$ then $x = 1, y = y_0$ or
$x = y = y_0$ is an end-point singularity of $G$, if not then no new
singularity is introduced since $f(y)$ is analytical in $y = y_0$.

To understand when minimal BT eventually fails we consider another simple
example which, however, is general enough to complete our discussion.
We start with a quadratic form
%--
\bqa
V(x,y) &=& H\,x^2 + 2\,K(y)\,x + L(y),
\eqa
%--
\bqa
H &=& m^2_d, \quad K = \frac{1}{2}\,(s-m^2_u-m^2_d)\,y - \frac{1}{2}\,M^2_z,
\quad
L = m^2_u\,y^2 + (M^2_w - s)\,y + M^2_z.
\eqa
%--
This is actually a $C_0$ function corresponding to a vertex correction for
$\wb\,$(off-shell)\,$ \to \baru d$. Applying minimal BT gives a $b(y)$ which
is a quadratic form in $y$. The two roots $b(y_{\pm}) = 0$ are coincident
when
%--
\bq
s = \frac{1}{2\,m^2_d}\,\Bigl[
 m^2_u M^2_z - M^2_w M^2_z + 2\, m^2_d M^2_w + m^2_d M^2_z
\pm
\lambda^{1/2}(M^2_z,m^2_d,m^2_d)\,\lambda^{1/2}(m^2_u,M^2_w,m^2_d)\Bigr],
\label{eqfs}
\eq
%--
which is the well-known anomalous threshold. Here $\lambda$ is the usual
K\"allen-function~\cite{Kallen:1972pu}. Let $y_{\ssL,\ssR}$ the solutions
of $V = 0$, which are now branch points in \eqn{expex}. These roots are
coincident for
%--
\bq
x = x_{\rm th} = \Bigl( m^2_u\,M^2_w - 2\,m^2_u\,M^2_z -
m^2_u\,s + m^2_d\,M^2_w - m^2_d\,s - M^2_w\,s + s^2\Bigr)\,
\lambda^{-1}(s,m^2_u,m^2_d),
\label{eqfx}
\eq
%--
and one can easily verify that with $x$ and $s$ given by \eqn{eqfx} and
\eqn{eqfs} respectively we have
%--
\bq
y_{\pm} = y_{\ssL} = y_{\ssR}.
\eq
%--
In this case, if $x_{\rm th} \in [0,1]$ and $y_{\pm} \in [0,1]$ we cannot
distort anymore the $y$ integration contour to avoid
$y = y_{\pm}$ without crossing a cut of the logarithm. However we can still
go on with our procedure if we distort the $x$ integration contour in order
to avoid $x = x_{\rm th}$. This will be possible unless, for some specific
configuration of the external parameters, we have $x_{\rm th} = 0$ or
$x_{\rm th} = 1$. In our particular example this would require an
unrealistic relation between $\wb,\zb$ and up, down masses. Even in this
case we still have a chance of continuing the derivation. Indeed we have used
logarithms of quadratic forms which give a pair of branch points which can
pinch the integration contour exactly where we would like to deform it.
However, when only logarithms are present and not logarithms squared, we
split the logarithm as
%--
\bq
\ln m^2_u + \eta(y - y_{\ssL},y - y_{\ssR}) +
\sum_{i=\ssL,\ssR}\,\ln(y-y_i),
\eq
%--
where $\eta$ is the usual Veltman's function,
$\eta(a,b) = \ln(ab) - \ln a - \ln b$,
and logarithms are defined with a cut along the negative real axis.
The final expression for the diagram will be the sum of terms proportional
to $\eta$-functions and terms of the form $\ln(y-y_i)/b(y)$.
The former do not create a problem in distorting, when needed, the
integration contour.
For the latter, however, if one of the roots of $b$, or both, is
real and internal to the interval $[0,1]$ then the distortion for the
corresponding term must be examined with care. If the imaginary part of
$y_i$ is positive(negative) then we move the contour into the
negative(positive) imaginary half-plane, so that the cut of the logarithm
will never be crossed. If $y_i$ is real and negative no problem will arise
and we can distort in any of the two ways. If its is real and positive then
the distortion is fixed according to the $i\,\delta$ prescription.
The reason why we cannot apply this argument to cases where we have
the presence of $\ln^2$-terms is simple: after the splitting of
the logarithm we will encounter terms proportional to $\ln(y-y_{\ssL})
\ln(y-y_{\ssR})$. When the roots are complex conjugated we will have
cuts both in the positive and in the negative imaginary half-planes and
the possibility of distorting the contour ceases when they approach the
real axis.
Note that these quadratic terms are always present for ultraviolet divergent
diagrams.
%--
\section{Special BT relations\label{sbtr}}
%--
In this section we present some explicit examples of the Bernstein-Tkachov
functional relations. Let $V(x)$ be a quadratic form of a vector $x$,
%--
\bq
V(x) = x^t\,H\,x + 2\,K^t\,x + L,
\eq
%--
then the solution to the problem of determining the polynomial ${\cal P}$ is
as given in \eqn{rol}.
There are special BT functional relations: typically, when we have a
quadratic
in two variables $y,z$ of the form $V = h z^2 + 2\,k_1 z + 2\,k_2 y + l$
then the matrix $H$ is singular. Here we use
%--
\bq
\Bigl[ 1 - \frac{y}{\mu+1}\,\partial_y - \frac{1}{2\,(\mu+1)}\,\lpar
z + \frac{k_1}{h}\rpar\,\partial_z\Bigr]\,V^{\mu+1} =
\frac{h l - k^2_1}{h}\,V^{\mu}.
\label{hsing}
\eq
%--
For a quadratic in one variable, $V = h y^2 + 2\,k y + l$ we use
%--
\bq
\Bigl[ 1 - \frac{1}{2\,(\mu+1)}\,\lpar y + \frac{k}{h}\rpar\,\frac{d}{dy}
\Bigr]\,V^{\mu+1} = \frac{hl - k^2}{h}\,V^{\mu},
\eq
%--
while for $V = k y + l$ we use
%--
\bq
\Bigl[ 1 - \frac{1}{\mu+1}\,y\,\frac{d}{dy}\Bigr]\,V^{\mu+1} = l\,V^{\mu},
\quad
\Bigl[ 1 + \frac{1}{\mu+1}\,\frac{1 - V}{k}\,\frac{d}{dy}
\Bigr]\,V^{\mu+1} = V^{\mu}.
\eq
%--
\section{The $S_{4}$ family\label{sfourf}}
\begin{figure}[th]
\vspace{0.5cm}
\[
  \vcenter{\hbox{
  \begin{picture}(150,0)(0,0)
  \Line(0,0)(50,0)
  \ArrowArc(75,0)(25,0,90)
  \ArrowArc(75,0)(25,-180,0)
  \ArrowArc(75,0)(25,90,180)
  \ArrowArc(50,25)(25,-90,0)
  \Line(100,0)(150,0)
  \Text(50,25)[cb]{$\scriptstyle q_1$}
  \Text(100,25)[cb]{$\scriptstyle q_2$}
  \Text(75,-35)[cb]{$\scriptstyle q_2+p$}
  \Text(80,-5)[cb]{$\scriptstyle q_1-q_2$}
  \end{picture}}}
\]
\vspace{0.5cm}
\caption[]{The two-loop diagram $S^{121}$ of \eqn{defsaba}. Arrows indicate
the momentum flow.}
\label{tops4}
\end{figure}
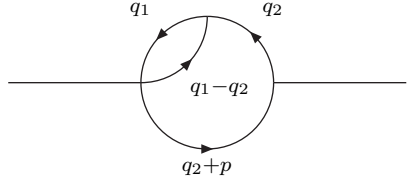

\begin{figure}[th]
\vspace{0.5cm}
\[
  \vcenter{\hbox{
  \begin{picture}(150,0)(0,0)
  \Line(50,0)(100,0)
  \CArc(75,0)(25,0,90)
  \CArc(75,0)(25,-180,0)
  \CArc(75,0)(25,90,180)
  \Line(0,-25)(150,-25)
  \end{picture}}}
\]
\vspace{0.5cm}
\caption[]{The second diagram of the $S_4$-family, \eqn{interms}, which is
evaluated in terms of $S^{111}$ (ref.~\cite{Passarino:2001wv}) or
$S_{33}$ diagrams (\eqn{defstt}) at zero external momentum.}
\label{topsp4}
\end{figure}
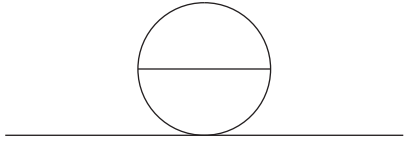

%--
There are two diagrams in the $S_4$-family, both with $\alpha = \gamma = 1$
and with $\beta = 2$. The first is $S^{121}$ and is given in \fig{tops4}.
It is overall logarithmically divergent and the $\alpha\gamma$ sub-diagram
is also logarithmically divergent.
We have another diagram with the same number of internal lines, given in
\fig{topsp4}.
%--
%--
This second diagram is given by the following integral:
%--
\bqa
(1;m_1|1;m_2|2,m_3,m_4) &=&
\frac{\mu^{2\ep}}{\pi^4}\,\int\,
{{d^nq_1d^nq_2}\over {
\lpar q^2_1+m^2_1\rpar\,
\lpar\,\lpar q_1-q_2\rpar^2+m^2_2\rpar\,
\lpar q^2_2 + m^2_3\rpar\,
\lpar q^2_2 + m^2_4\rpar}}.
\eqa
%--
Therefore, this case is trivially reduced to known diagrams. If $m_3 \ne
m_4$ then 
%--
\bq
(1;m_1|1;m_2|2,m_3,m_4) =
\frac{1}{m^2_4-m^2_3}\,\Bigl[ S^{111}(0;m_1,m_2,m_3) -
S^{111}(0;m_1,m_2,m_4)
\Bigr],
\label{interms}
\eq
%--
\bq
\pi^4\,S^{111} = \mu^{2\ep}\,\int\,\frac{d^nq_1\,d^nq_2}
{\lpar q^2_1+m^2_1\rpar\,\,\lpar\,\lpar q_1-q_2+p\rpar^2+m^2_2\rpar\,
\lpar q^2_2 + m^2_3\rpar}.
\eq
%--
otherwise we will have
%--
\bq
(1;m_1|1;m_2|2,m_3,m_4) = (\ep - 1)\,S_{33}(0;m_1,m_2,m_3 = m_4),
\eq
%--
where $S_{33}$ has been introduced in ref.~\cite{Passarino:2001wv} as
the following integral:
%--
\bqa
\pi^4\,S_{33} &=& \frac{\mu^{2\ep}}{\ep-1}\,\int\,\frac{d^nq_1\,d^nq_2}
{\lpar q^2_1+m^2_1\rpar\,
\lpar\,\lpar q_1-q_2+p\rpar^2+m^2_2\rpar\,
\lpar q^2_2+m^2_3\rpar^2},
\label{defstt}
\eqa
%--
\vskip 5pt
%--
The $S^{121}$ topology is given in terms of the following integral:
%--
\bqa
\pi^4\,S^{121} &=& \mu^{2\ep}\,\int\,\frac{d^nq_1\,d^nq_2}
{\lpar q^2_1+m^2_1\rpar\,
\lpar\,\lpar q_1-q_2\rpar^2+m^2_2\rpar\,
\lpar q^2_2 + m^2_3\rpar\,
\lpar\lpar q_2 + p\rpar^2 + m^2_4\rpar}.
\label{defsaba}
\eqa
%--
One way to evaluate $S^{121}$ starts with the introduction of a Feynman
parameter $x$ for the $q_1$-loop; successively, we integrate over $q_1$ and,
finally, combine the $q_2$-dependent propagators with parameters $y,z$.
After performing the $q_2$-integration we obtain
%--
\bqa
S^{121} &=& -\,2\,\lpar\frac{\tHss}{\pi\,s}\rpar^{\ep}\,\egam{\ep}\,
\intfx{x}\intfxy{y}{z}\,\Bigl[ x(1-x)\Bigr]^{-\ep/2}\,\lpar 1-y\rpar^{\ep/2}
\nl
{}&\times& \Bigl[ \lpar \mu^2_x - z^2\rpar\,\chi_{\aba}^{-1-\ep} +
\frac{2-\ep/2}{\ep}\,\chi_{\aba}^{-\ep}\Bigr],
\label{defs4}
\eqa
%--
where the quadratic form $\chi_{\aba}(x,y,z)$ is defined by
%--
\bq
\chi_{\aba}(x,y,z) =  z^2 - \mu^2_{34}\,z + (\mu^2_3 - \mu^2_x)\,y +
\mu^2_x,
\eq
%--
and where we have introduced additional auxiliary quantities,
%--
\bq
\mu^2_x = \frac{ (1-x) \mu^2_1 + x \mu^2_2}{x(1-x)} =
\frac{\mu^2_{12}}{x(1-x)}, \quad
\mu^2_{34} = 1 + \mu^2_3 - \mu^2_4.
\label{defchi4}
\eq
%--
In \eqn{defchi4} we have assumed that $p^2= -s$ and $s > 0$. Scaled masses
are, therefore, defined to be $m^2= \mu^2\,s$. If, instead, $p^2$ is positive,
we will have to change the procedure as described in the next subsection.
%--
\subsubsection{The case $p^2= -t$ with $t < 0$}
%--
Starting from the quadratic form
%--
\bq
U = (y-z)\,m^2_3 + z\,m^2_4 + (1-y)\,m^2_x + z(1-z)\,p^2,
\eq
%--
with $p^2 = -t$ and $t$ negative we redefine scaled masses to be
$m^2= -\mu^2\,t$ and obtain $U = -\,t\,\chi$,
%--
\bq
\chi = - z^2 + \mu^2_{43}\,z + (\mu^2_3-\mu^2_x)\,y + \mu^2_x,
\quad \mbox{with} \quad \mu^2_{43} = 1 + \mu^2_4 - \mu^2_3.
\eq
%--
The signs in this quadratic form will have an effect on the {\em raising}
procedure, as described below in \subsect{sabameto}.
%--
\subsection{Evaluation of $S^{121}$: method I\label{sabameto}}
%--
In the following sections we will be able to present several derivations
for $S^{121}$ and to compare the corresponding numerical results.
In order to evaluate $S^{121}$ we may use the following relation:
%--
\bq
\chi_{\aba}^{-1-\ep}(x,y,z) = - \,\frac{4}{b_{\aba}}\,\Bigl[
1 + \frac{1}{\ep}\,(P_y \partial_y + P_z \partial_z)\Bigr]
\chi_{\aba}^{-\ep}(x,y,z),
\eq
%--
with polynomials $P_{y,z}$ given by
%--
\bq
P_y = y,  \qquad
P_z = \frac{1}{2}\,(z - \frac{1}{2}\,\mu^2_{34}),  \qquad
b_{\aba} = \mu^4_{34} - 4\,\mu^2_x.
\label{l04}
\eq
%--
If $p^2 = -t$ and $t$ is negative, the definition of $b_{\aba}$ will change
accordingly,
%--
\bq
b_{\aba} \to \mu^4_{43} + 4\,\mu^2_x > 0,
\eq
%--
so that $b_{\aba}$ is never zero in the $t$-channel, as expected.

We will also use an additional quadratic form, $X_{\aba}$, which is related
to $\chi_{\aba}$ (\eqn{defchi4}), by
%--
\bqa
X_{\aba}(x,y,z) &=& x(1-x)\,\chi_{\aba}(x,y,z)  \nl
{}&=&  x (1-x) \Bigl[ - z (1-z) + \mu^2_4 z + \mu^2_3 (y-z)\Bigr] +
                \Bigl[(1-x) \mu^2_1 + x \mu^2_2\Bigr] (1-y),
\label{defxi4}
\eqa
%--
The next ingredient needed in the evaluation of the diagram is integration
by parts which is performed with the help of the following relations:
%--
\bqa
\intfz\,z^n\partial_z\,\ln^m X_{\aba}(x,y,z) &=&
 y^n\,\ln^m X_{\aba}(x,y,y) - \delta_{0,n}\,\ln^m X_{\aba}(x,y,0)  \nl
{}&-& n\,\intfz\,z^{n-1}\,\ln^m X_{\aba}(x,y,z),
\eqa
%--
\bqa
{}&{}&
\intfy\,y^n (1-y)^{\ep/2}\,\partial_y\,\ln^m X_{\aba}(x,y,z) =
 - z^n\,(1-z)^{\ep/2}\,\ln^m( X_{\aba}(x,z,z))  \nl
{}&-& \intfy\,\Bigl[ n y^{n-1} (1-y)^{\ep/2} - \frac{\ep}{2}\,y^n
(1-y)^{\ep/2-1}\Bigr]\,\ln^m X_{\aba}(x,y,z).
\eqa
%--
Furthermore, we will use
%--
\bqa
{}&{}&
\intfy\, y^n (1-y)^{\ep/2-1}\,\ln^m X_{\aba}(x,y,z) =
\frac{2}{\ep}\,(1-z)^{\ep/2}\,\ln^m X_{\aba}(x,1,z)  \nl
{}&+& \intfy\,(1-y)^{\ep/2-1}\,\Bigl[ y^n\ln^m X_{\aba}(x,y,z) -
\ln^m X_{\aba}(x,1,z)\Bigr].
\eqa
%--
Before discussing the finite part of $S^{121}$ we consider its ultraviolet
poles. As usual, the pole parts can be computed analytically.
The double pole of $S^{121}$ is easily derived and the corresponding
expression, where in the spirit of the $\MSB$ scheme we include a bunch of 
constants, reads as follows:
%--
\bq
S^{121}_{\rm double-pole} = - \frac{2}{\ep^2} - \Delta^2_{\ssU\ssV},
\eq
%--
where, for the constants, we obtain
%--
\bq
\Delta_{\ssU\ssV} = \gamma + \ln\pi + \ln\frac{s}{\tHss}.
\eq
%--
Here, $\gamma= 0.577216$ is the Euler's constant and $\tHs$ is the arbitrary
unit of mass. The single-pole requires some extra work to be discussed in 
\subsect{spfpsaba}.
%--
\subsection{Single-pole and finite part of $S^{121}$\label{spfpsaba}}
%--
For computing the single ultraviolet pole of $S^{121}$ we use the relation
%--
\bq
\frac{\mu^2_x-z^2}{\chi_{\aba}} = 1 - (z\partial_z + y\partial_y)\,\ln
\chi_{\aba},
\label{basicaba}
\eq
%--
and derive the following result:
%--
\bq
S^{121}_{\rm single-pole} = \frac{R_4}{\ep} + \mbox{constants},
\eq
%--
where we also include a bunch of constant into the definition and where
%--
\bq
R_4 = 2\,(\gamma + \ln\pi + \ln\frac{s}{\tHss}) - 1 +
2\,\intfxx{x}{y}\,\ln\chi_{\aba}(x,1,y).
\eq
%--
Since $\chi_{\aba}(x,1,y) = y^2 - \mu^2_{34}\,y + \mu^2_3$, we find
%--
\bq
R_4 = - 2\,B^{\rm fin}_0(-s;m_3,m_4),
\eq
%--
which is the finite part of the one-loop two-point function corresponding to
internal masses $m_3$ and $m_4$. This is exactly the result expected, from
general arguments, to compensate the overlapping divergency coming from the
associated subtraction diagram of \fig{counters4}. This result proves
cancellation of poles with logarithmic residues, leaving only harmless
poles, i.e. \ those with as residue a polynomial of finite order in the 
external momenta.
%--
\begin{figure}[th]
\vspace{0.5cm}
\[
  \vcenter{\hbox{
  \begin{picture}(150,0)(0,0)
  \Line(0,0)(50,0)
  \CArc(75,0)(25,0,90)
  \CArc(75,0)(25,-180,0)
  \CArc(75,0)(25,90,180)
  \Line(100,0)(150,0)
  \Text(50,-5)[cb]{{\Large$\times$}}
  \end{picture}}}
\]
\vspace{0.5cm}
\caption[]{The subtraction diagram, containing a one-loop counter-term
(represented by a $\times$) associated with the two-loop diagram
$S^{121}$ of \fig{tops4}.}\label{counters4}
\end{figure}
%--
Summarizing our findings for the ultraviolet divergent part of $S^{121}$, we
obtain
%--
\bq
S^{121}_{\rm single-pole} = \frac{1}{\ep}\,(2\,\Delta_{\ssU\ssV} - 1) +
\Delta_{\ssU\ssV} + 2\,(\frac{1}{\ep} - \Delta_{\ssU\ssV})\,\intfx{y}\,
\ln\chi_{\aba}(x,1,y),
\eq
%--
where one should note that $\chi_{\aba}(x,1,y)$ is $x$-independent.

For the finite part of $S^{121}$ we have to consider the following quadratic
forms that arise after integration by parts:
%--
\bqa
\chi_{\aba}(x,y,z) &=& (y-z) \mu^2_3 + z \mu^2_4 + (1-y) \mu^2_x - z (1-z),
\nl
\chi_{\aba}(x,y,y) &=& y \mu^2_4 + (1-y) \mu^2_x - y (1-y),  \nl
\chi_{\aba}(x,y,0) &=& y \mu^2_3 + (1-y) \mu^2_x,  \nl
\chi_{\aba}(x,1,y) &=& (1-y) \mu^2_3 + y \mu^2_4 - y (1-y).
\eqa
%--
Before continuing the evaluation of the finite part of $S^{121}$, however,
we need to understand its analytical structure. The tools for extracting all
the singularities of the diagrams are, as usual, represented by the
corresponding set of Landau equations.
%--
\subsection{Landau equations for $S^{121}$\label{lesaba}}
%--
Before starting the evaluation of this topology it is important to
know as much as possible about its singularities as a function of $p^2$ and
of the internal masses. The corresponding Landau equations are given by
%--
\[
\ba{ll}
\alpha_1\,(q^2_1+m^2_1) = 0 & \;\;
\alpha_2\,((q_1-q_2)^2+m^2_2) = 0, \\
\alpha_3\,(q^2_2+m^2_3) = 0 & \;\;
\alpha_4\,((q^2_2+p)^2+m^2_4) = 0,
\label{land1}
\ea
\]
%--
and also by
%--
\bq
\alpha_1 q_{1\mu} + \alpha_2 (q_1-q_2)_{\mu} = 0, \qquad
- \alpha_2 (q_1-q_2)_{\mu} + \alpha_3 q_{2\mu} + \alpha_4\,(q_2+p)_{\mu} =
0.
\label{land2}
\eq
%--
The leading Landau singularity occurs for $\alpha_i \ne 0, \forall i$.
We multiply the two equations \eqn{land2} by $q_{1\mu}, q_{2\mu}$ and
$p_{\mu}$ respectively. This gives an homogeneous system of six equations.
If all $\alpha_i$ are different from zero, the singularity will occur for
%--
\[
\ba{ll}
q^2_1 = - m^2_1 & \;\; q^2_2 = - m^2_3, \\
\spro{q_1}{q_2} = \frac{1}{2}\,(m^2_2 - m^2_1 - m^2_3) & \;\;
\spro{p}{q_2} = \frac{1}{2}\,( s + m^2_3 - m^2_4).
\ea
\]
%--
The Landau equations become as follows:
%--
\bqa
{}&{}& - 2\,m^2_1\,\alpha_1 + (m^2_3 - m^2_1 - m^2_2)\,\alpha_2 = 0,
\nl
{}&{}& - (m^2_3 + m^2_1 - m^2_2)\,\alpha_1
+  (m^2_3 - m^2_1 + m^2_2)\,\alpha_2 = 0,
\nl
{}&{}& \spro{p}{q_1}\,\alpha_1 + \Bigl[ \spro{p}{q_1} + \frac{1}{2}\,
(m^2_4 - m^2_3 - s)\Bigr]\,\alpha_2 = 0,
\nl
{}&{}& (m^2_1 + m^2_2 - m^2_3)\,\alpha_2 +
(m^2_2 - m^2_1 - m^2_3)\,\alpha_3 +
\Bigl[ 2\,\spro{p}{q_1} + m^2_2 - m^2_1 - m^2_3\Bigr]\,\alpha_4 = 0,
\nl
{}&{}& (m^2_1 - m^2_2 - m^2_3)\,\alpha_2 -
2\,m^2_3\,\alpha_3 + (s - m^2_3 - m^2_4)\,\alpha_4 = 0,
\nl
{}&{}& \Bigl[ s + m^2_3 - m^2_4 - 2\,\spro{p}{q_1}\Bigr]\,
\alpha_2 + (s + m^2_3 - m^2_4)\,\alpha_3 -
(s - m^2_3 + m^2_4)\,\alpha_4 = 0.
\eqa
%--
There are two compatibility conditions for the first three equations that
can be used to derive
%--
\bq
\spro{p}{q_1} = \frac{m_1}{2\,(m_1+m_2)}\,\Bigl[
 s + (m_1 + m_2)^2 - m^2_4\Bigr].
\eq
%--
Furthermore we must have a relation among masses, i.e.
%--
\bq
m^2_3 = (m_1+m_2)^2.
\eq
%--
From these considerations we derive that
%--
\bq
\alpha_1 = \frac{m_2}{m_1}\,\alpha_2,  \quad
\alpha_2 = \frac{1}{m_2\,(m_1+m_2)}\,\Bigl\{
- (m_1+m_2)^2\,\alpha_3 + \frac{1}{2}\,
\Bigl[ s - (m_1+m_2)^2 - m^2_4\Bigr]\,\alpha_4\Bigr\},
\eq
%--
is a solution for the system if the following condition is satisfied:
%--
\bq
s = ( m_1 + m_2 \pm m_4)^2.
\eq
%--
Note that we have obtained the leading Landau singularity, i.e.\ all
$\alpha_i \ne 0$, as
%--
\bq
s = ( m_1 + m_2 \pm m_4)^2, \qquad m^2_3 = (m_1+m_2)^2.
\eq
%--
Naturally we have that
%--
\bq
s = ( m_1 + m_2 \pm m_4)^2, \qquad \alpha_3 = 0,
\eq
%--
are sub-leading Landau singularities corresponding to the normal and pseudo
thresholds of the reduced diagram where the line corresponding to the
propagator $q^2_2 + m^2_3$ is contracted to a point. Similarly,
%--
\bq
s = ( m_3 \pm m_4)^2, \qquad \alpha_1 = \alpha_2 = 0,
\eq
%--
are the sub-leading ones corresponding to the reduced diagram where the
$q_1$ sub-diagram is shrunk to a point.
%--
\subsection{One iteration for $S^{121}$\label{oisaba}}
%--
Having discussed the analytical properties of the diagram we continue the
derivation of an integral representation which is particularly suited for
numerical treatment.
After one iteration of the {\em raising} procedure the result for $S^{121}$
is still very compact. First we introduce a combination of logarithms,
%--
\bq
L_{\pm}(x,y,z) = \mp \ln X(x,y,z) \pm \Bigl[
\ln(1-y) + \ln x + \ln(1-x)\Bigr]
\eq
%--
and obtain for the finite part of $S^{121}$
%--
\bqa
S^{121}_{\rm fin} &=& \frac{1}{B_{\aba}}\,\intfx{x}\,\Bigl[
\intfxy{y}{z}\,I^{\aba}_3 + \intfx{y}\,I^{\aba}_2 +
I^{\aba}_1\Bigr]
\nl
{}&+& 2\,\intfx{x}\intfxy{y}{z}\,\ln X_{\aba}(x,y,z)\,
\Bigl[ L_+(x,y,z) - \frac{1}{2}\Bigr] - \frac{21}{4} -
\frac{1}{2}\,\zeta(2).
\label{varint}
\eqa
%--
where $B_{\aba}$ follows from \eqn{l04} and becomes
%--
\bq
B_{\aba} = x(1-x)\,b_{\aba} = x(1-x)\,\mu^4_{34} - 4\,\mu^2_{12},
\eq
%--
with $\mu^2_{12}$ and $\mu^2_{34}$ defined in \eqn{defchi4}.
and the various integrands appearing inside \eqn{varint} are as follows:
%--
\bqa
I^{\aba}_3 &=& -\,\ln X_{\aba}(x,y,z)\,\Bigl\{ 4\,(\mu^2_{12} - X\,z^2) -
2\,L_+(x,y,z)\,
\Bigl[ 3\,\mu^2_{12} - X z\,(5\,z - \mu^2_{34})\Bigr]\Bigr\}  \nl
{}&-& 4\,(X\,z^2 - \mu^2_{12})\,\Bigl(\frac{\ln X_{\aba}(x,y,z)}{y-1}
\Bigr)_+
\label{isa}
\eqa
%--
\bqa
I^{\aba}_2 &=& -\,4\,(X\,y^2 - \mu^2_{12})\,\ln
X_{\aba}(x,1,y)\,L_-(x,1,y) -
( \mu^2_{34} + 2\,y)\,(X\,y^2 - \mu^2_{12})  \nl
{}&\times& \ln X_{\aba}(x,y,y)\,L_+(x,y,y) - \mu^2_{12}\mu^2_{34}\,
\ln X_{\aba}(x,y,0)\,L_+(x,y,0),
\label{isb}
\eqa
%--
\bqa
I^{\aba}_1 &=& -\ln X\,(\frac{1}{3}\,X - 2\,\mu^2_{12}) - 3\,\mu^2_{12} +
\frac{25}{36}\,X.
\label{isc}
\eqa
%--
Here $\zeta(z)$ is the Riemann zeta-function~\cite{bat}, $X = x(1-x)$ and
two additional quantities have been introduced:
%--
\bq
\mu^2_{12} = (1-x)\,\mu^2_1 + x\,\mu^2_2, \qquad
\mu^2_{34} = 1 + \mu^2_3 - \mu^2_4.
\eq
%--
The `+'-distribution is defined, as usual, by its action on a generic test
function $f(x)$:
%--
\bq
\intfx{x}\,\Bigl(\frac{f(x)}{x-1}\Bigr)_+ = \intfx{x}\,
\frac{f(x) - f(1)}{x-1}.
\label{plusd}
\eq
%--
If we stop the chain of iterations at this level we will have the advantage
of a result that contains only one denominator, namely $b_{\aba}$. A second
iteration, described in the \sect{twoitsaba}, makes the total integrand
smoother at the price of introducing additional denominators.
If $p^2 = -t$ with $t$ negative we define scaled masses as $m^2= -\mu^2 t$
and derive
%--
\bqa
I^{\aba}_3 &=& \Bigl\{ - 4\,(Xz^2+\mu^2_{12}) +
                 2\,\Bigl[ X(5\,z^2 - \mu^2_{43} z)  + 3\,\mu^2_{12}\Bigr]\,
L_+(x,y,z)\Bigr\}\,\ln X_{\aba}(x,y,z)  \nl
{}&+& 4\,(X z^2 + \mu^2_{12})\,\lpar\frac{\ln X_{\aba}(x,y,z)}{y-1}\rpar_+,
\nl
I^{\aba}_2 &=& ( X y^2 + \mu^2_{12})\,(\mu^2_{43} + 2\,y)\,\ln
X_{\aba}(x,y,y)\,
L_+(x,y,y) - \mu^2_{43}\,\mu^2_{12}\,\ln X_{\aba}(x,y,0)\,L_+(x,y,0)
\nl
{}&+& 4\,(X y^2 + \mu^2_{12})\,\ln X_{\aba}(x,1,y)\,L_-(x,1,y),
\nl
I^{\aba}_1 &=& (\frac{1}{3}\,X + 2\,\mu^2_{12})\,\ln X -
\frac{25}{36}\,X - 3\,\mu^2_{12},
\eqa
%--
instead of \eqns{isa}{isc}. As usual, the iterative procedure of {\em
raising} powers in the integrand produces apparent singularities that make 
the final result unstable in known regions of the $x$-integration. In the next
subsection we will explain how to cure these instabilities.
%--
\subsection{Distorting the integration contour for $S^{121}$\label{dicsaba}}
%--
Obviously, there are numerical instabilities when $x^0_{\pm}$, the zeros of
$b_{\aba}$ (\eqn{l04}), are real and internal to the integration domain. This 
can be avoided by distorting the $x$ integration contour in \eqn{varint} into 
the complex plane. However, this possibility ceases when $x^0_{\ssL,\ssR}$, 
the roots of the quadratic form $X_{\aba}(x,y,z)$ (\eqn{defxi4}) pinch the 
real $x$-axis.
Given that $X_{\aba}(x,y,z) = a x^2 + b x + c$, this situation will occur
when we simultaneously have
%--
\bq
x^0_- = x^0_+ = - \frac{b}{2\,a} \qquad \mbox{and} \qquad b^2 = 4\,ac.
\eq
%--
The condition for coincidence, i.e. \ $x^0_- = x^0_+ \in [0,1]$, is
%--
\bq
\mu^2_{34} = 2\,(\mu_1 + \mu_2).
\eq
%--
The remaining two conditions, namely $\Imb x^0_{\ssL,\ssR} = 0$ and
$x^0_{\ssL} = x^0_{\ssR} =  x^0_- = x^0_+$, require
%--
\bq
( \mu^2_3-\mu^2_+) y + ( z - \mu_+)^2 = 0,
\eq
%--
where $\mu_+ = \mu_1 + \mu_2$. There are, therefore, two solutions for $z$,
given by
%--
\bq
z = \mu_+ \pm \Bigl[ (\mu^2_+ - \mu^2_3)\,y\Bigr]^{1/2},
\eq
%--
and, as long as we can distort the $z$ integration contour to avoid these
points then the $x$ contour will not be pinched. This is always possible 
unless a new pinch will occur, which indeed is the case if $\mu_3 = \mu_+$.
Note that $y = 0$ and $z = \mu_+$ is outside the physical region where
$0 \le z \le y$. By inserting $\mu_3 = \mu_+$ inside the relation
$\mu^2_{34} = 2\,(\mu_1 + \mu_2)$ we obtain $\mu_+ \pm \mu_4 = 1$ and the
two correspond to the leading Landau singularity.

There are also logarithms of quadratic forms in two variables, $x$ and $y$.
For $X_{\aba}(x,y,y)$ the $x$-contour is pinched for
%--
\bq
y = \mu_+ + \frac{1}{2}\,\Bigl\{ \mu^2_+ - \mu^2_3 \pm \Bigl[
(\mu^2_3 - \mu^2_+)\,(\mu^2_3 - \mu^2_+ - 4\,\mu_+)\Bigr]^{1/2}\Bigr\}.
\eq
%--
Therefore, the $y$ contour will be pinched if $\mu^2_3 = \mu^2_+$ or
$\mu^2_3 = \mu^2_+ + 4\,\mu_+$. Only the former is physical,
corresponding to $y = \mu_+$, while the latter corresponds to $y= - \mu_+$.

For $X_{\aba}(x,y,0)$ the $x$-pinching occurs for $y = - \mu^2_+/(\mu^2_3 -
\mu^2_+)$ so that it is enough to distort the $y$-integration contour
avoiding this point. Finally, for $X_{\aba}(x,1,y)$ the solution is
%--
\bq
y = \mu_+ \pm (\mu^2_+ - \mu^2_3)^{1/2},
\eq
%--
showing, again a pinch for $\mu_3 = \mu_+$. Around these points we must
introduce an alternative derivation.
%--
\subsection{Around the leading Landau singularity: method II\label{allsmt}}
%--
When we are in the regions of external parameters that correspond to the
leading Landau singularity of $S^{121}$, an alternative approach must be
used. We easily derive the following relation:
%--
\bq
S^{121} = (\ep - 1)\,\intfx{x}\,S_{33}\lpar x^2 p^2;m_1,m_2,M_x\rpar,  \quad
\mbox{with} \quad
M^2_x = - p^2 x^2 + (p^2 + m^2_4 - m^2_3) x + m^2_3,
\label{sabamii}
\eq
%--
where $S_{33}$ (\eqn{defstt}) has been introduced in
ref.~\cite{Passarino:2001wv}. The finite part of $S_{33}$ is
%--
\bqa
(\ep - 1)\,S^{\rm fin}_{33}\lpar x^2 p^2;m_1,m_2,M_x\rpar &=&
-\,\intfxx{y}{z}\,\Bigl[\ln\xi(y,z) +
\lpar \frac{\ln\xi(y,z)}{z-1}\rpar_+\Bigr]  \nl
{}&+& \ln \frac{M^2_x}{s}\,\lpar 2 - \ln\frac{M^2_x}{s}\rpar -
\frac{7}{2} - \frac{1}{2}\,\zeta(2),
\eqa
%--
and $\xi$ is given by
%--
\bqa
\xi(y,z;\mu_1,\mu_2,\mu_3) &=& z y (y-z)\,\Bigl[ (1 - z)\,x^2 - 
\mu^2_3\Bigr] + (1-z)\,\Bigl[ ( \mu^2_2 - \mu^2_1)\,y + \mu^2_1\Bigr],
\label{quad}
\eqa
%--
and must be evaluated at $\mu^2_3 = M^2_x/s$.
%--
\subsection{Two iterations for $S^{121}$\label{twoitsaba}}
%--
In this section we return to the method described in \sect{sabameto}; an
additional iteration of the procedure requires the consideration of new
quadratic forms in two or one variables. In particular, when we apply
integration by parts, some care is needed in dealing with terms containing
$\ln (1-y)$. For instance, we will use
%--
\bqa
\intfx{y} \intfz \ln\chi_{\aba}(x,y,z) \ln(1-y) &=&
   \intfx{z} \intfy \ln\chi_{\aba}(x,y,z) \ln(1-y),  \nl
\intfx{z} \intfy y^n \ln\chi_{\aba}(x,y,z) \ln(1-y) &=&
    \intfx{z} \intfy y^n \ln\frac{\chi_{\aba}(x,y,z)}{\chi_{\aba}(x,1,y)}\,
\ln(1-y) \nl
{}&+&  \intfx{z} \ln\chi_{\aba}(x,1,z) \int_{\scriptstyle 0}^{1-z}\,dy
(1-y)^n \ln y.
\eqa
%--
With only two variables we use
%--
\bqa
\intfx{y} y^n \ln\chi_{\aba}(x,y,y) \ln(1-y) &=&
       \intfx{y} \,y^n\,\ln\frac{\chi_{\aba}(x,y,y)}{\chi_{\aba}(x,1,1)}\,
\ln(1-y)  \nl
{}&-&  \frac{1}{n+1}\,\ln\chi_{\aba}(x,1,1) \sum_{j=1}^{n+1}\,\frac{1}{j},
\nl
\intfx{y} y^n \ln\chi_{\aba}(x,y,0) \ln(1-y) &=&
\intfx{y}\,y^n\, \ln\frac{\chi_{\aba}(x,y,0)}{\chi_{\aba}(x,1,0)}\,
\ln(1-y)  \nl
{}&-& \frac{1}{n+1}\,\ln\chi_{\aba}(x,1,0) \sum_{j=1}^{n+1}\frac{1}{j}.
\eqa
%--
The second iteration of the {\em rasing} procedure is based on the following
set of identities:
%--
\bqa
\ln\chi_{\aba}(x,y,y) &=& -\,\frac{4}{b_{\aban{1}}}
\Bigl\{ \frac{1}{2}\, \lpar y + Y_1\rpar \partial_y \chi_{\aba}(x,y,y)
\Bigl[ 1 - \ln\chi_{\aba}(x,y,y)\Bigr] \nl
{}&+& \chi_{\aba}(x,y,y) \ln\chi_{\aba}(x,y,y)\Bigr\},  \nl
\ln^2\chi_{\aba}(x,y,y) &=& \frac{8}{b_{\aban{1}}}
\Bigl\{ \frac{1}{2} \lpar y + Y_1\rpar \partial_y
\Bigl[ \chi_{\aba}(x,y,y) - \chi_{\aba}(x,y,y)\ln\chi_{\aba}(x,y,y)  \nl
{}&+& \frac{1}{2} \chi_{\aba}(x,y,y)\ln^2\chi_{\aba}(x,y,y)\Bigr] -
\frac{1}{2} \chi_{\aba}(x,y,y)\ln^2\chi_{\aba}(x,y,y)\Bigr\}.
\eqa
%--
Furthermore, we will use an additional relation:
%--
\bqa
\intfx{y} y^n \partial_y \chi_{\aba}(x,y,y)\ln^m\chi_{\aba}(x,y,y) &=&
\chi_{\aba}(x,1,1)\,\ln^m\chi_{\aba}(x,1,1)  \nl
{}&-& \delta_{n,0}\chi_{\aba}(x,0,0)\ln^m\chi_{\aba}(x,0,0)
\nl
{}&-& n\,\intfx{y}\,y^{n-1} \chi_{\aba}(x,y,y)\ln^m\chi_{\aba}(x,y,y).
\eqa
%--
Here we have introduced the following quantities:
%--
\bq
Y_1 = - \frac{1}{2}\,\lpar 1 - \mu^2_4 + \mu^2_x\rpar, \qquad
b_{\aban{1}} = \lambda(1,\mu^2_4,\mu^2_x),
\label{l1}
\eq
%--
with special cases given by
$\chi_{\aba}(x,1,1) = \mu^2_4$and $\chi_{\aba}(x,0,0) = \mu^2_x$.
Similarly, for $\chi_{\aba}(x,1,y)$, we obtain
%--
\bqa
\ln\chi_{\aba}(x,1,y) &=& -\,\frac{4}{b_{\aban{2}}}
\Bigl\{ \frac{1}{2} \lpar y + Y_2\rpar \partial_y \chi_{\aba}(x,1,y) \nl
{}&\times& \Bigl[ 1- \ln\chi_{\aba}(x,1,y)\Bigr] + \chi_{\aba}(x,1,y)
\ln\chi_{\aba}(x,1,y)\Bigr\},
\nl
\ln^2\chi_{\aba}(x,1,y) &=& \frac{8}{b_{\aban{2}}}
\Bigl\{ \frac{1}{2} \lpar y + Y_2\rpar \partial_y
\Bigl[ \chi_{\aba}(x,1,y)- \chi_{\aba}(x,1,y) \ln\chi_{\aba}(x,1,y)  \nl
{}&+& \frac{1}{2} \chi_{\aba}(x,1,y) \ln^2\chi_{\aba}(x,1,y)\Bigr] -
\frac{1}{2} \chi_{\aba}(x,1,y) \ln^2\chi_{\aba}(x,1,y)\Bigr\}.
\eqa
%--
Furthermore, the following identities hold:
%--
\bqa
\nl
\intfx{y} y^n \partial_y \chi_{\aba}(x,1,y) \ln^m\chi_{\aba}(x,1,y) &=&
 \chi_{\aba}(x,1,1) \ln^m\chi_{\aba}(x,1,1)  \nl
{}&-& \delta_{n,0} \chi_{\aba}(x,1,0) \ln^m\chi_{\aba}(x,1,0)
\nl
{}&-& n\,\intfx{y} y^{n-1} \chi_{\aba}(x,1,y) \ln^m\chi_{\aba}(x,1,y),
\nl
\intfx{y} y^n \partial_y \chi_{\aba}(x,1,y) &=&
 \chi_{\aba}(x,1,1) - \delta_{n,0} \chi_{\aba}(x,1,0)  \nl
{}&-& n\,\intfx{y} y^{n-1} \chi_{\aba}(x,1,y),
\eqa
%--
where we have introduced a new $Y$-factor,
%--
\bq
Y_2 = - \frac{1}{2}\,\lpar 1 - \mu^2_4 + \mu^2_3\rpar,  \qquad
b_{\aban{2}} = \lambda(1,\mu^2_3,\mu^2_4),
\label{l2}
\eq
%--
with special cases given by
$\chi_{\aba}(x,1,1) = \mu^2_4, \chi_{\aba}(x,0,0)= \mu^2_x$ and also
$\chi_{\aba}(x,1,0) = \mu^2_3$.
%--
Finally we have to consider terms containing $\chi_{\aba}(x,y,0)$ which give
%--
\bqa
\ln\chi_{\aba}(x,y,0) &=&
   \chi_{\aba}(x,y,0) \ln\chi_{\aba}(x,y,0) - \Bigl[ 1 - \chi_{\aba}(x,y,0)
\Bigr]\partial_y \frac{\chi_{\aba}(x,y,0)}{b_{\aban{3}}}  \nl
{}&+&  \Bigl[ 1- \chi_{\aba}(x,y,0)\Bigr] \partial_y \,\Bigl[
\frac{\chi_{\aba}(x,y,0)}{b_{\aban{3}}} \ln\chi_{\aba}(x,y,0)\Bigr],
\nl
\ln^2\chi_{\aba}(x,y,0) &=& - 2 \Bigl\{
   - \frac{1}{2} \chi_{\aba}(x,y,0) \ln^2\chi_{\aba}(x,y,0) -
\Bigl[1-\chi_{\aba}(x,y,0)\Bigr]
\partial_y \frac{\chi_{\aba}(x,y,0)}{b_{\aban{3}}}  \nl
{}&+&   \Bigl[1-\chi_{\aba}(x,y,0)\Bigr] \partial_y \,\Bigl[
\frac{\chi_{\aba}(x,y,0)}{b_{\aban{3}}} \ln\chi_{\aba}(x,y,0)\Bigr]  \nl
{}&-&  \frac{1}{2} \Bigl[1-\chi_{\aba}(x,y,0)\Bigr] \partial_y \,\Bigl[
\frac{\chi_{\aba}(x,y,0)}{b_{\aban{3}}}
\ln^2\chi_{\aba}(x,y,0)\Bigr]\Bigr\},
\eqa
%--
where the special case is
$\chi_{\aba}(x,y,0) = y \mu^2_3 + (1-y) \mu^2_x$.
For this combination we use
%--
\bqa
\intfx{y} y^n \partial_y \chi_{\aba}(x,y,0) \ln^m\chi_{\aba}(x,y,0) &=&
 \chi_{\aba}(x,1,0) \ln^m\chi_{\aba}(x,1,0)  \nl
{}&-& \delta_{n,0} \chi_{\aba}(x,0,0) \ln^m\chi_{\aba}(x,0,0)
\nl
{}&-& n\,\intfx{y} y^{n-1} \chi_{\aba}(x,y,0) \ln^m\chi_{\aba}(x,y,0),
\eqa
%--
\bqa
\intfx{y} y^n \partial_y \chi_{\aba}(x,y,0) &=&
 \chi_{\aba}(x,1,0) - \delta_{n,0} \chi_{\aba}(x,0,0) - n\,\intfx{y} y^{n-1}
\chi_{\aba}(x,y,0).
\eqa
%--
Here we have defined
%--
\bq
b_{\aban{3}} = \mu^2_3 - \mu^2_x.
\label{l3}
\eq
%--
For terms involving $\chi_{\aba}(x,y,z)$ the second step will be as follows:
%--
\bqa
\ln\chi_{\aba}(x,y,z) &=& -\,\frac{4}{b_{\aba}}\Bigl\{
y \partial_y \chi_{\aba}(x,y,z) \Bigl[ 1 - \ln\chi_{\aba}(x,y,z) \Bigr]
\nl
{}&+& \frac{1}{2}  ( z + Y_2 )\,\partial_z  \chi_{\aba}(x,y,z) \,\Bigl[ 1 -
\ln\chi_{\aba}(x,y,z)\Bigr]
\nl
{}&+& \chi_{\aba}(x,y,z)  \ln\chi_{\aba}(x,y,z)\Bigr\},
\nl
\ln^2\chi_{\aba}(x,y,z) &=& \frac{2}{b_{\aba}}  \Bigl\{
y \partial_y  \chi_{\aba}(x,y,z)\,\Bigl[ 1  -
\ln\chi_{\aba}(x,y,z) + \frac{1}{2} \ln^2\chi_{\aba}(x,y,z)\Bigr] \nl
{}&+& \frac{1}{2}  ( z + Y_2 )\,\partial_z \chi_{\aba}(x,y,z)
\Bigl[ 1 - \ln\chi_{\aba}(x,y,z) +\frac{1}{2} \ln^2\chi_{\aba}(x,y,z)\Bigr]
\nl
{}&-& \frac{1}{2}  \chi_{\aba}(x,y,z)  \ln^2\chi_{\aba}(x,y,z)\Bigr\}.
\eqa
%--
Integration by parts can now be performed according to
%--
\bqa
\intfz  z^n  \partial_z  \chi_{\aba}(x,y,z)  \ln^m\chi_{\aba}(x,y,z) &=&
y^n\,\chi_{\aba}(x,y,y)  \ln^m\chi_{\aba}(x,y,y) \nl
{}&-& \delta_{n,0}  \chi_{\aba}(x,y,0)  \ln^m\chi_{\aba}(x,y,0)  \nl
{}&-&n\,\intfz z^{n-1}  \chi_{\aba}(x,y,z)  \ln^m\chi_{\aba}(x,y,z),
\eqa
%--
\bqa
\intfz  z^n  \partial_z  \chi_{\aba}(x,y,z) &=&
y^n\,\chi_{\aba}(x,y,y)-\delta_{n,0}  \chi_{\aba}(x,y,0) -
n\,\intfz  z^{n-1}  \chi_{\aba}(x,y,z).
\eqa
%--
Furthermore, we exchange to order of integration,
%--
\bq
\intfx{y}  \intfz  \partial_y = \intfx{z}  \intfy  \partial_y,
\eq
%--
to perform the remaining integration by parts according to
%--
\bqa
\intfy  y^n  \partial_y  \chi_{\aba}(x,y,z)  \ln^m\chi_{\aba}(x,y,z) &=&
\chi_{\aba}(x,1,z)  \ln^m\chi_{\aba}(x,1,z)  \nl
{}&-& z^n\,\chi_{\aba}(x,z,z)
\ln^m\chi_{\aba}(x,z,z) \nl
{}&-& n\,\intfy  y^{n-1}  \chi_{\aba}(x,y,z)  \ln^m\chi_{\aba}(x,y,z),
\nl
\intfy  y^n  \partial_y  \chi_{\aba}(x,y,z) &=&
\chi_{\aba}(x,1,z) - z^n\,\chi_{\aba}(x,z,z)  \nl
{}&-& n\,\intfy  y^{n-1}
\chi_{\aba}(x,y,z).
\eqa
%--
To summarize, the following quadratic forms will appear in the final result
as arguments of the logarithms:
%--
\bqa
X_{\aba}(x,y,z) &=& x(1-x)\Bigl[ (y-z) \mu^2_3 + z \mu^2_4 - z (1-z)\Bigr] +
 (1-y) \Bigl[ (1-x) \mu^2_1 + x \mu^2_2\Bigr],
\nl
X_{\aba}(x,y,y) &=& x(1-x)\Bigl[ y \mu^2_4 -y (1-y)\Bigr] + (1-y)
\Bigl[ (1-x) \mu^2_1 + x \mu^2_2\Bigr],
\nl
X_{\aba}(x,y,0) &=& x(1-x) y \mu^2_3 + (1-y) \Bigl[ (1-x) \mu^2_1 + x
\mu^2_2\Bigr],
\nl
X_{\aba}(x,1,y) &=& (1-y) \mu^2_3 + y \mu^2_4 - y (1-y).
\eqa
%--
Collecting the various terms we obtain a final answer that can be cast into
the following form:
%--
\bq
S^{121} = \intfxx{x}{y}\,\Bigl\{
\intfz \sum_{n=0}^2\,\frac{I_{3n}}{\lambda^n_0} +
\Bigl[ \sum_{n=0}^2\frac{I_{2n}}{\lambda^n_0} +
\frac{1}{\lambda_0}\sum_{n=1}^3\frac{I_{21n}}{\lambda_n} +
\sum_{n=1}^3\frac{I_{20n}}{\lambda_n}\Bigr]\Bigr\}.
\label{twoiter}
\eq
%--
Since we are not dealing with an analytical result and given that the number
of terms in \eqn{twoiter} is $\ord{1.6 \times 10^3}$ we will not present its
final explicit form in this paper.
%--
\subsubsection{Zeros of $b$-functions}
%--
In the final result for $S^{121}$ we encounter negative powers of four
different $b$-functions whose zeros are as follows:
%--
\begin{itemize}

\item[--] $b_{\aba}$ of \eqn{l04}. There are two roots for $x$, i.e.\
%--
\bq
x^0_{\pm} = \frac{2}{\mu^4_{34}}\,\Bigl[ \mu^2_1 - \mu^2_2 + \frac{1}{4}\,
\mu^4_{34} \pm \lambda^{1/2}\lpar \frac{\mu^4_{34}}{4},\mu^2_1,\mu^2_2
\rpar\Bigr],
\eq
%--
where $\mu^2_{34} = 1 + \mu^2_3 - \mu^2_4$. These roots are real and
internal
to the interval $[0,1]$ if $\mu^4_{34} \le 4\,(\mu_1-\mu_2)^2$ or
$\mu^4_{34} \ge 4\,(\mu_1+\mu_2)^2$.
%--
\item[--] $ b_{\aban{1}}$ of \eqn{l1}. There are three possibilities:
\begin{enumerate}
%--
\item $\mu_1 + \mu_2 - \mu_4 \ge 1$, therefore $b_{\aban{1}}$ can never be
zero.
%--
\item $\lpar 1 - \mu_4\rpar^2 \le \,\lpar \mu_1+\mu_2\rpar^2 \le \,
\lpar 1 + \mu_4\rpar^2$, when there are two values of $x$ where
$b_{\aban{1}} = 0$,
%--
\bq
x^{1+}_{\pm} = \frac{1}{2\,\lpar 1 + \mu_4\rpar^2}\,\,\Bigl[
\,\lpar 1 + \mu_4\rpar^2 + \mu^2_1 - \mu^2_2 \pm \lambda^{1/2}\,\lpar
\,\lpar 1 + \mu_4\rpar^2,\mu^2_1,\mu^2_2\rpar\Bigr].
\eq
%--
\item $\lpar \mu_1 + \mu_2\rpar^2 \le \,\lpar 1 - \mu_4\rpar^2$,
when we have four values of $x$ where $b_{\aban{1}} = 0$. The new pair of
points is given by
%--
\bq
x^{1-}_{\pm} = \frac{1}{2\,\lpar 1 - \mu_4\rpar^2}\,\,\Bigl[
\,\lpar 1 - \mu_4\rpar^2 + \mu^2_1 - \mu^2_2 \pm \lambda^{1/2}\,\lpar
\,\lpar 1 - \mu_4\rpar^2,\mu^2_1,\mu^2_2\rpar\Bigr].
\label{defxm}
\eq
%--
\end{enumerate}
%--
\item[--] $b_{\aban{2}}$ of \eqn{l2}. This is $x$-independent and vanishes
for
$s = (m_3-m_4)^2$ or $s = (m_3+m_4)^2$. The latter corresponds to the
normal threshold relative to the two-particle cut of the
diagram~\cite{Cutkosky:1960sp}.
%--
\item[--] $b_{\aban{3}}$ of \eqn{l3}. There are two roots for $x$, i.e.\
%--
\bq
x^3_{\pm} = \frac{1}{2\,\mu^2_3}\,\Bigl[ \mu^2_3 + \mu^2_1 - \mu^2_2
\pm \lambda^{1/2}\lpar \mu^2_3,\mu^2_1,\mu^2_2\rpar\Bigr],
\eq
%--
These roots are real and internal to the interval $[0,1]$ if
$\mu^2_3 \le (\mu_1-\mu_2)^2$ or $\mu^2_3 \ge (\mu_1+\mu_2)^2$.
\end{itemize}
%--
\subsubsection{The logarithms}
%--
To discuss the logarithms and their branch points we write
%--
\bq
\ln X_{\aba}(x,y,z) = \ln (a x^2 + b x + c - i\,\delta),
\eq
%--
\bq
a = z (1-z) - (y-z) \mu^2_3 - z \mu^2_4,  \quad
b = - a + (1-y) ( \mu^2_2 - \mu^2_1),  \quad
c = (1-y) \mu^2_1.
\eq
%--
Let us denote the two roots of the quadratic by $x^0_{\ssL,\ssR}$.
There are three sub-cases to be discussed,
%--
\begin{itemize}
%--
\item[--] $X_{\aba}(x,y,y)$ where $a = y(1-y) - y \mu^2_4$ and with roots
$x^1_{\ssL,\ssR}$,
\item[--] $X_{\aba}(x,1,y)$, where $a = y(1-y) - (1-y) \mu^2_3 - y \mu^2_4$,
$b= -a$ and $c = 0$, with roots $x^2_{\ssL,\ssR}$,
\item[--] $X_{\aba}(x,y,0)$ where $a = - y \mu^2_3$, and with roots
$x^3_{\ssL,\ssR}$.
\end{itemize}
%--
\subsection{Evaluation of $S^{121}$: methods III and IV\label{esabamtf}}
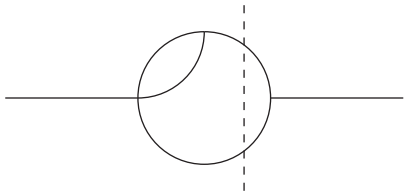
\begin{figure}[th]
\vspace{0.5cm}
\[
  \vcenter{\hbox{
  \begin{picture}(150,0)(0,0)
  \Line(0,0)(50,0)
  \CArc(75,0)(25,0,90)
  \CArc(75,0)(25,-180,0)
  \CArc(75,0)(25,90,180)
  \CArc(50,25)(25,-90,0)
  \Line(100,0)(150,0)
  \DashLine(90,35)(90,-35){3.}
  \end{picture}}}
\]
\vspace{0.5cm}
\caption[]{The two-particle cut of diagram $S^{121}$ of \fig{tops4}}
\label{s4tpc}
\end{figure}

%--
$S^{121}$ can be cast into different forms, all suitable for numerical
integration.
Let us consider again \eqn{defs4} and change variable, $y= 1-y'$. We obtain
%--
\bqa
S^{121} &=& -\,2\,
\lpar\frac{\tHss}{\pi\,s}\rpar^{\ep}\,\egam{\ep}\,\intfx{x}
\intfx{y}\intfzp\,\Bigl[ x(1-x)\Bigr]^{-\ep/2}\,y^{\ep/2}  \nl
{}&\times& \Bigl[ \lpar \mu^2_x - z^2\rpar\,\chi_{\aba}^{-1-\ep} +
\frac{2-\ep/2}{\ep}\,\chi_{\aba}^{-\ep}\Bigr],
\label{defns4}
\eqa
%--
where the quadratic form $\chi_{\aba}$ is now given by
%--
\bq
\chi_{\aba}(x,1-y,z) = z^2 - \mu^2_{34}\,z + (\mu^2_x-\mu^2_3)\,y + \mu^2_3.
\eq
%--
At this point we apply a special differential relation:
%--
\bqa
(\mu^2_x-\mu^2_3)\,\chi_{\aba}^{-1-\ep} &=& -\,\frac{1}{\ep}\,
\partial_y\,\chi_{\aba}^{-\ep},  \nl
\Bigl[ 1 + \frac{y}{\ep}\,\partial_y + \frac{1}{2\,\ep}\,(z -
\frac{1}{2}\,\mu^2_{34})\,
\partial_z\Bigr]\,\chi_{\aba}^{-\ep} &=& -
\frac{1}{4}\,b'_{\aba}\,\chi_{\aba}^{-1-\ep},
\eqa
%--
where now $b'_{\aba}$ has the advantage of being $x$-independent, namely we
get
%--
\bq
b'_{\aba} = \lambda(1,\mu^2_3,\mu^2_4).
\label{defnl0}
\eq
%--
This second method is closer to the original BT approach in the sense that
the denominator, after the {\em raising} procedure, is a function of the
external
parameters but does not depend on residual Feynman parameters. The zeros
of $b'_{\aba}$, \eqn{defnl0}, are the normal and pseudo thresholds
corresponding to the two-particle cut of \fig{s4tpc}. They are non-leading
Landau singularities. The fact that $b'_{\aba}$ is $x$ independent means
that, with this method, we can perform the $x$-integration without having to
distort the integration contour. After integration by parts we obtain
%--
\bqa
S^{121}_{\rm fin} &=& -\,2\,\Bigl\{ \frac{1}{b'_{\aba}}\,
\Bigl[ \intfxx{x}{y}\,\Bigl(\intfzp\,J^{\aba}_{31} + J^{\aba}_{21}\Bigr) -
\frac{1}{2}\, \mu^2_3 - \frac{1}{72}\Bigr]  \nl
{}&+& \intfxx{x}{y}\Bigl(\intfzp\,J^{\aba}_{30} + J^{\aba}_{20}\Bigr) +
\frac{5}{8}\, + \frac{1}{4}\,\zeta(2)\Bigr]\Bigr\},
\eqa
%--
where the various integrands are as follows:
%--
\bqa
J^{\aba}_{31} &=& \,2\,(\mu^2_3 - z^2)\,\ln\chi_{\aba}(x,1-y,z) -
           (3\,\mu^2_3 + \mu^2_{34}\,z - 5\,z^2)\,\ln\chi_{\aba}(x,1-y,z)\,
           {\cal L}_-(x,y,z),
\nl
J^{\aba}_{21} &=& \Bigl[ - \frac{1}{2}\,\mu^2_{34}\mu^2_3 +
\frac{1}{2}\,\mu^2_{34} + \mu^2_3 - 1 +
( 1 - \mu^2_{34} + \mu^2_3)\,y + (1 + \frac{1}{2}\,\mu^2_{34})\,y^2 - y^3
\Bigr]  \nl
{}&\times& \ln\chi_{\aba}(x,1-y,1-y) \,
{\cal L}_-(x,y,1-y) + \frac{1}{2}\,\mu^2_3\mu^2_{34}\,
\ln\chi_{\aba}(x,1-y,0)\,{\cal L}_-(x,y,0),
\nl
J^{\aba}_{30} &=& \ln\chi_{\aba}(x,1-y,z)\,\Bigl[ \frac{1}{2} -
{\cal L}_-(x,y,z)\Bigr] - \frac{1}{2}\,\lpar\frac{\ln\chi{\aba}(x,1-y,z)}{y}
\rpar_+,
\nl
J^{\aba}_{20} &=&  \frac{1}{2}\,\Bigl[
\ln\chi_{\aba}(x,1-y,1-y)\,{\cal L}_-(x,y,1-y) + \ln\chi_{\aba}(x,1,y)\,
{\cal L}_+(x,1,y)\Bigr].
\label{mIII}
\eqa
%--
In \eqn{mIII} we have introduced special combinations of logarithms,
%--
\bqa
{\cal L}_-(x,y,z) &=& \ln y - \ln x - \ln(1-x) - \ln\chi_{\aba}(x,1-y,z),
\nl
{\cal L}_+(x,1,y) &=& \ln\chi_{\aba}(x,1,y) + \ln x + \ln(1-x) - \ln(1-y).
\eqa
%--
If requested, the procedure can be iterated without introducing
numerical instabilities. However, the method fails at the non-leading Landau
singularities corresponding to the one-loop sub-diagram with the largest
number of internal lines. Therefore, in these regions we have to modify the
procedure. Here, $\mu^2_{34} = \pm 2\,\mu_3$ and $\chi_{\aba} =
(z \mp \mu_3)^2 + (\mu^2_x-\mu^2_3)\,y$; correspondingly we change variable,
$z' = z \mp \mu_3$ and obtain
%--
\bq
\intfx{y}\int_{\scriptstyle \mp \mu_3}^{\scriptstyle 1 \mp \mu_3 - y}\,
dz\,y^{\ep/2}\Bigl[ \lpar \mu^2_x - (z\pm \mu_3)^2\rpar\,
\chi_{\aba}^{-1-\ep} + \frac{2-\ep/2}{\ep}\,\chi_{\aba}^{-\ep}\Bigr],  \quad
\chi_{\aba} = z^2 + (\mu^2_x-\mu^2_3)\,y.
\eq
%--
After a new change of variable, $z^2 = t$, we use $M(t,y)= t +
(\mu^2_x-\mu^2_3)\,y$ to derive
%--
\bq
\Bigl\{(\mu^2_x-\mu^2_3)\,;\,1\Bigr\}\,M^{-1-\ep} =
\Bigl\{\partial_y\,;\,\partial_z\Bigr\}\,\Bigl[ ln M - \frac{\ep}{2}\,
\ln^2 M + \ord{\ep^2}\Bigr], 
\eq
%--
and arrive at the final result.
%--
%--
For the $S^{121}$ diagram we have also another method (IV) which makes use
of the following general property:
%--
\bq
(\mu^2_x - z^2)\,\chi_{\aba}^{-1-\ep}(x,y,z) = \Bigl[ 1 + \frac{1}{\ep}\,
(y\partial_y + z\partial_z)\Bigr]\,\chi_{\aba}^{-\ep}(x,y,z).
\eq
%--
After integration by parts we obtain:
%--
\bqa
S^{121}_{\rm fin} &=& -\,\Bigl\{ \intfxx{x}{y}\intfz\,
\lpar\frac{\ln\chi_{\aba}(x,y,z)}{y-1}\rpar_+  \nl
{}&-& \intfxx{x}{y}\,\ln\chi_{\aba}(x,1,y)\,L_1(x,1,y)
 + \frac{3}{2} + \frac{1}{2}\,\zeta(2)
\Bigr\},
\eqa
%--
with $L_1$ given by
%--
\bq
L_1(x,y,z) = \ln (1-y) - \ln x - \ln(1-x) - \ln\chi_{\aba}(x,y,z).
\eq
%--
Method IV represents the simplest results for $S^{121}$  but it is not
generalizable to more complex topologies.
%--
\subsection{Derivative of $S^{121}$ and infrared poles.\label{dsabaip}}
%--
The approach described in this paper is primarily intended for evaluation
of massive multi-loop diagrams. However, QED and QCD will be part of any
realistic calculation and they usually lead to infrared singularities.
Any method aimed to a numerical evaluation of diagrams must be able
to handle the infrared problem. For one-loop diagrams we have seen that
the method is able to extract the infrared pole in dimensional
regularization with a residue and a finite part that can be treated 
numerically~\cite{Passarino:2001wv}. We have to extend the treatment of 
infrared divergences to multi-loop and we will start the discussion with one 
specific example: we define the on-shell derivative of a two-point function,
where possibly some of the internal masses are zero, as the 
$\partial/\partial p^2$ derivative evaluated at the mass shell of one of the 
non-zero internal masses.
This paper concerns infrared configurations only and the general case will be 
presented together with vertices~\cite{preparation}.
 
Consider now the on-shell derivative of $S^{121}$. Indeed, for the simplest 
topology $S^{111}$ the on-shell derivative is infrared finite, see Appendix A.

\begin{figure}[th]
\vspace{0.5cm}
\[
  \vcenter{\hbox{
  \begin{picture}(150,0)(0,0)
  \Line(0,0)(50,0)
  \CArc(75,0)(25,0,90)
  \PhotonArc(75,0)(25,-180,0){2}{7}
  \CArc(75,0)(25,90,180)
  \PhotonArc(50,25)(25,-90,0){2}{7}
  \Line(100,0)(150,0)
  \Text(50,25)[cb]{$\ssW$}
  \Text(100,25)[cb]{$\ssW$}
  \Text(75,-38)[cb]{$\ph$}
  \Text(80,-5)[cb]{$\ph$}
  \end{picture}}}
\]
\vspace{0.5cm}
\caption[]{A two-loop diagram contribution to the $\wb$-boson self-energy.}
\label{wwf}
\end{figure}
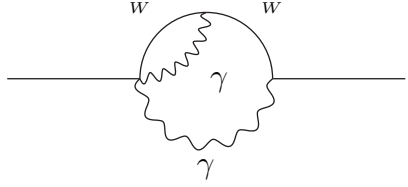

It is important to recall that a necessary condition for the presence of
infrared divergences is that the Landau equations are fulfilled.
In the case of $S^{121}$ we see that $s = (m_1 + m_2 \pm m_4)^2$ and
$m^2_3 = (m_1 + m_2)^2$ are satisfied by $m_2 = m_4 = 0$, $ m_1 = m_3 = m$
and $s = m^2$. However $S^{121}$ itself is not infrared divergent but
self-energy diagrams enter into the calculation of wave function
renormalization factor and, therefore, also their derivative with respect to
$p^2$ is needed. This derivative, as in the one-loop case, shows an infrared
pole when computed on-shell.
Therefore, we consider the case $m_2 = m_4 = 0$ and $m_1 = m_3 = m$, a
typical example of which is shown in \fig{wwf}.
%--
%--
For these values of the internal masses we get
%--
\bqa
S^{121} &=& -\,2\,\lpar\frac{\tHss}{\pi}\rpar^{\ep}\,\egam{\ep}\,
\intfx{x}\intfxy{y}{z}\,\Bigl[ x(1-x)\Bigr]^{-\ep/2}\,\lpar 1-y\rpar^{\ep/2}
\nl
{}&\times& \Bigl[ \lpar m^2_x + z^2 p^2\rpar\,U_{\aba}^{-1-\ep} +
\frac{2-\ep/2}{\ep}\,U_{\aba}^{-\ep}\Bigr],
\eqa
%--
where the quadratic form $U_{\aba}$ is
%--
\bq
U_{\aba} = - p^2 z^2 + (p^2-m^2) z + m^2 (1 - \frac{1}{x}) y +
\frac{m^2}{x}.
\eq
%--
First we take the derivative of $S^{121}$ with respect to $p^2$ and next we
go on-shell by putting $p^2 = - m^2$.
%--
\subsubsection{On-shell derivative of $S^{121}$: method I}
%--
Proceeding as described above we obtain the derivative as
%--
\bqa
S^{121}_p &=& -\,\frac{2}{m^2}\,
\lpar\frac{\tHss}{\pi\,m^2}\rpar^{\ep}\,\egam{\ep}\,
\intfx{x}\intfxy{y}{z}\,\Bigl[ x(1-x)\Bigr]^{-\ep/2}\,(1-y)^{\ep/2}  \nl
{}&\times& \Bigl\{ z(1-z)\,\Bigl[ (1+\ep)\,(z^2-\frac{1}{x})\,
\chi^{-2-\ep}_{\aba} + \frac{\ep}{2}\,\chi^{-1-\ep}_{\aba}\Bigr] +
z\,(3\,z - 2)\,\chi^{-1-\ep}_{\aba}\Bigr\}|_{\rm os},
\label{sabap}
\eqa
%--
where we have introduced
%--
\bq
\chi_{\aban{\rm os}} = ( z - 1 )^2 + \frac{1}{X}\,( 1 - y ), \qquad
X = \frac{x}{1-x}.
\eq
%--
We introduce new variables $y = 1-y'$ and $z= 1-z'$, so that
%--
\bq
U^{\rm os}_{\aba} = m^2\,\chi_{\aban{\rm os}} = m^2\,(z^2 + \frac{y}{X}),
\eq
%--
and consider the term
%--
\bq
\Bigl[ (1-z)^2 - \frac{1}{x}\Bigr]\,\Bigl( \chi_{\aban{\rm os}}
\Bigr)^{-2-\ep}.
\eq
%--
To go further, we can use the following relation:
%--
\bq
{\cal D}_i\,(z^2 + \frac{y}{X})^{-1-\ep} = (1+\ep)\,d_i\,
(z^2 + \frac{y}{X})^{-2-\ep},
\eq
%--
\bq
{\cal D} \equiv \Bigl(\partial_y - \frac{z}{2}\,\partial_z\,,\,\partial_z\,,
\,\partial_y\Bigr), \quad
d \equiv \Bigl(z^2 - \frac{1}{X}\,,\,-\,2\,z\,,\,-\frac{1}{X}\Bigr).
\eq
%--  
%\bqa
%\frac{1}{1+\ep}\,(\partial_y - \frac{z}{2}\,\partial_z)\,
%(z^2 + \frac{y}{X})^{-1-\ep} &=& (z^2 - \frac{1}{X})\,
%(z^2 + \frac{y}{X})^{-2-\ep},
%\nl
%\frac{1}{1+\ep}\,\partial_z (z^2 + \frac{y}{X})^{-1-\ep} &=&
%-\,2\,z\,(z^2 + \frac{y}{X})^{-2-\ep},
%\nl
%\frac{X}{1+\ep}\,\partial_y (z^2 + \frac{y}{X})^{-1-\ep} &=&
%-\,(z^2 + \frac{y}{X})^{-2-\ep}.
%\eqa
%--
After integration by parts we use
%--
\bq
\Bigl(\partial_z\,;\,\partial_y\Bigr)\,(z^2 + \frac{y}{X})^{-\ep} =
\ep\,\Bigl(-\,2\,x\,;\,-\frac{1}{X}\Bigr)\,(z^2 + \frac{y}{X})^{-1-\ep}.
\eq
%--
%\bqa
%\frac{1}{\ep}\,\partial_z\,(z^2 + \frac{y}{X})^{-\ep} &=&
%-\,2\,z\,(z^2 + \frac{y}{X})^{-1-\ep},
%\nl
%\frac{X}{\ep}\,\partial_y\,(z^2 + \frac{y}{X})^{-\ep} &=&
%-\,(z^2 + \frac{y}{X})^{-1-\ep}.
%\eqa
%--
In this way we derive
%--
\bq
S^{121}_p = -\,\frac{2}{m^2}\,
\lpar\frac{\tHss}{\pi\,m^2}\rpar^{\ep}\,\egam{\ep}\,
\intfx{x}\,\Bigl[ x(1-x)\Bigr]^{-\ep/2}\,{\cal S},
\eq
%--
where ${\cal S}$ is a combination of one and two dimensional integrals.
Due to the presence of an infrared pole in the on-shell derivative we must
keep $\ep \ne 0$. Next we change variable, $y = z y'$, and introduce an
auxiliary integral
%--
\bq
I(\mu,\nu) = \intfxx{y}{z}\,y^{\mu}\,z^{\nu}\,(z+\frac{y}{X})^{-\ep}.
\eq
%--
Using the relation
%--
\bq
\Bigl[ 1 + \frac{1}{\ep}\,(y\,\partial_y + z\,\partial_z)\Bigr]\,
(z+\frac{y}{X})^{-\ep} = 0,
\eq
%--
we are able to derive
%--
\bq
I(\mu,\nu) = \frac{1}{2+\mu+\nu-\ep}\,\intfx{y}\,\Bigl[ y^{\mu}\,
(1+\frac{y}{X})^{-\ep} + y^{\nu}\,(y+\frac{1}{X})^{-\ep}\Bigr].
\label{Imunu}
\eq
%--
In this way the evaluation of ${\cal S}$ is simply reduced to the
calculation of a one-dimensional integral which can be cast into the
following form:
%--
\bq
{\cal S} = \intfx{y}\,{\cal S}_y,
\eq
%--
with an integrand of the form
%-
\bqa
{\cal S}_y &=&
\Bigl[ - \frac{1}{4}\,y^{-1-\ep/2} + \frac{1}{4}\,
( \frac{4}{\ep} - 2 + \frac{3}{3-\ep} )\,y^{1-\ep/2} 
+ \frac{1}{4}\,( 3 - \frac{2}{\ep} + \frac{1}{1-\ep} - 
\frac{3}{3-\ep})\,y^{-\ep/2} \Bigr]\,(1+X\,y)^{-\ep}  \nl
{}&+& \frac{1}{2}\,X^{1+\ep}\,( y^{1-\ep/2} -
y^{2-\ep/2})\,(1+X\,y)^{-1-\ep}
+ \frac{1}{4}\,\frac{1}{1-\ep}\,y^{-1+\ep/2}\,(1+\frac{y}{X})^{-\ep}.
%--
%\frac{1}{2}\,X^{1+\ep}\,y^{1-\ep/2}\,( 1 + Xy)^{-1-\ep}
%-\,\frac{1}{2}\,X^{1+\ep}\,y^{2-\ep/2}\,( 1 + Xy)^{-1-\ep}
%\nl
%{}&+& \frac{1}{4}\,\frac{\ep}{1-\ep}\,y^{-1+\ep/2}\,(1+\frac{y}{X})^{-\ep}
%+ \frac{1}{4}\,y^{-1+\ep/2}\,(1+\frac{y}{X})^{-\ep}
%\nl
%{}&+& \frac{1}{2\,\ep}\,( 1 - \frac{3}{3-\ep})\,y^{\ep/2}\,
%(1+\frac{y}{X})^{-\ep}
%- \frac{1}{4}\,\frac{\ep}{3-\ep}\,y^{\ep/2}\,(1+\frac{y}{X})^{-\ep}
%\nl
%{}&-& \frac{1}{4}\,( 1 -
%\frac{5}{3-\ep})\,y^{\ep/2}\,(1+\frac{y}{X})^{-\ep}
%- \frac{1}{4}\,y^{-1-\ep/2}\,(1 + Xy)^{-\ep}
%\nl
%{}&+& \frac{3}{2\,\ep}\,( 1 - \frac{1}{3-\ep})\,y^{1-\ep/2}\,(1 +
%Xy)^{-\ep}
%- \frac{1}{4}\,\frac{\ep}{3-\ep}\,y^{-\ep/2}\,(1 + Xy)^{-\ep}
%\nl
%{}&-& \frac{1}{2}\,( 1 - \frac{5}{2}\,\frac{1}{3-\ep})\,y^{1-\ep/2}\,
%(1 + Xy)^{-\ep}
%- \frac{1}{2\,\ep}\,y^{-\ep/2}\,(1 + Xy)^{-\ep}
%\nl
%{}&+& \frac{1}{4}\,\frac{\ep}{1-\ep}\,y^{-\ep/2}\,(1 + Xy)^{-\ep}
%+ \frac{3}{4}\,y^{-\ep/2}\,(1 + Xy)^{-\ep}.
\eqa
%--
First we consider the integrals that produce a pole at $\ep = 0$, therefore
%--
\bqa
\intfx{y}\,y^{\pm\ep/2-1}\,(1+\alpha\,y)^{-\ep} &=&
\pm\,\frac{2}{\ep} - \ep\,\intfx{y}\,\frac{\ln(1+\alpha\,y)}{y} + \ord{\ep^2},
\quad \alpha = X,X^{-1}.
\eqa
%--
%\bqa
%\intfx{y}\,y^{\ep/2-1}\,(1+\frac{y}{X})^{-\ep} &=&
%\frac{2}{\ep} - \ep\,\intfx{y}\,\frac{\ln(1+y/X)}{y} + \ord{\ep^2},
%\nl
%\intfx{y}\,y^{-\ep/2-1}\,(1 + X y)^{-\ep} &=&
%-\,\frac{2}{\ep} - \ep\,\intfx{y}\,\frac{\ln(1+Xy)}{y} + \ord{\ep^2}.
%\eqa
%--
Successively we multiply the result by $\bigl[ x (1-x)\bigr]^{-\ep}$ and
expand around $\ep = 0$. The remaining integrals can be computed in terms of
a long list of master integrals that will be reported in Appendix B.
The $x$-integration gives
%--
\bqa
\intfx{x}\,\Bigl[ x (1-x)\Bigr]^{-\ep/2}\,{\cal S} &=&
S_{-1}\,\ep^{-1} + S_0 + S_1\,\ep,
\eqa
%--
with coefficients
%--
\bq
S_{-1} = 1,  \qquad
S_0 = 2,  \qquad
S_{1} =  \frac{7}{2} - \frac{1}{4}\,\zeta(2).
\eq
%--
To obtain our final result we have used the following relation:
%--
\bq
\li{2}{-\,\frac{1}{X}} = \li{2}{x} + \ln x \ln (1-x) - \frac{1}{2}\,
\ln^2 x - \zeta(2),
\eq
%--
and the following integrals:
%--
\bq
\intfx{x}\,\Bigl\{\frac{\ln(1-x)}{x}\,;\,\li{2}{x}\,;\,\ln^2(1-x)\,;\,
\ln(1-x)\Bigr\} = \Bigl\{ -\zeta(2)\,;\,\zeta(2)-1\,;\,2\,;\,-1\Bigr\}.
\eq
%\intfx{x} \ln (1-x) x^{-1} &=& -\,\zeta(2), \quad
%\intfx{x} \li{2}{x} = \zeta(2)-1,  \nl
%\intfx{x} \ln^2(1-x) &=& 2, \quad
%\intfx{x} \ln(1-x) = -1.
%\eqa
%--
The final result for the on-shell derivative reads as follows:
%--
\bq
S^{121}_p = \frac{1}{m^2}\,\Bigl(\frac{\tHss}{\pi\,m^2}\Bigr)^{\ep}\,
\Bigl[ -\,\frac{2}{\ep^2} - 2\,(2 - \gamma)\,\frac{1}{\ep} - 7 +
\gamma\,(4 -\gamma) - \frac{1}{2}\,\zeta(2) + \ord{\ep}\Bigr].
\label{irder}
\eq
%--
\subsubsection{On-shell derivative of $S^{121}$: method II}
%--
An alternative derivation of the on-shell derivative of $S^{121}$ will
also be presented. We can use the following relation
%--
\bq
\Big[ 1 + \frac{1}{1+\ep}\,(y\partial_y + z\partial_z) \Big]\,
\chi_{\aban{\rm os}}^{-1-\ep} =
- (z^2 - \frac{1}{x})\,\chi_{\aban{\rm os}}^{-2-\ep},
\eq
%--
and derive an alternative result
%--
\bqa
S^{121}_p &=& \frac{\egam{1+\ep}}{m^2}\,
\lpar\frac{\tHss}{\pi\,m^2}\rpar^{\ep}
\intfx{x}\intfxy{y}{z}\,\,\Bigl[ x(1-x)\Bigr]^{-\ep/2}\,(1-y)^{\ep/2-1}\,
z(1-z)\,\chi_{\aban{\rm os}}^{-1-\ep}.
\eqa
%--
Instead of changing variables we can use another important identity,
%--
\bq
z(1-z)\,\chi_{\aban{\rm os}}^{-1-\ep} =
\frac{1}{2\ep}\,z\partial_z\,
\chi_{\aban{\rm os}}^{-\ep}.
\eq
%--
Once this relation is applied and the integration by parts performed we 
introduce new variables, $y = 1 - y'$ and $z = 1 - z'$, so that
%--
\bqa
S^{121}_p &=& \frac{1}{2m^2}\,
\lpar\frac{\tHss}{\pi\,m^2}\rpar^{\ep}\,\egam{\ep}\,
\intfx{x}\,\Bigl[ x(1-x)\Bigr]^{-\ep/2}\,({\cal S}_1 - {\cal S}_2),
\eqa
%--
\bq
{\cal S}_1 = \intfx{y}\,(1-y)\,y^{-\ep/2-1}\,(y + X')^{-\ep}, \quad
{\cal S}_2 = \intfx{y}\intfzs\,y^{\ep/2-1}\,(z^2 + X'\,y)^{-\ep},
\eq
%--
where we used $X' = X^{-1} = 1/x-1$. Next we change variable, $y = z y'$, and 
utilize the auxiliary integral $I(\mu,\nu)$ of \eqn{Imunu}, which gives for 
${\cal S}_2$ the following result:
%--
\bqa
{\cal S}_2 &=& \frac{1}{1-\ep}\,\intfx{y}\,\Bigl[ y^{\frac{\ep}{2}-1}\,
(1 + X' y)^{-\ep} + y^{-\frac{\ep}{2}}\,(y + X')^{-\ep}\Bigr]
\eqa
%--
From well-known properties of the hypergeometric function~\cite{bat} we get
%--
\bqa
{\cal S}_1
&=& - \frac{2}{\ep} \, X^{\ep} \,
      \hyper{\ep}{-\frac{\ep}{2}}{1-\frac{\ep}{2}}{-X}
    - \frac{2}{2-\ep} \, X^{\ep} \,
      \hyper{\ep}{1-\frac{\ep}{2}}{2-\frac{\ep}{2}}{-X}  \nl
&=& - \frac{2}{\ep} \, x^{\ep} \,
      \hyper{\ep}{1}{1-\frac{\ep}{2}}{x}
    - \frac{2}{2-\ep} \, x^{\ep} \,
      \hyper{\ep}{1}{2-\frac{\ep}{2}}{x}  \nl
{\cal S}_2
&=&   \frac{2}{\ep(1-\ep)} \,
      \hyper{\ep}{\frac{\ep}{2}}{1+\frac{\ep}{2}}{-X'}
    + \frac{2}{(1-\ep)(2-\ep)} \, X^{\ep} \,
      \hyper{\ep}{1-\frac{\ep}{2}}{2-\frac{\ep}{2}}{-X}  \nl
&=&   \frac{2}{\ep(1-\ep)} \, x^{\ep}
      \hyper{\ep}{1}{1+\frac{\ep}{2}}{1-x}
    + \frac{2}{(1-\ep)(2-\ep)} \, x^{\ep} \,
      \hyper{\ep}{1}{2-\frac{\ep}{2}}{x}
\eqa
%--
After introducing the hypergeometric series, we are able to perform the $x$
integration explicitly. Finally we expand around $\ep=0$ by using:
%--
\bq
\egam{n+m} =
\egam{n}\,\Bigl\{ 1 + (\sum_{j=1}^{n-1}\,\frac{1}{j}-\gamma)\,m +
\ord{m^2}\Bigr\}
\eq
%--
and arrive at the result of \eqn{irder}. To obtain \eqn{irder} in method II
we have used the following series:
%--
\bq
\sum_{n=1}^{\infty} \frac{1}{n(n+1)} = 1,
\qquad
\sum_{n=1}^{\infty} \frac{1}{n(n+1)} \sum_{j=1}^{n}\,\frac{1}{j} = \zeta(2),
\qquad
\sum_{n=1}^{\infty} \frac{1}{n^2(n+1)} = \zeta(2) - 1.
\eq
%--
\subsubsection{Another on-shell derivative of $S^{121}$}
\begin{figure}[th]
\vspace{0.5cm}
\[
  \vcenter{\hbox{
  \begin{picture}(150,0)(0,0)
  \Line(0,0)(50,0)
  \PhotonArc(75,0)(25,0,90){2}{7}
  \CArc(75,0)(25,-180,0)
  \CArc(75,0)(25,90,180)
  \CArc(50,25)(25,-90,0)
  \Line(100,0)(150,0)
  \Text(50,25)[cb]{$\ssW$}
  \Text(100,25)[cb]{$\gamma$}
  \Text(75,-38)[cb]{$\ssW$}
  \Text(80,-5)[cb]{$\ssW$}
  \end{picture}}}
\]
\vspace{0.5cm}
\caption[]{A two-loop diagram contribution to the $\wb$-boson self-energy.}
\label{wwfb}
\end{figure}
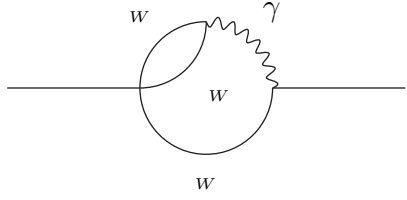

%--
There is another configuration of interest shown in \fig{wwfb}
%--
%--
\bq
\mbox{b)} \qquad m_3 = 0, \quad \mbox{and} \quad m_i = m, \,\forall\,i\ne 3,
\eq
%--
which corresponds to the non-leading Landau singularity $\alpha_1 =
\alpha_2 = 0$ and $s = m^2$. We obtain
%--
\bq
S^{121}_{pb} = \lpar\frac{\tHss}{\pi}\rpar^{\ep}\,\egam{\ep+1}\,
\intfx{x}\intfx{y}\,\int_0^{1-y}\,dz\,\Bigl[ x(1-x)\Bigr]^{-\ep/2}\,
y^{\ep/2-1}\,z\,(1-z)\,U^{-1-\ep}_{\aban{\rm os}}(x,y,z),
\label{defabapb}
\eq
%--
where we have introduced
%--
\bq
U_{\aban{\rm os}}(x,y,z) = m^2\,z^2 + m^2_x\,y, \qquad
m^2_x = \frac{m^2}{x(1-x)}.
\eq
%--
The evaluation of $S^{121}_{pb}$ will be postponed till \sect{dsacaip}
where it will be discussed with other similar functions.
%--
\subsection{Tensor integrals of the $S^{121}$ family\label{tisabaf}}
%--
Let us define the following function:
%--
\bq
\pi^4\,S^{1-i/2,2,1-j/2}_4(\mu_1\dots \mu_i\,|\,\nu_1\dots \nu_j)
= \mu^{2\ep}\,\intmomsii{n}{q_1}{q_2}  \nl
\frac{q^{\mu_1}_1\dots q^{\mu_i}_1\,q^{\nu_1}_2\dots q^{\nu_j}_2}
{(q^2_1+m^2_1)\dots ((q_2+p)^2+m^2_4)},
\label{s4irre}
\eq
%--
where we assume $i+j \le 3$ and where $\alpha$ and $\beta$ have been
changed in order to account for the effective degree of convergence.
For tensor integrals we always indicate explicitly the total number of
internal lines.
In any realistic calculation the first step is about simplifying numerators
and denominators as much as possible. For example we will use results as the
following one:
%--
\bqa
{}&{}&
2\,\frac{\mu^{2\ep}}{\pi^4}\,\intmomsii{n}{q_1}{q_2}
\frac{\spro{p}{q_2}}{(q^2_1+m^2_1)\dots ((q_2+p)^2+m^2_4)} =
\Bigl[ S^{111}(0,m_1,m_2,m_3) \nl
{}&-&
S^{111}(p^2,m_1,m_2,m_4) + (m^2_3 - m^2_4 - p^2)\,
S^{121}_4(p^2,m_1,m_2,m_3,m_4)\Bigr].
\eqa
%--
After possible simplifications few irreducible integrals, of the type
appearing in \eqn{s4irre}, will remain. Their structure will be as follows:
%--
\bqa
{}&{}&
S^{1-i/2,2,1-j/2}_4(p\dots p\,|\,p\dots p) = \lpar\frac{\tHss}{\pi\,s}
\rpar^{\ep}\,\egam{\ep}\intfxx{x}{y}\intfz  \nl
{}&\times& \Bigl[ x (1-x)\Bigr]^{-\ep/2}\,(1-y)^{\ep/2}\,\sum_{n=-1}^{+1}\,
Q^{ij}_n(\ep,x,y,z)\,\chi_{\aba}^{n-\ep}(x,y,z).
\eqa
%--
The coefficients $Q$ are:
%--
\[
\ba{lll}
Q^{10}_{-1} = -2\,xz\,(\mu^2_x - z^2) & \;\;
Q^{10}_{0} = xz\,(1 - \frac{6}{\ep})  & \;\;
Q^{10}_{+1} = 0,  \\
Q^{01}_{-1} = -2\,z\,(\mu^2_x - z^2) & \;\;
Q^{01}_{0} = z\,(1 - \frac{6}{\ep})  & \;\;
Q^{01}_{+1} = 0,  \\
\ea
\]
%--
and also
%--
\bqa
Q^{20}_{-1} &=& (\mu^2_x - z^2)\,\Bigl[ -2\,x^2z^2 + \frac{2}{\ep-2}\,
x (1-x) (\mu^2_x - z^2)\Bigr],  \nl
Q^{20}_0 &=& ( \mu^2_x - z^2 )\,\Bigl[\frac{x^2}{\ep} - 2\,
\frac{\ep-4}{\ep(\ep-2)}\,x (1-x)\Bigr] + z^2\,\Bigl[(1 -
\frac{8}{\ep})\,x^2 -
4\,\frac{x(1-x)}{\ep(\ep-2)}\Bigr],  \nl
Q^{20}_{+1} &=& -\,(\frac{3}{\ep} - \frac{5}{2}\,\frac{1}{\ep-1})\,\Bigl[
x^2 - x (1-x)\,\frac{\ep-4}{\ep-2}\Bigr],
\eqa
%--
\bqa
Q^{02}_{-1} &=& -2\,z^2\,(\mu^2_x - z^2),  \quad
Q^{02}_0 = z^2\,(1-\frac{8}{\ep}) + \,(\mu^2_x - z^2)\,\frac{1}{\ep},
\nl
Q^{02}_{+1} &=& - \frac{3}{\ep} + \frac{5}{2}\,\frac{1}{\ep-1},  \quad
Q^{11}_{-1} = -2\,x\,z^2\,(\mu^2_x - z^2),  \nl
Q^{11}_0 &=& x\,z^2\,(1-\frac{8}{\ep}) + x\,(\mu^2_x - z^2)\,\frac{1}{\ep},
\quad
Q^{11}_{+1} = - x\,(\frac{3}{\ep} - \frac{5}{2}\,\frac{1}{\ep-1}).
\eqa
%--
Note that all terms proportional to $\chi_{\aba}^{-1-\ep}$, i.e. the 
$Q^{ij}_{-1}$ contain a factor $\mu^2_x - z^2$ which, in turn, allows us to
raise the negative power through \eqn{basicaba}.

Several of the properties that we have shown for $S^{121}$ hold for a more
general class of multi-loop diagrams, $G^{1\ssN\,1}$ with $N \ge 2$.
%--
\section{The $G^{1\ssN\,1}$ topology\label{gonot}}
%--
$S^{121}$ is a special example of a general class of two-loop diagrams with
$N+2$ internal lines which are overall convergent for $N > 2$ and contain
the $\alpha\gamma$ logarithmically divergent sub-diagram. We have
%--
\bq
\pi^4\,G^{1\ssN\,1} = \mu^{2\ep}\,\intmomsii{n}{q_1}{q_2}\,
\lpar q^2_1 + m^2_1\rpar^{-1}\lpar\lpar q_1-q_2\rpar^2 + m^2_2\rpar^{-1}
\prod_{i=0}^{\ssN-1}\,\lpar\lpar q_2 + k_i\rpar^2 +
m^2_{i+3}\rpar^{-1},
\eq
%--
where the momenta $k_i$ are linear combinations of the external momenta
$p_j$, $k_0 = 0$ and $k_i = p_1 + \dots + p_i$.
We introduce Feynman parameters $z_i$ for the propagators of the $q_2$ loop,
%--
\bq
dz_{\ssL} = \prod_{i=0}^{\ssN-1}\,dz_i\,\delta\,\lpar 1-z_{\ssL}\rpar,
\qquad
z_{\ssL} = \sum_{i=0}^{\ssN-1}\,z_i
\eq
%--
and also a $\{z_i\}$-dependent momentum and mass
%--
\bq
P_{\mu} = \sum_{i=0}^{\ssN-1}\,z_i k_{i\mu}, \qquad
M^2 = \sum_{i=0}^{\ssN-1}\,z_i\,\lpar k^2_i + m^2_{i+3}\rpar,
\eq
%--
and obtain
%--
\bqa
G^{1\ssN\,1} &=& - 2\,\lpar\frac{\tHss}{\pi}\rpar^{\ep}\,
\frac{\egam{N-1+\ep}}{\ep}
\int d[P] \Bigl[ \lpar m^2_x + P^2 y^2\rpar\,\chi^{1-\ssN-\ep} +
\frac{1}{2}\,\frac{4-\ep}{N-2+\ep}\,\chi^{2-\ssN-\ep}\Bigr],
\eqa
%--
where $x$ and $y$ parameters have been introduced to combine all propagators
arising after the $q_2$ integration. Additional quantities are as follows:
%--
\bqa
\chi &=& - P^2 y^2 + ( M^2 - m^2_x ) y + m^2_x,  \nl
\int d[P] &=& \intfxx{x}{y}\int\,dz_{\ssL}\,\Bigl[ x(1-x)\Bigr]^{-\ep/2}
y^{\ssN-1} (1-y)^{\ep/2}.
\eqa
%--
Next we change variables, $z_i = z'_i/y$, and obtain
%--
\bqa
G^{1\ssN\,1} &=& - 2\,\lpar\frac{\tHss}{\pi}\rpar^{\ep}\,
\frac{\egam{N-1+\ep}}{\ep}
\intfxx{x}{y}\int_0^{+\infty}\,\prod_{i=0}^{\ssN-1} dz_i\,
\Bigl[ x(1-x)\Bigr]^{-\ep/2}(1-y)^{\ep/2}\,\delta(y - z_{\ssL})  \nl
{}&\times& \Bigl[ \lpar m^2_x - z^t H z\rpar\,\chi^{1-\ssN-\ep} +
\frac{1}{2}\,\frac{4-\ep}{N-2+\ep}\,\chi^{2-\ssN-\ep}\Bigr],
\eqa
%--
where the quadratic form $\chi$ has been rewritten as
%--
\bq
\chi = z^t H z + 2\,K^t z + L,  \quad
H_{ij} = -\,\spro{k_i}{k_j}, \quad
K_i = \frac{1}{2}\,\lpar k^2_i + m^2_{i+3}\rpar, \quad
L = m^2_x\,(1-y).
\eq
%--
\section{The $S_{5}$ family\label{sfivet}}
\begin{figure}[th]
\vspace{0.5cm}
\[
  \vcenter{\hbox{
  \begin{picture}(150,0)(0,0)
  \Line(0,0)(50,0)
  \CArc(75,0)(25,0,90)
  \CArc(75,0)(25,-180,0)
  \CArc(75,0)(25,90,180)
  \Line(75,25)(75,-25)
  \Line(100,0)(150,0)
  \end{picture}}}
\]
\vspace{0.5cm}
\caption[]{The two-loop diagram $S^{221}$ of \eqn{defsbab}.}\label{tops5p}
\end{figure}
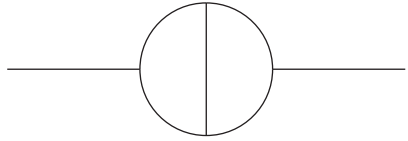

%--
There are two diagrams in the $S_5$ family, $S^{131}$ given in \fig{tops5}
and $S^{221}$, given in \fig{tops5p}.
$S^{131}$ is overall logarithmically divergent and contains the
logarithmically divergent sub-diagram $\alpha\gamma$. $S^{221}$ and all its
sub-diagrams are convergent. Their evaluation will be discussed in the
following sections.
%--
\subsection{$S^{131}$ topology\label{sacat}}
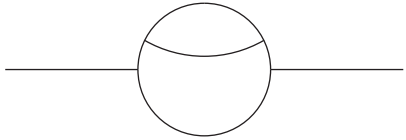
\begin{figure}[th]
\vspace{0.5cm}
\[
  \vcenter{\hbox{
  \begin{picture}(150,0)(0,0)
  \Line(0,0)(50,0)
  \CArc(75,0)(25,0,90)
  \CArc(75,0)(25,-180,0)
  \CArc(75,0)(25,90,180)
  \CArc(75,50)(45,-120,-60)
  \Line(100,0)(150,0)
  \end{picture}}}
\]
\vspace{0.5cm}
\caption[]{The two-loop diagram $S^{131}$ of \eqn{defsaca}.}\label{tops5}
\end{figure}

%--
The $S^{131}$ topology, explicitly shown in \fig{tops5}, is given by
%--
%--
\bqa
\pi^4\,S^{131} &=& \mu^{2\ep}\,\int\,
{{d^n q_1d^n q_2}\over {
\lpar q^2_1+m^2_1\rpar\,
\lpar\,\lpar q_1-q_2\rpar^2+m^2_2\rpar\,
\lpar q^2_2 + m^2_3\rpar\,
\lpar\lpar q_2 + p\rpar^2 + m^2_4\rpar\,
\lpar q^2_2 + m^2_5\rpar}}.
\label{defsaca}
\eqa
%--
If $m_3 \ne m_5$ then $S^{131}$ is the difference of two $S^{121}$ diagrams,
%--
\bq
S^{131} = \frac{1}{m^2_5-m^2_3}\,\Bigl[
S^{121}(p^2;m_1,m_2,m_3,m_4) - S^{121}(p^2;m_1,m_2,m_5,m_4)\Bigr],
\eq
%--
otherwise it is a special case of $G^{131}$ with $k_0 = k_2 = 0$, and
$k_1 = - p$. The quadratic $\chi_{\aca}$ will be
%--
\bq
\chi_{\aca} = z^2 - (1 - \mu^2_4 + \mu^2_3) z + (\mu^2_3 - \mu^2_x) y +
\mu^2_x,
\eq
%--
where we have used $z_2 = y-z_1-z_0$ and, moreover, $z\equiv z_1$.
Limits of integration are $0 \le z_0 \le y-z_1$ and $0 \le z_1 \le y$.
Therefore, for $m_3 = m_5$ we obtain
%--
\bqa
S^{131} &=& - \frac{2}{s}\,\lpar\frac{\tHss}{\pi\,s}\rpar^{\ep}\,
\frac{\egam{2+\ep}}{\ep}\,\intfxx{x}{y}\intfz\,
\Bigl[ x(1-x)\Bigr]^{-\ep/2}(1-y)^{\ep/2}\,(y-z)  \nl
{}&\times& \Bigl[ \lpar \mu^2_x - z^2\rpar\,\chi_{\aca}^{-2-\ep} +
\frac{2-\ep/2}{1+\ep}\,\chi_{\aca}^{-1-\ep}\Bigr],
\label{eqmass}
\eqa
%--
The ultraviolet singularities can be easily derived by taking the derivative
with respect to $m^2_3$ of $S^{121}$, therefore we obtain no double-pole,
as expected, and a single pole given by
%--
\bq
- 2\,(\frac{1}{\ep} - \Delta_{\ssU\ssV})\,
\frac{\partial}{\partial m^2_3}\,
\intfxx{x}{y}\, \ln\chi_{\aba}(x,1,y) = - \frac{2}{s}\,(\frac{1}{\ep} -
\Delta_{\ssU\ssV})\,\intfx{y}\, \frac{1-y}{\chi_{\aba}},
\eq
%--
Note that the residue of the single pole is what we expect in order to
compensate the overlapping divergence coming from the associated subtraction
diagram of \fig{counters5}. The integral can be computed, giving
%--
\bqa
\intfx{y}\, \frac{1-y}{\chi_{\aba}} &=& \frac{1}{2}\,\Bigl\{
\ln\frac{\mu^2_3}{\mu^2_4} + \frac{\mu^2_{43}}{\lambda^{1/2}
(1,\mu^2_3,\mu^2_4)}\,\Bigl[ \ln(1-\frac{1}{y_+}) - \ln(1-\frac{1}{y_-})
\Bigr]\Bigr\},  \nl
y_{\pm} &=& \frac{1}{2}\,\Bigl[ \mu^2_{34} \pm \lambda^{1/2}
(1,\mu^2_3,\mu^2_4)\Bigr],
\eqa
%--
where $\lambda \to \lambda + i\,\delta$ and $\delta \to 0_+$.
If we introduce $y_m = 1/2 \mu^2_{34}$ and $Y_{\pm} = 1 - 1/y_{\pm}$, then
it follows that
%--
\bq
\ln Y_+ - \ln Y_- = \ln\frac{Y_+}{Y_-} + \eta(Y_-,\frac{Y_+}{Y_-}),
\quad
\ln\frac{Y_+}{Y_-} = \lambda^{1/2}(1,\mu^2_3,\mu^2_4)\,
\frac{1}{y_m\,(y_m-1)} + \ord{\lambda}.
\eq
%--
Therefore, for $\lambda \to 0_+(0_-)$ we observe a square root singularity
in the imaginary(real) part. 
Recalling that the counter-term, in the MS-scheme, is $-2\,i\pi^2/\ep$ and
that the remaining integral is a $C_0$-function with $p_1= -p_2= p$ we 
obtain the following result for the subtraction diagram of \fig{counters5}:
%--
\bq
S^{{\underline 1}3{\underline 1}} = \frac{2}{s\,\ep}\,\Bigl[
\intfx{y}\, \frac{1-y}{\chi_{\aba}} + \ord{\ep}\Bigr].
\eq
%--
\begin{figure}[th]
\vspace{0.5cm}
\[
  \vcenter{\hbox{
  \begin{picture}(150,0)(0,0)
  \Line(0,0)(50,0)
  \CArc(75,0)(25,0,90)
  \CArc(75,0)(25,-180,0)
  \CArc(75,0)(25,90,180)
  \Line(100,0)(150,0)
  \Text(75,18)[cb]{{\Large$\times$}}
  \end{picture}}}
\]
\vspace{0.5cm}
\caption[]{The subtraction diagram, containing a one-loop counter-term
(represented by a $\times$) associated with the two-loop diagram
$S^{131}$ of \fig{tops5}.}
\label{counters5}
\end{figure}
%--
\subsection{Landau equations for $S^{131}$ when $m_5 = m_3$\label{lesacasc}}
%--
The Landau equations for this topology and with $m_5 = m_3$ are as follows:
%--
\bqa
{}&{}&\alpha_1\,(q^2_1+m^2_1) = 0, \qquad
\alpha_2\,((q_1-q_2)^2+m^2_2) = 0, \nl
{}&{}& \alpha_3\,(q^2_2+m^2_3) = 0, \qquad
\alpha_4\,((q_2+p)^2+m^2_4) = 0,  \nl
{}&{}& \alpha_5\,(q^2_2+m^2_5) = 0,
\label{nnland1}
\eqa
%--
and also
%--
\bqa
{}&{}& \alpha_1 q_{1\mu} + \alpha_2 (q_1-q_2)_{\mu} = 0,
\nl
{}&{}& - \alpha_2 (q_1-q_2)_{\mu} + \alpha_3 q_{2\mu} +
\alpha_4\,(q_2+p)_{\mu} + \alpha_5\,q_{2\mu}= 0.
\label{nnland2}
\eqa
%--
The leading Landau singularity occurs for $\alpha_i \ne 0, \forall i$.
We multiply the two equations \eqn{nnland2} by $q_{1\mu}, q_{2\mu}$ and
$p_{\mu}$ respectively. This gives an homogeneous system of six equations.
If all $\alpha_i$ are different from zero, the singularity will occur for
%--
\[
\ba{ll}
q^2_1 = - m^2_1 & \;\; q^2_2 = - m^2_3, \\
\spro{q_1}{q_2} = \frac{1}{2}\,(m^2_2 - m^2_1 - m^2_3) & \;\;
\spro{p}{q_2} = \frac{1}{2}\,( s + m^2_3 - m^2_4).
\ea
\]
%--
The equations become as follows:
%--
\bqa
{}&{}& -2\,m^2_1\,\alpha_1 -
(m^2_1+m^2_2-m^2_3)\,\alpha_2 = 0,
\nl
{}&{}&  (m^2_1-m^2_2+m^2_3)\,\alpha_1 +
(m^2_1-m^2_2-m^2_3)\,\alpha_2 = 0,
\nl
{}&{}& 2\,\spro{p}{q_1}\,\alpha_1 +
(2\,\spro{p}{q_1}-s-m^2_3+m^2_4)\,\alpha_2 = 0,
\nl
{}&{}& (m^2_1+m^2_2-m^2_3)\,\alpha_2 -
(m^2_1-m^2_2+m^2_3)\,(\alpha_3 + \alpha_5) +
(2\,\spro{p}{q_1}-m^2_1+m^2_2-m^2_3)\,\alpha_4 = 0,
\nl
{}&{}& (m^2_1-m^2_2-m^2_3)\,\alpha_2 -
2\,m^2_3\,(\alpha_3+\alpha_5) +
(s-m^2_3-m^2_4)\,\alpha_4 = 0,
\nl
{}&{}& (-\,2\,\spro{p}{q_1}+s+m^2_3-m^2_4)\,\alpha_2 +
(s+m^2_3-m^2_4)\,(\alpha_3+\alpha_5) -
(s-m^2_3+m^2_4)\,\alpha_4 = 0.
\eqa
%--
Compatibility between the first three equations requires the conditions
%--
\bq
m^2_3 = ( m_1+m_2)^2,  \qquad
\spro{p}{q_1} = \frac{m^2_1}{m^2_1-m^2_2+m^2_3}\,(s+m^2_3-m^2_4).
\eq
%--
As a consequence, it follows that
%--
\bq
\alpha_1 = \frac{m_2}{m_1}\,\alpha_2,  \quad
\alpha_2 = \frac{1}{m_2\/(m_1+m_2)}\,\Bigl\{
- (m_1+m_2)^2\,(\alpha_3 + \alpha_5) +
\frac{1}{2}\,\Bigl[ s - (m_1+m_2)^2 - m^2_4 \Bigr]\,\alpha_4\Bigr\},
\eq
%--
is a solution, if and only if $s = ( m_1 + m_2 \pm m_4)^2$ and 
$m_3 = m_1 + m_2$,
therefore, representing the leading Landau singularity. We are now in a
position to attempt the evaluation of this topology.
%--
\subsection{Evaluation of $S^{131}$ for $m_5 = m_3$\label{esacasp}}
%--
In \eqn{eqmass} we have a power $-2-\ep$ but for the first iteration we
prefer to use the identity
%--
\bq
\Bigl[ 1 + \frac{1}{1+\ep}\,(y\partial_y + z\partial_z)\Bigr]\,
\chi_{\aca}^{-1-\ep} =
\lpar \mu^2_x - z^2\rpar\,\chi_{\aca}^{-2-\ep},
\eq
%--
and obtain
%--
\bqa
S^{131} &=& - \frac{1}{s}\,\lpar\frac{\tHss}{\pi\,s}\rpar^{\ep}\,
\egam{1+\ep}\,\intfxx{x}{y}\intfz\,
\Bigl[ x(1-x)\Bigr]^{1+\ep/2}(1-y)^{\ep/2-1}  \nl
{}&\times& (y-z)\,X_{\aca}^{-1-\ep}(x,y,z),
\eqa
%--
with $X = x(1-x)\,\chi$. It is more convenient to change variables,
$y= 1- y'$ and write
%--
\bqa
S^{131} &=& - \frac{1}{s}\,\lpar\frac{\tHss}{\pi\,s}\rpar^{\ep}\,
\egam{1+\ep}\,\intfxx{x}{y}\intfzp\,
\Bigl[ x(1-x)\Bigr]^{1+\ep/2}\,y^{\ep/2-1}  \nl
{}&\times& (1-y-z)\,X_{\aca}^{-1-\ep}(x,1-y,z),
\eqa
%--
where now the quadratic form is
%--
\bq
X_{\aca}= x(1-x)\,\chi_{\aca}, \qquad
\chi_{\aca}(x,1-y,z) = z^2 - (1+\mu^2_3-\mu^2_4)\,z +
(\mu^2_x -\mu^2_3)\,y + \mu^2_3,
\eq
%--
and where \eqn{hsing} can be used. We introduce
%--
\bq
b_{\aca} = \lambda(1,\mu^2_3,\mu^2_4).
\eq
%--
The finite part of $S^{131}$ becomes
%--
\bq
S^{131}_{\rm fin} = -\,\frac{2}{s\,b_{\aca}}\,
\sum_{n=0}^{3}\,I^{\aca}_n,
\eq
%--
\bqa
I^{\aca}_3 &=& \intfxx{x}{y}\intfzp\,\Bigl[ 3\,\ln X_{\aca}(x,y,z) +
( 2\,z - 1 - \frac{1}{2}\,\mu^2_{34})\,\lpar\frac{\ln X_{\aca}(x,y,z)}{y}
\rpar_+\Bigr],  \nl
I^{\aca}_2 &=& \intfxx{x}{y}\,\Bigl\{ \ln X_{\aca}(x,0,y)\,\Bigl[
2\,(1-y) + ( 2\,y - 1 - \frac{1}{2}\,\mu^2_{34})\,L_{\aca}(x,y)\Bigr]  \nl
{}&+& \frac{1}{2}\,\mu^2_{34}\,\Bigl[ \lpar\frac{\ln
X_{\aca}(x,y,0)}{y}\rpar_+
- \ln X_{\aca}(x,y,0)\Bigr]\Bigr\},  \nl
I^{\aca}_1 &=& \frac{1}{2}\,\mu^2_{34}\,\intfx{x}\,\Bigl[ \ln(x-x^2) -
\ln X_{\aca}(x,0,0)\Bigr]\,\ln X_{\aca}(x,0,0),  \nl
I^{\aca}_0 &=& \frac{7}{2},
\eqa
%--
where $L_{\aca}$ is given by
%--
\bq
L_{\aca}(x,y)= \ln(1-y) + \ln x + \ln(1-x) - \ln X_{\aca}(x,0,y).
\eq
%--
The obvious advantage of this result is that $b_{\aca}$ does not depend on
$x$.
However, $b_{\aca} = 0$ for $s = (m_3\pm m_4)^2$. In this case we have
a quadratic form $z^2 + b\,z + c\,y + d$ where $4\,d = b^2$. Therefore, we
consider first the case $\mu^2_{34} = - 2\,\mu_3$; we have
%--
\bq
\chi_{\aca}(x,y,z) = ( z + \mu_3)^2 + (\mu^2_x - \mu^2_3)\,y.
\eq
%--
It is convenient to change variable, $z = z'-\mu_3$, so that the integral
to be evaluated becomes
%--
\bqa
J &=& \intfx{y}\intfzp\,y^{\ep/2-1}\,(1-y-z)\,\chi_{\aca}^{-1-\ep}(x,y,z)
\nl
{}&=& \intfx{y}\,\int_{\mu_3}^{1+\mu_3-y}\,dz\,y^{\ep/2-1}\,
(1 + \mu_3 - y - z)\,(z^2+cy)^{-1-\ep},
\label{defJ}
\eqa
%--
with $c = \mu^2_x-\mu^2_3$. After changing variables, $z^2 = t$ we use
%--
\bq
(t+cy)^{-1-\ep} = -\,\frac{1}{\ep}\,\partial_t\,(t+cy)^{-\ep},
\eq
%--
and integrate by parts. The final result, which is valid for
$\sqrt{s} = m_4 - m_3$ and $m_4 \ge m_3$ can be cast into the following
form:
%--
\bqa
J &=& -\,\frac{1}{\ep}\,\Bigl[ \frac{\ln\mu^2_3}{\mu_3} - 2\,(1+\mu_3)\,
\int_{\mu_3}^{1+\mu_3}\,dy\,\frac{\ln y}{y^2}\Bigr]  \nl
{}&-&
\frac{1}{2}\,\intfx{y}\,\int_{\mu_3}^{1+\mu_3-y}\,\frac{dz}{z^2}\,\Bigl[
\ln(z^2+cy) - (1+\mu_3)\,\lpar\frac{\ln(z^2+cy)}{y}\rpar_+\Bigr]  \nl
{}&+& (1+\mu_3)\,\int_{\mu_3}^{1+\mu_3}\,dy\,\Bigl[
\frac{\ln y\ln(1+\mu_3 -y)}{y^2} - \frac{\ln^2 y}{y^2}\Bigr]
\nl
{}&+& \frac{1}{2\,\mu_3}\,\intfx{y}\,\Bigl[
\ln(\mu^2_3+cy) - \lpar\frac{\ln(\mu^2_3+cy)}{y}\rpar_+\Bigr] +
\frac{1}{2\,\mu_3}\,\ln^2\mu^2_3.
\eqa
%--
The case $\sqrt{s} = m_3-m_4$, with $m_3 \ge m_4$ is obtained after the
substitution $z = 1 - z'$ in \eqn{defJ}. We obtain
%--
\bqa
J &=& \intfx{y}\intfzs\,y^{\ep/2-1}\,(z-y)\,\chi_{\aca}^{-1-\ep}(x,y,1-z)
\nl
{}&=& \intfx{y}\,\int_{y+\mu_4}^{1+\mu_4}\,dz\,y^{\ep/2-1}\,
(z - \mu_4 - y)\,(z^2+cy)^{-1-\ep}, \nl
\chi_{\aca}(x,y,1-z) &=& z^2 - \mu^2_{43}\,z + (\mu^2_x - \mu^2_3)\,y +
\mu^2_4,
\eqa
%--
and the case considered here corresponds to $\mu^2_{43} = -2\,\mu_4$.
Therefore the pseudo-threshold $s = (m_3 - m_4)^2$ is covered.

Always for $m_5 = m_3$ we have a second method which is based on the
observation that the following identity holds:
%--
\bq
S^{131} = -\,\intfx{x}\,(1-x)\,\frac{\partial}{\partial M^2_x}\,
(\ep-1)\,S_{33}(x^2p^2;m_1,m_2,M_x),
\eq
%--
with $\mu^2_x = M^2_x/s$ and $\mu^2_x = x^2 - \mu^2_{34}\,x + \mu^2_3$.
The idea is to change variable $x \to \mu^2_x$ in the integral and perform
integration by parts. Unless numerical differentiation is used, this method
will fail at $b_{\aca} = 0$ and, therefore, it will be no longer discussed.
%--
\subsubsection{Evaluation around $s= (m_3+m_4)^2$}
%--
In this region we use
%--
\bq
\mu^2_3 = \frac{1}{4}\,(\mu^4_{34} - \lambda), \qquad
\lambda \equiv \lambda(1,\mu^2_3,\mu^2_4),
\eq
%--
and write the master integral $J$ of \eqn{defJ} as
%--
\bq
J = \intfx{y}\int_{-z_m}^{1-z_m-y}\,dz\,y^{\ep/2-1}\,
(1-z_m-y-z)\,(z^2 + c y-\frac{1}{4}\,\lambda)^{-1-\ep},
\eq
%--
where $z_m = \mu^2_{34}/2$ and $c = \mu^2_x-\mu^2_3$. As a first
step we will set $\lambda = 0$ whenever possible; $J$ is the sum of three
contributions, the first one being
%--
\bq
J_1 = - \intfx{y}\int_{-z_m}^{1-z_m-y}\,dz\,y^{\ep/2-1}\,z\,
(z^2 + c y)^{-1-\ep}.
\eq
%--
Here we use the identity
%--
\bq
-\,\frac{1}{\ep}\,\partial_z\,(z^2 + c y)^{-\ep} = 2\,z\,
(z^2 + c y)^{-1-\ep},
\eq
%--
which gives, after integration by parts, the following result:
%--
\bqa
J_1 &=& \frac{2}{\ep}\,\ln\frac{z_m}{1-z_m} + 2\,\Bigl[ \ln^2z_m -
\ln^2(1-z_m)\Bigr]  \nl
{}&+& \frac{1}{2}\,\intfx{y}\,\Bigl[ \frac{\ln\eta(y,-z_m)}{y} -
\frac{\ln\eta(y,1-z_m-y)}{y}\Bigr]_+,
\eqa
%--
where $\eta(y,z) = z^2 + c y$. The second contribution, $J_2$ is
defined by
%--
\bq
J_2 = - \intfx{y}\int_{-z_m}^{1-z_m-y}\,dz\,y^{\ep/2}\,
(z^2 + c y)^{-1-\ep}.
\eq
%--
Here we will us another identity, i.e.
%--
\bq
-\,\frac{1}{\ep}\,\partial_y\,(z^2 + c y)^{-\ep} =
c \,(z^2 + c y)^{-1-\ep},
\eq
%--
giving, after integration by parts, the following result:
%--
\bq
J_2 = -\,\frac{1}{c}\,\int_{-z_m}^{1-z_m}\,dz\,\Bigl[
\ln\eta(1-z_m-z,z) - \ln\eta(0,z)\Bigr].
\label{J2res}
\eq
%--
The last contribution is the most difficult to obtain. We start from
the definition of $J_3$,
%--
\bq
J_3 = - \intfx{y}\int_{-z_m}^{1-z_m-y}\,dz\,y^{\ep/2-1}\,
(z^2 + c y)^{-1-\ep}.
\eq
%--
The $y$ integration range is further divided into two regions, $[0,1-z_m]$
and $[1-z_m,1]$. In the former one $1-z_m-y$ is positive and we write
%--
\bq
J_{31} = \int_0^{1-z_m}\,dy\,y^{\ep/2-1}\,\sum_{i=1}^{2}\,\int_0^{Y_i}\,
dz\,(z^2 + c y)^{-1-\ep},
\eq
%--
where $Y_1= z_m$ and $Y_2 = 1- z_m - y$. Therefore $J_{31}$ is a combination
of new integrals denoted by
%--
\bq
H_i = \int_0^{Y_i}\,dz\,(z^2 + c y)^{-1-\ep} = ( c y)^{-1-\ep} Y_i\,
\hyper{1+\ep}{\frac{1}{2}}{\frac{3}{2}}{-\,\frac{Y^2_i}{c y}}.
\eq
%--
Using a well known property of the hypergeometric functions~\cite{bat} we
obtain
%--
\bqa
{}&{}&
\hyper{1+\ep}{\frac{1}{2}}{\frac{3}{2}}{-\,\frac{Y^2_i}{c y}} =
B_2\,\lpar \frac{c}{Y^2_i}\,y\rpar^{1/2}  \nl
{}&+& B_1\,\lpar \frac{c}{Y^2_i}\,y\rpar^{1+\ep}\,
\hyper{1+\ep}{\frac{1}{2}+\ep}{\frac{3}{2}+\ep}{-\,\frac{c}{Y^2_i}\,y},
\eqa
%--
where the coefficients $B_{1,2}$ are as follows:
%--
\bqa
B_1 &=& {{\egam{3/2}\egam{-1/2-\ep}}\over {\egam{1/2}\egam{1/2-\ep}}} =
- \frac{1}{1 + 2\,\ep},  \nl
B_2 &=& {{\egam{3/2}\egam{1/2+\ep}}\over {\egam{1+\ep}}} =
\frac{1}{2}\,\pi\,
\Bigl\{ 1 + \ep\,\bigl[ \gamma + \psi(\frac{1}{2})\Bigr] +
\ord{\ep^2}\Bigr\},
\eqa
%--
where $\psi(z)$ is the Euler psi-function~\cite{bat}.
In this way the functions $H_i$ become
%--
\bq
H_i = B_1\,Y^{-1-2\ep}_i\,
\hyper{1+\ep}{\frac{1}{2}+\ep}{\frac{3}{2}+\ep}{-\,\frac{c}{Y^2_i}\,y} +
B_2\,( c y)^{-1/2-\ep}.
\label{defHi}
\eq
%--
At this point we observe that the term proportional to $B_2$ will give rise
to a divergency after the $y$-integration. If $\lambda$ is kept different
from zero then \eqn{defHi} becomes
%--
\bq
H_i = B_1\,Y^{-1-2\ep}_i\,
\hyper{1+\ep}{\frac{1}{2}+\ep}{\frac{3}{2}+\ep}
{-\,\frac{c y - \lambda/4}{Y^2_i}} +
B_2\,( cy - \frac{1}{4}\,\lambda)^{-1/2-\ep}.
\label{resHi}
\eq
%--
Consider now the $y$-integration for the second term: we obtain
%--
\bqa
H_{i2} &=& \int_0^{1-z_m}\,dy\,y^{\ep/2-1}\,
(c y - \frac{1}{4}\,\lambda)^{-1/2-\ep}  \nl
{}&=& 2\,\lpar -\frac{1}{4}\,\lambda\rpar^{-1/2-\ep}\,\frac{(1-z_m)^{\ep/2}}
{\ep}\,\hyper{\frac{1}{2}+\ep}{\frac{\ep}{2}}{1+\frac{\ep}{2}}
{4\,\frac{c}{\lambda}(1-z_m)}  \nl
{}&=& 2\,\frac{b_1}{\ep}\,(1-z_m)^{-1/2-\ep/2}\,c^{-1/2-\ep}\,
\hyper{\frac{1}{2}+\ep}{\frac{1+\ep}{2}}{\frac{3+\ep}{2}}
{\frac{\lambda}{4\,c(1-z_m)}}  \nl
{}&+& 2\,\frac{b_2}{\ep}\,c^{-\ep/2}\,\lpar - \frac{1}{4}
\lambda\rpar^{-1/2-\ep/2},
\eqa
%--
where the coefficients $b_{1,2}$ are
%--
\bq
b_1 = {{\egam{-1/2-\ep/2}}\over {\egam{1/2-\ep/2}}}\,\frac{\ep}{2},
\quad
b_2 = {{\egam{1+\ep/2}\egam{1/2+\ep/2}}\over {\egam{1/2+\ep}}}.
\eq
%--
For the term proportional to $b_1$ there is no UV pole and we may set
$\ep = 0$. Therefore we get
%--
\bqa
H_{i2} &=& -\,2(1-z_m)^{-1/2}\,c^{-1/2}\,
\hyper{\frac{1}{2}}{\frac{1}{2}}{\frac{3}{2}}{\frac{\lambda}{4\,c(1-z_m)}} +
2\,\frac{b_2}{\ep}\,c^{-\ep/2}\,\lpar - \frac{1}{4}
\lambda\rpar^{-1/2-\ep/2}  \nl
{}&=& -\,4\,\lambda^{-1/2}\,\arcsin\Bigl[ \frac{\lambda}{4\,c(1-z_m)}
\Bigr]^{1/2} + 2\,(-\,\lambda)^{-1/2}\,\Bigl[ \frac{2}{\ep} - \ln c -
\ln(-\,\frac{\lambda}{4}) - \gamma - \psi(\frac{1}{2})\Bigr].
\eqa
%--
Setting $\lambda = 0$ whenever possible produces the following result:
%--
\bq
H_{i2} = -\,2(1-z_m)^{-1/2}\,c^{-1/2} +
2\,(-\,\lambda - i\,\delta)^{-1/2}\,\Bigl[ \frac{2}{\ep} - \ln c -
\ln(-\,\frac{\lambda}{4} - i\,\delta) - \gamma - \psi(\frac{1}{2})\Bigr],
\eq
%--
where $\delta \to 0_+$. Returning to \eqn{resHi} we consider the part
proportional to $B_1$ and the corresponding $y$-integral which becomes
%--
\bq
H_{i1} = \int_0^{1-z_m}\,dy\,y^{\ep/2-1}\,f_i(\ep,y)  \quad
f_i(\ep,y) = Y^{-1-2\ep}_i\,
\hyper{1+\ep}{\frac{1}{2}+\ep}{\frac{3}{2}+\ep}{-\,\frac{c}{Y^2_i}\,y}.
\eq
%--
We split the integral into two parts, according to the following recipe:
%--
\bqa
H_{i1} &=& \int_0^{1-z_m}\,dy\,\Bigl\{ y^{\ep/2-1}\,f_i(\ep,0) +
y^{\ep/2}\,\Bigr[\frac{f_i(\ep,y)}{y}\Bigr]_+\Bigr\}  \nl
{}&=& \frac{2}{\ep}\,(1-z_m)^{\ep/2}\,f_i(\ep,0) +
\int_0^{1-z_m}\,dy\,\Bigr[\frac{f_i(0,y)}{y}\Bigr]_+ + \ord{\ep}.
\label{defHi1}
\eqa
%--
The function $f_i$ is defined by
%--
\bqa
f_i(\ep,0) &=& Y^{-1-2\ep}_i(y=0),  \qquad
f_i(0,0) = \frac{1}{Y_i(y=0)},  \nl
f_i(0,y) &=& \frac{1}{Y_i}\,\hyper{1}{\frac{1}{2}}{\frac{3}{2}}
{-\,\frac{c y}{Y^2_i}} = \frac{1}{2}\,(-\,c y)^{-1/2}\,
\ln {{Y_i + (-\,cy)^{1/2}}\over {Y_i - (-\,c y)^{1/2}}}.
\eqa
%--
Since $J_{31}$ is the following combination,
%--
\bq
J_{31} = \sum_{i=1}^{2}\,( B_1\,H_{i1} + B_2\,H_{i2}),
\eq
%--
we obtain the final result by further inserting
%--
\bqa
2\,\frac{B_1}{\ep}\,Y^{-1-2\ep}_1(y=0)\,(1-z_m)^{\ep/2} &=&
-\,\frac{2}{z_m}\,\Bigl[ \frac{1}{\ep} + \frac{1}{2}\,\ln(1-z_m) -
2\,\ln z_m - 2\Bigr],  \nl
2\,\frac{B_1}{\ep}\,Y^{-1-2\ep}_2(y=0)\,(1-z_m)^{\ep/2} &=&
-\,\frac{2}{1-z_m}\,\Bigl[ \frac{1}{\ep} - \frac{3}{2}\,\ln(1-z_m) -
 2\Bigr].
\eqa
%--
The second part of the function is $J_{32}$ defined by
%--
\bq
J_{32} = \int_{1-z_m}^{1}\,dy\,y^{\ep/2-1}\,\int_{-Y_2}^{Y_1}\,
dz\,(z^2 + c y)^{-1-\ep},
\eq
%--
where we can set $\ep = 0$ from the beginning. We obtain
%--
\bqa
J_{32} &=& -\,\frac{1}{c}\,\int_{1-z_m}^{1}\,\frac{dy}{y^2}\,
\sum_{i=1}^{2}(-1)^i\,|Y_i|\,\hyper{1}{\frac{1}{2}}{\frac{3}{2}}
{-\,\frac{Y^2_i}{c y}}
\nl
{}&=& \frac{1}{2}\,\int_{1-z_m}^{1}\,\frac{dy}{y}\,
\sum_{i=1}^{2}(-1)^i\,(-c y)^{-1/2}\,\ln {{(-c y)^{1/2} + |Y_i|}\over
{(- c y)^{1/2} - |Y_i|}}.
\eqa
%--
In all formulas the square roots must be understood with the replacement
$c y \to c y - i\,\delta$, where $\delta \to 0_+$.

Let us return to \eqn{defHi} and consider also the term proportional
to $B_1$ for $\lambda \ne 0$. In this case we will use again \eqn{defHi1}
but with
%--
\bq
f_i(\ep,y) = Y^{-1-2\ep}_i\,\hyper{1+\ep}{\frac{1}{2}+\ep}{\frac{3}{2}+\ep}
{-\,\frac{c y - \lambda/4}{Y^2_i}}.
\eq
%--
Due to the pole at $\ep = 0$ we must expand $f_i(\ep,0)$ to $\ord{\ep}$.
We derive
%--
\bq
\hyper{1+\ep}{\frac{1}{2}+\ep}{\frac{3}{2}+\ep}{y^2_i} =
\frac{\egam{3/2+\ep}}
{\egam{1/2+\ep}}\,\int_0^1\,dt\,t^{-1/2}(1-y^2_i\,t)^{-1}\,\Bigl\{
1 + \ep\,\Bigl[ \ln t - \ln(1-y^2_i\,t)\Bigr]\Bigr\},
\eq
%--
where we have introduced $y^2_i = \lambda/(4\,Y^2_i)$. Therefore, expanding
around $\ep = 0$, we obtain
%--
\bqa
{}&{}&
\hyper{1+\ep}{\frac{1}{2}+\ep}{\frac{3}{2}+\ep}{y^2_i} = \frac{1}{2\,x_i}\,
\ln\frac{1+x_i}{1-x_i} + \ep\,{\cal F},  \nl
{}&{}&
{\cal F} = -\,\intfx{x}\,\frac{\Bigl[ 2\,\ln x - \ln(1 - y^2_i x^2)\Bigr]}
{y^2_i x^2 - 1} + \frac{1}{x_i}\,\ln\frac{1+x_i}{1-x_i},
\eqa
%--
a result which is based on the standard integral representation of the
hypergeometric function~\cite{bat}.
The integral can be computed exactly; introducing $x_i = \sqrt{y^2_i} =
\sqrt{\lambda}/(2\,Y_i)$ we obtain
%--
\bqa
{\cal F} &=& -\,\frac{1}{x_i}\,\Bigl[ \li{2}{x_i} - \li{2}{-\,x_i}\Bigr] -
\frac{1}{4\,x_i}\,\Bigl[ \ln^2(1+x_i) - \ln^2(1-x_i)\Bigr]  \nl
{}&-&
\frac{1}{2\,x_i}\,\Bigl[ \ln 2\,\ln(1+x_i) + \li{2}{\frac{1+x_i}{2}} -
\li{2}{\frac{1}{2}}\Bigr]  \nl
{}&+&\frac{1}{2\,x_i}\,\Bigl[ \ln 2\,\ln(1-x_i) + \li{2}{\frac{1-x_i}{2}} -
\li{2}{\frac{1}{2}}\Bigr] + \frac{1}{x_i}\,\ln\frac{1+x_i}{1-x_i},
\label{defcalF}
\eqa
%--
where $Li_2(1/2) = (\zeta(2)-\ln^2 2)/2$.
To obtain \eqn{defcalF} we have used the following integrals~\footnote{
We disagree with Eq.(1) p. 34 of ref.~\cite{Devoto:1984tc}}:
%--
\bq
\intfx{x}\,\frac{\ln x}{x \pm\, x^{-1}_i} = \li{2}{\mp\, x_i},
\quad
\intfx{x}\,\frac{\ln(1 \pm \,x_i x)}{1 \pm\, x_i x} = \pm \,
\frac{1}{2\,x_i}\,\ln^2(1 \pm\, x_i),
\eq
%--
\bq
\intfx{x}\,\frac{\ln(1 \pm\, x_i x)}{1 \mp\, x_i x} = \mp\,
\frac{1}{x_i}\,\Bigl[ \ln 2\,\ln(1 \mp\, x_i) -
\li{2}{\frac{1 \mp\, x_i}{2}} + \li{2}{\frac{1}{2}}\Bigr].
\eq
%--
Furthermore, $\li{2}{1/2}= (\zeta(2) - \ln^2 2)/2$.
The implementation of this result requires distorting the $x$-integration
contour for $J_2$ (\eqn{J2res}) in order to avoid the points
%--
\bq
x_{\pm} = \frac{1}{2\,\mu^2_3}\,\Bigl[ \mu^2_1 - \mu^2_2 + \mu^2_3 \pm
\lambda^{1/2}(\mu^2_3,\mu^2_1,\mu^2_2)\Bigr].
\eq
%--
These two points coincide for $\mu^2_3 = (\mu_1+\mu_2)^2$. In practical
cases this may happens for $m_1 = m_3$ and $m_2 = 0$ where we have
%--
\bq
\mu^2_x = \frac{\mu^2_1}{x}, \quad c = \mu^2_x - \mu^2_3 =
\mu^2_1\,\frac{1-x}{x}.
\eq
%--
Actually there is no problem at $x = 1 (c = 0)$ since the result
for $J_2$ is
%--
\bq
J_2 = \frac{1}{\mu^2_1}\,\frac{x}{1-x}\,\int_{-z_m}^{1-z_m}\,dz\,
\Bigl\{ \ln z^2 - \ln\Bigl[ z^2 + \mu^2_1\,\frac{1-x}{x}\,(1-z_m-z)\Bigr]
\Bigr\}.
\eq
%--
Our result for the threshold region deviates from the main approach
of expressing the integrand as products of polynomials and logarithms
of polynomials and requires some analytical work before attempting
the numerical evaluation. However, in this way, we have been able to isolate
the square root singularity which shows up in the coefficient of the single
pole and will disappear after including the corresponding subtraction
diagram. The net result can be safely computed by numerical integration.
%--
\subsection{Derivative of $S^{131}$ and infrared poles\label{dsacaip}}
%--
Self-energy diagrams enter into the calculation of wave function
renormalization factor and, therefore, also their derivative with respect to
$p^2$ is needed. We concentrate on the on-shell derivative of $S^{131}$
where $m_5 = m_3$ and compute
%--
\bqa
{}&{}&
S^{131}_{p} \equiv \frac{\partial}{\partial p^2}\,
S^{131}(p^2;m_1,m_2,m_3,m_4,m_3) =  \nl
{}&-& \frac{1}{2\,p^2}\,\Bigl[ ( p^2+m^2_3-m^2_4)\,
S^{141}(p^2;m_1,m_2,m_3,m_4,m_3,m_4)  \nl
{}&+& S^{131}(p^2;m_1,m_2,m_3,m_4,m_3) -
S^{131}(p^2;m_1,m_2,m_3,m_4,m_4)\Bigr].
\eqa
%--
The result for the derivative is
%--
\bqa
S^{131}_{p} &=& -\,\frac{1}{s^2}\,
\lpar\frac{\tHss}{\pi\,s}\rpar^{\ep}\,\frac{1}{\ep}\,
\intfx{x}\,\Bigl[ x(1-x)\Bigr]^{-\ep/2}\,\intfxy{y}{z}\,(1-y)^{\ep/2}  \nl
{}&\times& \Bigl\{ \egam{3+\ep} z\,(y-z)\,(\mu^2_3-\mu^2_4-1)\,
\Bigl[ (\mu^2_x-z^2)\chi_{\aca}^{-3-\ep} + \frac{2-\ep/2}{2+\ep}\,
\chi_{\aca}^{-2-\ep}\Bigr]
\nl
{}&+& \egam{2+\ep}\,(y-2\,z)\,\Bigl[ (\mu^2_x-z^2)\,\chi_{\aca}^{-2-\ep} +
\frac{2-\ep/2}{1+\ep}\,\chi_{\aca}^{-1-\ep}\Bigr]\Bigr\}.
\label{sacap}
\eqa
%--
It is important to recall once more that a necessary condition for the 
presence of infrared divergences is that the Landau equations are fulfilled.
In the case of $S^{131}_p$ we see that $s = (m_1 + m_2 \pm m_4)^2$ and
$m^2_3 = (m_1 + m_2)^2$ are satisfied by the QED-like configuration,
$m_2 = m_4 = 0$, $ m_1 = m_3 = m$ and $s = m^2$ but, for instance,
they are not when $m_1 = m_3 = m$, $m_2 = 0$ and $s = m^2$ but
$m_4 = M$. This configuration, that corresponds to electron (or $\wb$-boson)
self-energy with $2 \equiv \ph$ and $4 \equiv \zb$, is therefore
free from infrared poles.

\begin{figure}[th]
\vspace{0.5cm}
\[
  \vcenter{\hbox{
  \begin{picture}(150,0)(0,0)
  \ArrowLine(0,0)(50,0)
  \ArrowArcn(75,0)(25,180,0)
  \PhotonArc(75,0)(25,-180,0){2}{7}
  \PhotonArc(75,50)(45,-120,-60){2}{7}
  \ArrowLine(100,0)(150,0)
  \end{picture}}}
\]
\vspace{0.5cm}
\caption[]{An $S^{131}$-type electron self-energy in QED.}
\label{wfQED}
\end{figure}
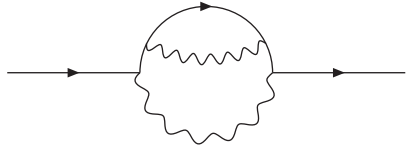

\begin{figure}[th]
\vspace{0.5cm}
\[
  \vcenter{\hbox{
  \begin{picture}(150,0)(0,0)
  \ArrowLine(0,0)(50,0)
  \PhotonArc(75,0)(25,120,180){2}{7}
  \ArrowArcn(75,0)(25,120,60)
  \PhotonArc(75,0)(25,0,60){2}{7}
  \ArrowArc(75,0)(25,-180,0)
  \ArrowArcn(75,45)(25,-60,-120)
  \ArrowLine(100,0)(150,0)
  \end{picture}}}
\]
\vspace{0.5cm}
\caption[]{A second $S^{131}$-type electron self-energy in QED. The innermost
loop can be a fermion of any flavor.}
\label{wfQEDb}
\end{figure}
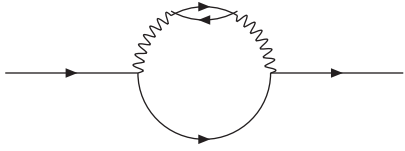

Let us consider a realistic case, \fig{wfQED} which represents one of the
contributions to the electron self-energy in QED.
%--
%--
In this case the configuration is
%--
\bq
\mbox{a)} \qquad m_1 = m_3 = m \quad \mbox{and} \quad m_2 = m_4 = 0
\eq
%--
and, therefore
%--
\bq
\chi_{\aca} = z^2 - (1+\mu^2)\,z + (\mu^2-\mu^2_x)\,y + \mu^2_x, \quad
\mbox{with} \quad \mu^2_x = \frac{\mu^2}{x},
\eq
%--
with $\mu^2= m^2/s$. The derivative is needed for on-shell electrons,
therefore $\mu^2= 1$ and
%--
\bq
\chi_{\acan{\rm os}} = z^2 -2\,z + (1 - \frac{1}{x})\,y + \frac{1}{x}.
\eq
%--
We therefore obtain
%--
\bqa
S^{131}_{p}|_{_{\rm os}} &=& -\,\frac{1}{m^4}\,
\lpar\frac{\tHss}{\pi\,m^2}
\rpar^{\ep}\,\frac{\egam{2+\ep}}{\ep}\,
\intfx{x}\,\Bigl[ x(1-x)\Bigr]^{-\ep/2}\,\intfxy{y}{z}\,(1-y)^{\ep/2}  \nl
{}&\times& (y-2\,z)\,\Bigl[ (\frac{1}{x}-z^2)\,\chi_{\acan{\rm os}}^{-2-\ep}
+
\frac{2-\ep/2}{1+\ep}\,\chi_{\acan{\rm os}}^{-1-\ep}\Bigr].
\eqa
%--
The well-known infrared singularity shows up in parametric space at 
$y = z = 1$.
Note that the diagram itself is not infrared divergent but only the
on-shell derivative, due to the factor $y-z$ in the integral of
\eqn{eqmass}. To continue our derivation, we use the identity
%--
\bq
\Bigl[ 1 + \frac{1}{1+\ep}\,(y\partial_y + z\partial_z)\Bigr]\,
\chi_{\acan{\rm os}}^{-1-\ep} =
\lpar \frac{1}{x} - z^2\rpar\,\chi_{\acan{\rm os}}^{-2-\ep},
\eq
%--
to write the on-shell derivative as
%--
\bqa
S^{131}_{p}|_{_{\rm os}} &=& -\,\frac{1}{m^4}\,
\lpar\frac{\tHss}{\pi\,m^2}\rpar^{\ep}\,\egam{\ep}\,
\intfx{x}\,\Bigl[ x(1-x)\Bigr]^{-\ep/2}\,\intfxy{y}{z}\,(1-y)^{\ep/2}  \nl
{}&\times& (y-2\,z)\,\Bigl[ 3 + \frac{\ep}{2} +
(y\partial_y + z\partial_z)\Bigr]\,\chi_{\acan{\rm os}}^{-1-\ep}.
\eqa
%--
After integration by parts we obtain
%--
\bqa
S^{131}_{p}|_{_{\rm os}} &=& -\,\frac{1}{2\,m^4}\,
\lpar\frac{\tHss}{\pi\,m^2}\rpar^{\ep}\,\egam{1+\ep}\,
\intfx{x}\,\Bigl[ x(1-x)\Bigr]^{-\ep/2}  \nl
{}&\times& \intfxy{y}{z}\,(y-2\,z)\,(1-y)^{\ep/2-1}
\chi_{\acan{\rm os}}^{-1-\ep}  \nl
{}&=& -\,\frac{1}{2\,m^4}\,
\lpar\frac{\tHss}{\pi\,m^2}\rpar^{\ep}\,\egam{1+\ep}\,
\intfx{x}\,\Bigl[ x(1-x)\Bigr]^{-\ep/2}  \nl
{}&\times& \intfx{y}\,\intfzs
\,(2\,z-y-1)\,y^{\ep/2-1}\,\Bigl[ z^2 + (\frac{1}{x}-1)\,y\Bigr]^{-1-\ep}.
\eqa
%--
The total result is the sum of three contributions. We write
%--
\bq
J = \intfx{y}\,\intfzs\,(2\,z-y-1)\,y^{\ep/2-1}\,\Bigl[ z^2 +
(\frac{1}{x}-1)\,y\Bigr]^{-1-\ep} = J_1-J_2-J_3,
\eq
%--
and introduce $X = 1/x-1$. For $J_1$ we obtain
%--
\bqa
J_1 &=& -\,\frac{1}{\ep}\,\intfx{y}\intfzs\,y^{\ep/2-1}\,\partial_z\,
(z^2 + X\,y)^{-\ep}  \nl
{}&=& -\,\frac{1}{\ep}\,\intfx{y}\,\Bigl[ y^{\ep/2-1}\,(1+X\,y)^{-\ep} -
y^{-\ep/2-1}\,(y+X)^{-\ep}\Bigr].
\eqa
%--
For computing the remaining two integrals we use the following relation
%--
\bq
\Bigl[ 1 + \frac{1}{1+\ep}\,(\frac{1}{2}\,z\partial_z + y\partial_y)\Bigr]\,
( z^2 + X\,y)^{-1-\ep} = 0.
\eq
%--
After integration by parts we obtain the following results:
%--
\bqa
(1-\ep)\,J_2 &=& \intfx{y}\,\Bigl[ y^{\ep/2}\,(1+X\,y)^{-1-\ep} +
y^{-\ep/2}\,(y+X)^{-1-\ep}\Bigr],
\nl
(1+\ep)\,J_3 &=& -\,\intfx{y}\,\Bigl[ y^{\ep/2-1}\,(1+X\,y)^{-1-\ep} +
y^{-\ep/2-1}\,(y+X)^{-1-\ep}\Bigr].
\eqa
%--
The $J_i$ integrals are successively folded to become
%--
\bq
I_i = \intfx{x}\,\Bigl[x(1-x)\Bigr]^{-\ep/2}\,J_i.
\eq
%--
We illustrate in details one particular case and refer to Appendix C for
further details:
%--
\bq
{\cal J} = \intfx{y}\,y^{-\ep/2-1}\,(y+X)^{-\ep-1}.
\eq
%--
This integral gives rise to an hypergeometric function~\cite{bat},
%--
\bq
{\cal J} = -\,\frac{2}{\ep}\,X^{-1-\ep}\,\hyper{1+\ep}{-\frac{\ep}{2}}
{1-\frac{\ep}{2}}{-\,\frac{1}{X}}.
\eq
%--
Using well-known properties of the hypergeometric functions~\cite{bat} we
obtain
%--
\bqa
{\cal J} &=& -\,\frac{2}{\ep}\,x^{1+\ep}\,
\hyper{1+\ep}{1}{1-\frac{\ep}{2}}{x}  \nl
{}&=& -\,\frac{2}{\ep}\,x^{1+\ep}\,\Bigl[
A_1\;\hyper{1+\ep}{1}{2+\frac{3}{2}\,\ep}{1-x}  \nl
{}&+&
A_2\;(1-x)^{-1-3/2\ep}\,\hyper{-\frac{3}{2}\,\ep}{-\frac{\ep}{2}}
{-\frac{3}{2}\,\ep}{1-x}\Bigr]  \nl
{}&=& -\,\frac{2}{\ep}\,x^{1+\ep}\,\Bigl[
A_1\;\hyper{1+\ep}{1}{2+\frac{3}{2}\,\ep}{1-x} +
A_2\,x^{\ep/2}(1-x)^{-1-3/2\ep}\Bigr],
\eqa
%--
where the coefficients $A_i$ are given by
%--
\bq
A_1 = {{\egam{1-\ep/2}\egam{-1-3/2\ep}}\over
         {\egam{-3/2\ep}\egam{-\ep/2}}} = \frac{\ep}{2+3\,\ep},
\quad
A_2 = {{\egam{1-\ep/2}\egam{1+3/2\ep}}\over {\egam{1+\ep}}}.
\eq
%--
We next compute
%--
\bq
{\cal I} = \intfx{x}\,\Bigl[x(1-x)\Bigr]^{-\ep/2}\,{\cal J} = {\cal I}_1 +
{\cal I}_2.
\eq
%--
The first integral is
%--
\bqa
{\cal I}_1 &=& -\,\frac{2}{\ep}\,{{\egam{1-\ep/2}\egam{1+3/2\ep}}
\over {\egam{1+\ep}}}\,\intfx{x}\,x^{1+\ep}(1-x)^{-1-2\ep}  \nl
{}&=& -\,\frac{2}{\ep}\,{{\egam{1-\ep/2}\egam{1+3/2\ep}
\egam{2+\ep}\egam{-2\ep}}\over {\egam{1+\ep}\egam{2-\ep}}} = -\,
\frac{2}{\ep}\,\egam{-2\ep} + {\cal R}.
\eqa
%--
In expanding around $\ep = 0$ we use
%--
\bqa
\egam{-2\ep} &=& - \frac{1}{2\,\ep} - \gamma - \Bigl[ \gamma^2 + \zeta(2)
\Bigr]\,\ep + \ord{\ep^2},  \nl
\egam{a+\lambda} &=& \egam{a}\,\Bigl\{ 1 + \psi(a)\,\lambda +
\frac{1}{2}\,\lambda^2\,\Bigl[\psi^2(a) + \psi'(a)\Bigr] +
\ord{\lambda^3}\Bigr\}.
\eqa
%--
It follows that
%--
\bq
{\cal I}_1 = \frac{1}{\ep^2} + \frac{2}{\ep} + 2 + \frac{11}{4}\,\zeta(2).
\eq
%--
The second integral will be
%--
\bq
{\cal I}_2 = - \frac{2}{2+3\,\ep}\,\sum_{n=0}^{\infty}\,
{{\egam{2+3/2\ep}}\over {\egam{1+\ep}}}\,
{{\egam{n+1+\ep}}\over {\egam{n+2+3/2\ep}}}\,\intfx{x}\,
x^{1+\ep/2}\,(1-x)^{n-\ep/2},
\eq
%--
where we have introduced the hypergeometric series. After integrating over
$x$ we can put $\ep = 0$ and obtain that
%--
\bq
{\cal I}_2 = 1 -\,\zeta(2).
\eq
%--
We also have the subtraction diagram of \fig{counters5}. The counter-term,
in the MS-scheme, is $-2\,i\pi^2/\ep$. After taking the derivative with
respect to $p^2$ we put $p^2= -m^2$ and obtain
%--
\bq
S^{{\underline 1}3{\underline 1}}_{p}|_{_{\rm os}} =
-\,\frac{2}{m^4}\,\lpar\frac{\tHss}{\pi\,m^2}\rpar^{\ep}\,
\frac{\egam{2+\ep/2}}{\ep}\,\intfx{x}\,x\,(1-x)^{-2-\ep}.
\label{irsub}
\eq
%--
In both cases one has to approach the mass-shell in a interval of the
dimension $n$ where the integral is convergent and continue the
result in the number of space-time dimensions. For instance, from
\eqn{irsub} we obtain
%--
\bq
S^{{\underline 1}3{\underline 1}}_{p}|_{_{\rm os}} =
-\,\frac{1}{m^4}\,\lpar\frac{\tHss}{\pi\,m^2}\rpar^{\ep}\,
\Bigl[\frac{2}{\ep^2} - ( 1 + \gamma)\,\frac{1}{\ep} +
1 + \frac{1}{2}\,\gamma + \frac{1}{4}\,\gamma^2 + \frac{1}{4}\,\zeta(2) +
\ord{\ep}\Bigr].
\eq
%--
Consider now another configuration, depicted in \fig{wfQEDb},
%--
\bq
\mbox{b)} \qquad m_1 = m_2 = m, \quad m_3 = m_5 = 0 \quad \mbox{and} \quad
m_4 = M.
\eq
%--
%--
It is immediately seen that b) does not correspond to the leading Landau
singularity but rather to the sub-leading one corresponding to $\alpha_1 =
\alpha_2 = 0$ and $s = M^2$. We start from 
%--
\bq
S^{131} = - \lpar\frac{\tHss}{\pi}\rpar^{\ep}\,
\egam{1+\ep}\,\intfxx{x}{y}\intfz\,
\Bigl[ x(1-x)\Bigr]^{-\ep/2}(1-y)^{\ep/2-1}\,
(y-z)\,U_{\aca}^{-1-\ep}(x,y,z),
\eq
%--
take the derivative and put $p^2= - M^2$ obtaining
%--
\bq
S^{131}_{pb} = \lpar\frac{\tHss}{\pi}\rpar^{\ep}\,
\egam{2+\ep}\,\intfxx{x}{y}\int_0^y\,dz\,
\Bigl[ x(1-x)\Bigr]^{-\ep/2}(1-y)^{\ep/2-1}\,(y-z)\,z(1-z)\,
U^{-2-\ep}_{\acan{\rm os}}(x,y,z),
\label{confbaca}
\eq
%--
where the quadratic form $U$ is
%--
\bq
U_{\acan{\rm os}}(x,y,z) = M^2\,z^2 + m^2_x\,(1-y), \qquad
m^2_x = \frac{m^2}{x(1-x)}.
\eq
%--
Now we make use of the identity
%--
\bq
z(1-z)\,U^{-2-\ep}_{\acan{\rm os}}(x,y,z) =
- \frac{1}{2\,(1+\ep)\,M^2}\,(1-z)\partial_z\,
U^{-1-\ep}_{\acan{\rm os}}(x,y,z)
\eq
%--
and integrate by parts obtaining
%--
\bq
S^{131}_{pb} = \frac{1}{2\,M^2}\,\lpar\frac{\tHss}{\pi}\rpar^{\ep}\,
\egam{1+\ep}\,\intfx{x}\,
\Bigl[ x(1-x)\Bigr]^{-\ep/2}\,({\cal S}_{b}+{\cal S}_{c})
\label{confbacaII}
\eq
%--
where we have introduced
%--
\bqa
{\cal S}_{b} &=&
(m_x^2)^{-1-\ep}\,\intfx{y}\,y\,(1-y)^{-2-\ep/2} =
\frac{4}{\ep\,(2+\ep)}\,(m_x^2)^{-1-\ep}  \nl
{\cal S}_{c} &=&
\intfxy{y}{z}\,(1-y)^{\ep/2-1}\,(2z-1-y)\,
U^{-1-\ep}_{\acan{\rm os}}(x,y,z).
\eqa
%--
In order to compute ${\cal S}_{c}$, we make the transformation $y = 1-y'$
and introduce the function
%--
\bq
E(\mu,\nu,\alpha) = \intfx{y}\,\int_0^{1-y}\,dz\,y^{\mu}\,z^{\nu}\,
U^{\alpha}_{\acan{\rm os}}(x,y,z)
\label{Emunu}
\eq
%--
which can be computed explicitly with the help of the identity
%--
\bq
\Bigl[ 1 - \frac{1}{\alpha}\,(y\,\partial_y + 
\frac{z}{2}\,\partial_z)\Bigr]
\,U^{\alpha}_{\acan{\rm os}}(x,y,z) = 0,
\eq
%--
giving the following result:
%--
\bq
E(\mu,\nu,\alpha) = \frac{1}{2\alpha + 2\mu + \nu + 3}\,\intfx{y}
y^{\mu}\,(1-y)^{\nu}\,(1+y)\,U^{\alpha}_{\acan{\rm os}}(x,y,1-y)
\eq
%--
Substituting this result into \eqn{confbacaII} and expanding around $\ep = 0$
we obtain
%--
\bq
S^{131}_{pb} = =
\frac{1}{2M^4}\,\lpar\frac{\tHss}{\pi\,M^2}\rpar^{\ep}\,
\egam{1+\ep}\,\intfx{x}\,\intfx{y}
\Bigl[ {\cal S}_{-2}\,\ep^{-2} + {\cal S}_{-1}\,\ep^{-1} +
{\cal S}_0\Bigr],
\eq
%--
where the coefficients ${\cal S}$ are given by
%--
\bqa
{\cal S}_{-2} &=& - 4,  \quad
{\cal S}_{-1} =
  2\,X\,\Bigl[ \frac{y}{\chi} - \Bigl(\frac{1}{y\,\chi}\Bigr)_+ \Bigr]
  + \frac{1}{3\,\mu^2},
\nl
{\cal S}_{\,0} &=&
  X\,\Bigl[   \frac{3+y}{\chi} - 2\,\frac{y}{\chi}\,L
           + 2\,\Bigl(\frac{1}{y\,\chi}\Bigr)_+ + 2\,L_+ \Bigr]
- 4 + \zeta(2) - \frac{1}{3\,\mu^2}\,( \frac{4}{3} - \ln \mu^2 ),
\eqa
%--
where we have introduced $\chi_{\acan{\rm os}} = X\,(1-y)^2 + 
\mu^2\,y$, $\mu^2 = m^2/M^2$, $X=x(1-x)$ and
%--
\bq
L = \ln\chi - \frac{1}{2}\ln X - \frac{1}{2}\,\ln y,  \quad
L_+ =
\left(\frac{\ln\chi}{y\,\chi}\right)_+  - 
\frac{1}{2}\,\left(\frac{1}{y\,\chi}\right)_+\,(\ln X + \ln y).
\eq
%--
Note that we can treat these coefficients numerically since $\chi$ is a
semi-positive defined quadratic form with zeros at $y = 1$ and $x = 0$ or
$x = 1$ which are compensated by the numerators (note that we have changed
again variable, from $y$ to $1-y$). A convenient trick
consists in extracting these factors in order to obtain a manifestly finite
expression. Consider, for instance, the following integral,
%--
\bq
{\cal I} = \intfxx{x}{y}\,\frac{x^2(1-x)^2}{\chi^2}.
\eq
%--
after some straightforward manipulation we obtain
%--
\bqa
{\cal I} &=& \intfxx{x}{y}\,\Bigl[ \frac{x^2 y (1-x\,y)^2}{\chi^2_1} +
\frac{x(1-x)}{\chi^2_2} + \frac{x^2 y (1 - x\,y)}{\chi^2_3}\Bigr],
\nl
\chi_1 &=& x\,(1-x\,y)\,(1 - y)^2 + \mu^2,  \nl
\chi_2 &=& \Bigl[ 1 - x\,y\,(1 - x)\Bigr]^2 + \mu^2\,y,  \nl
\chi_3 &=& x\,\Bigl[1 - y\,(1 - x\,y)\Bigr]^2 + \mu^2,
\eqa
%--
where it is immediately seen that the three quadratic forms are strictly
positive over the $[0,1]^2$ region.

By looking at \eqn{defabapb} we see that the evaluation of $S^{121}_{pb}$
is very similar to the one we have just performed. Using the same technique
we obtain
%--
\bqa
S^{121} &=& -\,2\,\lpar\frac{\tHss}{\pi}\rpar^{\ep}\,\egam{\ep}\,
\intfx{x}\intfxy{y}{z}\,\Bigl[ x(1-x)\Bigr]^{-\ep/2}\,\lpar 1-y\rpar^{\ep/2}
\nl
{}&\times& \Bigl[ \lpar m^2_x + z^2 p^2\rpar\,U_{\aba}^{-1-\ep} +
\frac{2-\ep/2}{\ep}\,U_{\aba}^{-\ep}\Bigr],
\eqa
%--
where the quadratic form now is
%--
\bq
U_{\aba} = - p^2 z^2 + (p^2+m^2) z + m_x^2\,(1-y)
\eq
%--
First we take the derivative with respect to $p^2$ and next we impose the
on-shell condition, $p^2 = - m^2$, obtaining
%--
\bqa
S^{121}_{pb} &=& -\,\frac{2}{m^2}\,
\lpar\frac{\tHss}{\pi\,m^2}\rpar^{\ep}\,\egam{\ep}\,
\intfx{x}\intfxy{y}{z}\,\Bigl[ x(1-x)\Bigr]^{-\ep/2}\,(1-y)^{\ep/2}  \nl
{}&\times& \Bigl\{ z(1-z)\,\Bigl[ (1+\ep)\,(z^2-X)\,
\chi^{-2-\ep}_{\aba} + \frac{\ep}{2}\,\chi^{-1-\ep}_{\aba}\Bigr]
+ z\,(3\,z - 2)\,\chi^{-1-\ep}_{\aba}\Bigr\}|_{\rm os}.
\eqa
%--
where the on-shell quadratic form is
%--
\bq
\chi_{\aban{\rm os}} = z^2 + X\,(1-y),
\qquad
X = \frac{1}{x(1-x)}.
\eq
%--
To proceed, we can use the following relation:
%--
\bq
\Big[ 1 + \frac{1}{1+\ep}\,(y\partial_y + z\partial_z) \Big]\,
\chi_{\aban{\rm os}}^{-1-\ep} =
- (z^2 - X)\,\chi_{\aban{\rm os}}^{-2-\ep}.
\eq
%--
In this way we derive
%--
\bqa
S^{121}_{pb} &=& \frac{\egam{\ep+1}}{m^2}\,
\lpar\frac{\tHss}{\pi\,m^2}\rpar^{\ep}
\intfx{x}\intfxy{y}{z}\,\,\Bigl[ x(1-x)\Bigr]^{-\ep/2}\,(1-y)^{\ep/2-1}\,
z(1-z)\,\chi_{\aban{\rm os}}^{-1-\ep}.
\eqa
%--
Now we can use
%--
\bq
z(1-z)\,(\chi_{\aban{\rm os}})^{-1-\ep} = - \frac{1}{2\ep}\,(1-z)\partial_z\,
\chi_{\aban{\rm os}}^{-\ep}.
\eq
%--
Once the integration by parts is performed we introduce $y'=1-y$
and it follows that
%--
\bqa
S^{121}_p &=& - \frac{\egam{\ep}}{2m^2}\,
\lpar\frac{\tHss}{\pi\,m^2}\rpar^{\ep}\,
\intfxx{x}{y}\,\Bigl[ x(1-x)\Bigr]^{-\ep/2}\,y^{\ep/2-1}\,
\left( y\,\chi_y^{-\ep} - \chi_0^{-\ep} + \int_0^{1-y} dz\,\chi^{-\ep}
\right)
\eqa
%--
where the various $\chi$'s are expressed as
%--
\bq
\chi = z^2 + X\,y,
\qquad
\chi_y = (1-y)^2 + X\,y,
\qquad
\chi_0 = X\,y.
\eq
%--
In order to perform the $z$-integration we use the function
$E(\mu,\nu,\alpha)$ introduced for $S_{pb}^{131}$ in \eqn{Emunu} and derive:
%--
\bq
\intfx{y}\int_0^{1-y} dz\,(1-y)^{\ep/2-1}\,
\chi^{-\ep} =
\frac{1}{1-\ep}\,\intfx{y}\,y^{\frac{\ep}{2}-1}\,
(1+y)\,\chi_y^{-\ep}
\eq
%--
After the integration of the $\chi_0$ term, we introduce
$U_y = x\,(1-x)\,\chi_y = x\,(1-x)\,(1-y)^2 + y$ and perform the
$\ep$-expansion, obtaining:
%--
\bq
S^{121}_{pb} = - \frac{1}{2m^2}\,\lpar\frac{\tHss}{\pi\,m^2}\rpar^{\ep}\,
\egam{1+\ep}\,\Bigl( {\cal S}_1 + {\cal S}_2 \Bigr),
\eq
%--
where the coefficients are
%--
\bq
{\cal S}_1 =
\frac{2}{\ep^2}\,\frac{\egams{1+\ep/2}}{\egam{2+\ep}},
\quad
{\cal S}_2 =
  \frac{2}{\ep^2} + \frac{6}{\ep}
- \intfxx{x}{y} \Big[ 2\,\ln U_y + \left( \frac{\ln U_y}{y} \right)_+ \Big]
+ 4 - \frac{1}{2}\,\zeta(2).
\eq
%--
\subsection{$S^{221}$ topology\label{sbabt}}
%--
%--
The evaluation of $S^{221}$ is more involved than ones already presented
for the other two-loop self-energies. We have
%--
\bqa
\pi^4\,S^{221} &=& \mu^{2\ep}\,\intmomsii{n}{q_1}{q_2}\,
(q^2_1+m^2_1)^{-1}
((q_1+p)^2+m^2_2)^{-1}
((q_1-q_2)^2+m^2_3)^{-1}  \nl
{}&\times&
(q^2_2+m^2_4)^{-1}
((q_2+p)^2+m^2_5)^{-1}.
\label{defsbab}
\eqa
%--
First of all the $\alpha-\gamma$ propagators are combined with parameters
$x_1,x_2$. After performing the $q_1$-integration we introduce two
additional parameters $y,z$, perform the $q_2$-integration and derive
%--
\bqa
S^{221} &=& - \,\lpar\frac{\tHss}{\pi}\rpar^{\ep}\,\egam{1+\ep}
\intfxy{x_1}{x_2}\,\Bigl[ x_2(1-x_2)\Bigr]^{-1-\ep/2}  \nl
{}&\times& \intfxy{y}{z}\,(1-y)^{\ep/2}\,U_{\bba}^{-1-\ep},  \nl
U_{\bba} &=& - \Bigl[ (1-y) p_x + z p\Bigr]^2 + z (p^2+m^2_5) + (y-z) m^2_4
+
(1-y) m^2_{xx},
\label{defs5}
\eqa
%--
where $x_{1,2}$-dependent mass and momentum have been introduced:
%--
\bqa
p_x &=& \frac{x_1-x_2}{1-x_2}\,p, \nl
m^2_{xx} &=& \frac{1}{x_2(1-x_2)}\,\Bigl[ (x_1-x_2)(1-x_1+x_2) p^2 +
(1-x_1) m^2_1 + (x_1-x_2) m^2_2 + x_2 m^2_3\Bigr].
\label{defmxx}
\eqa
%--
The quadratic form in $y,z$ has coefficients
%--
\bqa
H = \,\lpar
\ba{lr}
- p^2_x & \spro{p}{p_x} \\
\spro{p}{p_x} & - p^2
\ea
\rpar,
\qquad
K^t = \frac{1}{2}\,\lpar 2 p^2_x + m^2_4 - m^2_{xx}, p^2 -2 \spro{p}{p_x} +
m^2_5 - m^2_4\rpar,
\eqa
%--
\bq
L = m^2_{xx} - p^2_x - i\,\delta.
\eq
%--
Since $p_x$ and $p$ are simply related by
%--
\bq
p_x = X\,p, \qquad X = \frac{x_1-x_2}{1-x_2},
\eq
%--
the matrix H is singular. To overcome this problem we introduce new
variables $y = y'$, and $z = z' + X (y'-1)$, giving for the quadratic form 
$U_{\bba}$ the simple expression $U_{\bba} = a z^2 + b z + c y + d$ with 
coefficients
%--
\[
\ba{ll}
a = - p^2 & \;\;\;b = p^2 + m^2_5 - m^2_4,  \\
c = m^2_4 - m^2_{xx} + X (p^2 + m^2_5 - m^2_4) & \;\;\;
d =  m^2_{xx} - X (p^2 + m^2_5 - m^2_4).
\ea
\]
%--
After the transformation we will use the identity
%--
\bq
\Bigl[ 1 + \frac{y}{\ep}\,\partial_y + \frac{1}{2\,\ep}\,\lpar
z + \frac{b}{2\,a}\rpar\,\partial_z\Bigr]\,U_{\bba}^{-\ep} = -\,
\frac{B_{\bba}}{4\,a}\,U_{\bba}^{-1-\ep}.
\eq
%--
with $B_{\bba} = b^2 - 4 a d$. This factor reads as follows:
%--
\bq
B_{\bba} = \lambda(-p^2,m^2_4,m^2_5) + 4\,p^2\,(m^2_{xx} - m^2_4) - 4\,X
p^2\,
(p^2 + m^2_5 - m^2_4) = s\,b_{\bba}.
\label{defB}
\eq
%--
Before continuing in the evaluation of the diagram we study its
singularities.
%--
\subsection{Landau equations for $S^{221}$\label{lesbab}}
\begin{figure}[th]
\vskip 5pt
\[
  \vcenter{\hbox{
  \begin{picture}(150,100)(0,0)
  \Line(0,50)(50,50)
  \CArc(75,50)(25,-180,180)
  \Line(100,50)(150,50)
  \Line(50,50)(100,50)
  \Text(75,80)[cb]{$\scriptstyle 1$}
  \Text(75,55)[cb]{$\scriptstyle 3$}
  \Text(75,30)[cb]{$\scriptstyle 5$}
  \end{picture}}}
\quad \mbox{+ perm.} ; \quad
  \vcenter{\hbox{
  \begin{picture}(150,0)(0,0)
  \Line(0,0)(50,0)
  \CArc(75,0)(25,0,90)
  \CArc(75,0)(25,-180,0)
  \CArc(75,0)(25,90,180)
  \CArc(50,25)(25,-90,0)
  \Line(100,0)(150,0)
  \Text(50,20)[cb]{$\scriptstyle 1$}
  \Text(100,20)[cb]{$\scriptstyle 4$}
  \Text(75,-35)[cb]{$\scriptstyle 5$}
  \Text(72,0)[cb]{$\scriptstyle 3$}
  \end{picture}}}
\quad \mbox{+ perm.}
\]
\vspace{-0.5cm}
\caption[]{The reduced diagrams corresponding to the $S^{221}$ topology of
\fig{tops5p}. Their leading Landau singularities are the sub-leading ones
for
$S^{221}$.}
\label{s5redd}
\end{figure}
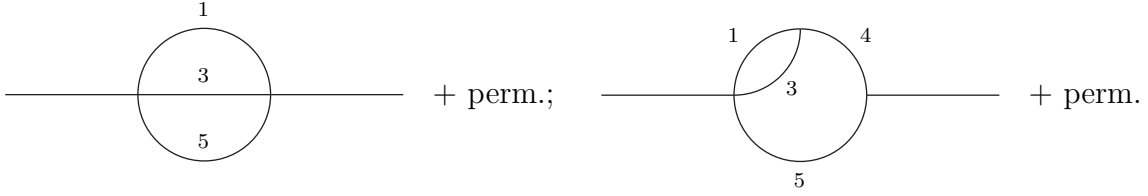

%--
The Landau equations corresponding to $S^{221}$ are
%--
\[
\ba{ll}
\alpha_1\,(q^2_1+m^2_1) = 0 & \;\;\alpha_2\,((q_1+p)^2+m^2_2) = 0, \\
\alpha_3\,((q_1-q_2)^2+m^2_3) = 0 & \;\; \alpha_4\,(q^2_2+m^2_4) = 0,  \\
\alpha_5\,((q_2+p)^2+m^2_5) = 0& \\
\label{nland1}
\ea
\]
%--
and also
%--
\bqa
{}&{}& \alpha_1 q_{1\mu} + \alpha_2 (q_1+p)_{\mu} +
\alpha_3\,(q_1-q_2)_{\mu} = 0,
\nl
{}&{}& - \alpha_3 (q_1-q_2)_{\mu} + \alpha_4 q_{2\mu} +
\alpha_5\,(q_2+p)_{\mu} = 0.
\label{nland2}
\eqa
%--
The leading Landau singularity occurs for $\alpha_i \ne 0, \forall i$.
We multiply the two equations \eqn{nland2} by $q_{1\mu}, q_{2\mu}$ and
$p_{\mu}$ respectively. This gives an homogeneous system of six equations.
If all $\alpha_i$ are different from zero, the singularity will occur for
%--
\[
\ba{ll}
q^2_1 = - m^2_1 & \;\; q^2_2 = - m^2_4, \\
\spro{q_1}{q_2} = \frac{1}{2}\,(m^2_3 - m^2_1 - m^2_4) & \;\;
\spro{p}{q_2} = \frac{1}{2}\,( s + m^2_4 - m^2_5),  \\
\spro{p}{q_1} = \frac{1}{2}\,( s + m^2_1 - m^2_2) \\
\ea
\]
%--
The equations become as follows:
%--
\bqa
{}&{}& - 2\,m^2_1\,\alpha_1 + (s-m^2_1-m^2_2)\,\alpha_2 -
(m^2_1+m^2_3-m^2_4)\,\alpha_3 = 0,
\nl
{}&{}& - (m^2_1-m^2_3+m^2_4)\,\alpha_1 +
(s-m^2_1+m^2_3-m^2_5)\,\alpha_2 -
(m^2_1-m^2_3-m^2_4)\,\alpha_3 = 0,
\nl
{}&{}& (s+m^2_1-m^2_2)\,\alpha_1 -
(s-m^2_1+m^2_2)\,\alpha_2 +
(m^2_1-m^2_2-m^2_4+m^2_5)\,\alpha_3 = 0,
\nl
{}&{}& (m^2_1+m^2_3-m^2_4)\,\alpha_3 -
(m^2_1-m^2_3+m^2_4)\,\alpha_4 +
(s-m^2_2+m^2_3-m^2_4)\,\alpha_5 = 0,
\nl
{}&{}& (m^2_1-m^2_3-m^2_4)\,\alpha_3 -
2\,m^2_4\,\alpha_4 +
(s-m^2_4-m^2_5)\,\alpha_5 = 0,
\nl
{}&{}& - (m^2_1-m^2_2-m^2_4+m^2_5)\,\alpha_3 +
(s+m^2_4-m^2_5)\,\alpha_4 -
(s-m^2_4+m^2_5)\,\alpha_5 = 0.
\label{lsystem}
\eqa
%--
In order to have non trivial solutions the determinants of the coefficients
must be equal to zero. In both cases the solution is given by
%--
\bqa
s &=& \frac{1}{2\,m^2_3}\,\Bigl\{ m^2_3\,(m^2_3-m^2_1-m^2_2-m^2_4-m^2_5) -
(m^2_1-m^2_4)\,(m^2_5-m^2_2)  \nl
{}&\pm& \Bigl[ \lambda(m^2_3,m^2_1,m^2_4)\,\lambda(m^2_3,m^2_2,m^2_5)
\Bigr]^{1/2}\Bigr\},
\eqa
%--
which represents the so-called anomalous threshold for this topology, the
leading Landau singularity. Actually the singularity occurs when the product
of the two lambda-functions is positive, i.e\ for
%--
\bq
m_3^2 \le m_{01}^2 \qquad \mbox{or} \qquad
m_{02}^2 \le m_3^2 \le m_{03}^2 \qquad \mbox{or} \qquad
m_3^2 \ge m_{04}^2
\eq
%--
where auxiliary masses are defines as
%--
\bqa
m_{01}^2 &=& \min\{(m_1-m_4)^2,(m_2-m_5)^2\},  \\
m_{02}^2 &=&
\min\{\max\{(m_1-m_4)^2,(m_2-m_5)^2\},\min\{(m_1+m_4)^2,(m_2+m_5)^2\}\},  \\
m_{03}^2 &=&
\max\{\max\{(m_1-m_4)^2,(m_2-m_5)^2\},\min\{(m_1+m_4)^2,(m_2+m_5)^2\}\},  \\
m_{04}^2 &=& \max\{(m_1+m_4)^2,(m_2+m_5)^2\},
\eqa
%--
The sub-leading Landau singularities follow from a study of the reduced
diagrams of \fig{s5redd}.
%--
%--
\subsection{Evaluation of $S^{221}$: method I\label{esbabmo}}
%--
To compute $S^{221}$ we must, first of all, considers the roots of
$b_{\bba}$ defined in \eqn{defB}. It follows that $b_{\bba} = 0$ for
%--
\bq
x^2_1 + (\mu^2_5 + \frac{1}{4}\,\lambda) x^2_2 - (1 + \mu^2_5 - \mu^2_4)
x_1x_2 - (1 + \mu^2_1 - \mu^2_2) x_1 + (1 - \mu^2_2 + \mu^2_3 - \mu^2_4 -
\frac{1}{4}\,\lambda) x_2 + \mu^2_1 = 0,
\eq
%--
where $\lambda = \lambda(1,\mu^2_4,\mu^2_5)$. Before discussing the general
case let us consider some simple example, e.g. \ $m_i = m, \forall i\ne 3$
and
three cases: $m_3 = 0$, $m_3 = m$ and also $m_3 = M$, where $M$ is some
large
mass, $M > m$. The curves corresponding to $B = 0$ are given in \fig{scurv}
where the integration domain, $0 \le x_1 \le 1$ and $0 \le x_2 \le x_1$ is
also shown.
%--
\bfi
\centerline{
\epsfig{file=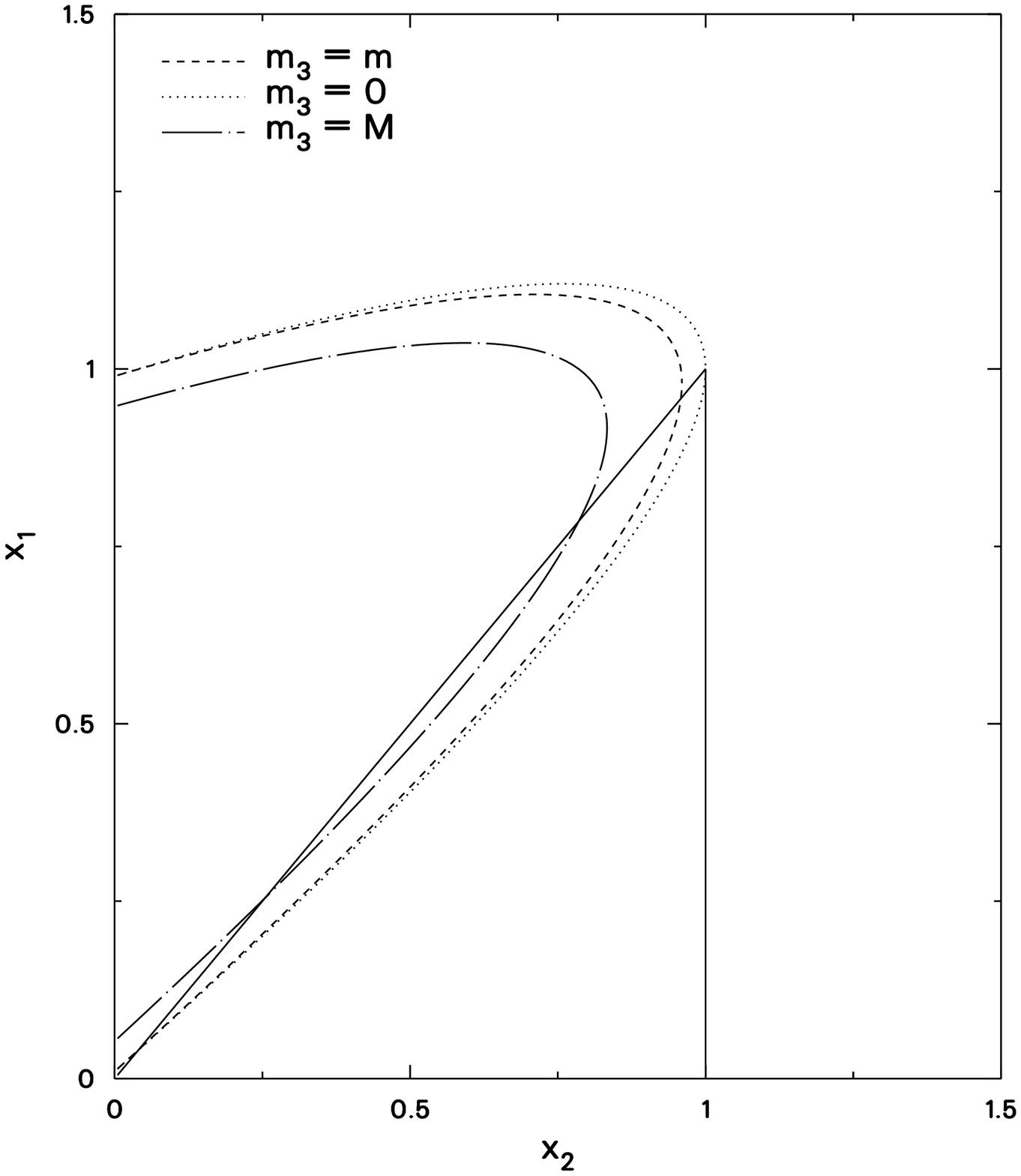,height=15cm,angle=0}}
\caption[]{The $x_1 - x_2$ plane with the curves $b_{\bba} = 0$ where
$b_{\bba}$ is defined in \eqn{defB}. Here $m_i = m, \forall i\ne 3$ and
$m_3 = 0, m, M$ with $M > m$.}
\label{scurv}
\efi
%--
Before {\em raising} powers we introduce $U_{\bba}= s\chi_{\bba}$, with
%--
\bq
\chi_{\bba} = z^2 - \mu^2_{45}\,z + (\mu^2_4 - \mu^2_{xx} -
X\,\mu^2_{45})\,y +
\mu^2_{xx} + X\,\mu^2_{45},
\eq
%--
where, as usual, $\mu^2_{ij} = 1 +\mu^2_i - \mu^2_j$. The {\em raising}
procedure requires
%--
\bqa
\chi_{\bba}^{-1-\ep} &=& -\,\frac{4}{b_{\bba}}\,\Bigl[ 1 + \frac{y}{\ep}\,
\partial_y + \frac{1}{2\,\ep}\,(z+Z)\partial_z\Bigr]\,\chi_{\bba}^{-\ep},
\eqa
%--
where the object of interest is
%--
\bq
b_{\bba} = \lambda(1,\mu^2_4,\mu^2_5) - 4\,(\mu^2_{xx} - \mu^2_4) - 4\,X\,
\mu^2_{45},  \quad
Z = -\,\frac{1}{2}\,\mu^2_{45},  \quad
X = \frac{x_1-x_2}{1-x_2},
\eq
%--
and $0 \le y \le 1$, with $X(1-y) \le z \le y+X(1-y)$. After a Laurent
expansion in $\ep$ we will have terms proportional to powers of
$\ln\chi_{\bba}$; following these operation we substitute back
$z = z' - X (y-1)$, so that the order of integration will be specified as
%--
\bq
\intfx{y}\intfz\intfx{x_2}\int_{x_2}^{\scriptstyle 1}\,dx_1.
\eq
%--
The argument of the logarithm is
%--
\bqa
\chi_{\bba} &=& \frac{X_{\bba}}{x_2(1-x_2)^2},  \qquad
X_{\bba} = A_{\xi} x^2_1 + B_{\xi} x_1 + C_{\xi},
\eqa
%--
where the coefficients $A_{\xi} - C_{\xi}$ are given by
%--
\bqa
A^0_{\xi} &=& (1-y)\,(1- x_2),  \quad
B^0_{\xi} = -(1-y)\,(1-x_2)\,(\mu^2_{54} x_2 +\mu^2_{12}), \nl
C^0_{\xi} &=& x_2 (1-x_2)^2\,(z^2 - \mu^2_{45} z + \mu^2_4 y) -
(1-y)(1-x_2)\,\Bigl[(\mu^2_4 - \mu^2_5) x^2_2 - \mu^2_{32} x_2 -
\mu^2_1\Bigr],
\eqa
%--
and $\mu^2_{ij} = 1 + \mu^2_i - \mu^2_j$.
Zeros of $b_{\bba}$ are avoided by distorting the $x_1$ integration contour
in a way that preserves the correct imaginary part of $\ln X_{\bba}$. As
done for the sunset diagram we consider the two roots of the equation $X_{\bba}
= 0$, say $x_{1\{\ssL,\ssR\}}$. If the roots are complex conjugated
we select a contour which starts at $x_1 = x_2$ and bypasses the zeros of
$b_{\bba}$ in the upper (or lower) half-plane to return to $x_1 = 1$.

If the roots are real we must distinguish between two cases:
when $A_{\xi}$ is positive then the cut is between $x_{1\ssL}$ and
$x_{1\ssR}$, while for negative $A_{\xi}$, the cut is $[-\infty,x_{1\ssL}]\,
\cup\,[x_{1\ssR},+\infty]$.
Furthermore, with $x_1 = \alpha + i\,\beta$, the imaginary part of the
logarithm is $\beta ( 2\,\alpha A_{\xi} + B_{\xi})$. This consideration
immediately tells us the sign of the imaginary part when $x_1$ approaches
the real axis on the cut. This sign is crucial when we distort the contour in
the complex plane since, for $x_1$ real and on the cut, the $i\,\delta$
prescription, gives $-\,i\pi$ for the imaginary part.

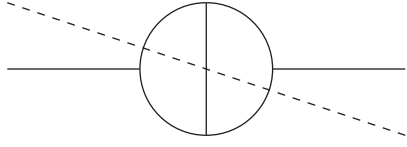
\begin{figure}[th]
\vspace{0.5cm}
\[
  \vcenter{\hbox{
  \begin{picture}(150,0)(0,0)
  \Line(0,0)(50,0)
  \CArc(75,0)(25,0,90)
  \CArc(75,0)(25,-180,0)
  \CArc(75,0)(25,90,180)
  \Line(75,25)(75,-25)
  \Line(100,0)(150,0)
  \DashLine(0,25)(150,-25){3.}
  \end{picture}}}
\]
\vspace{0.5cm}
\caption[]{The three-particle cut of $S^{221}$.}\label{tpc}
\end{figure}

Let us consider one of the normal thresholds, $s = (m_1+m_3+m_5)^2$, as we
have done for other diagrams. This corresponds to the three-particle cuts
of \fig{tpc}. Let $x_{1\{\ssL,\ssR\}}$ be the two points where $X_{\bba} =
0$ and $x_{1\pm}$ the two roots of $b_{\bba} = 0$. Whenever $x_{1\pm}$ are on
the real axis we want to distort the integration contour.

In \fig{irs5} we give the real and imaginary parts of these four points
as a function of $x_2$ for fixed $y$ and different values of $z$.
For decreasing $x_2$ the branch points and $x_{1\pm}$ are complex conjugated
and no distortion is needed. For values of $z$ near to $y$ the branch points
become real when $x_{1\pm}$ are still complex and, again, distortion is not
needed till the value of $x_2$ where $x_{1\pm}$ are real. However, for some
value of $z$, that we call $z_{\rm th}(y)$ the following happens: the two
branch points pinch the real $x_1$-axis exactly at $x_{1+} \equiv
x_{1-}$ and distortion is not allowed anymore. A part for having a larger
number of Feynman parameters, the situation is completely similar to the
one described for the sunset diagram~\cite{Passarino:2001wv}.
%--
%--
%--
\bfi
\centerline{
\epsfig{file=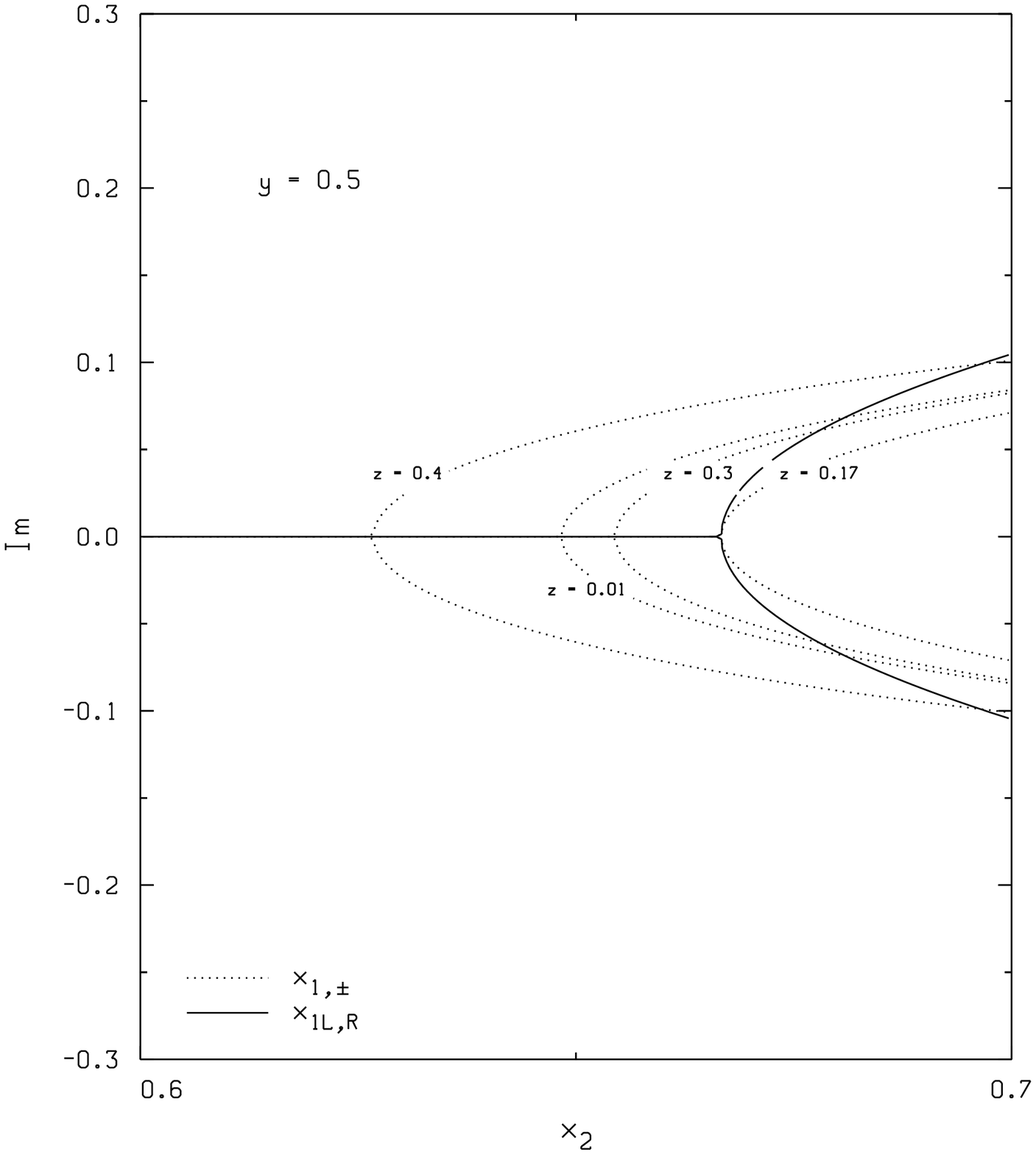,height=12cm,angle=0}
\epsfig{file=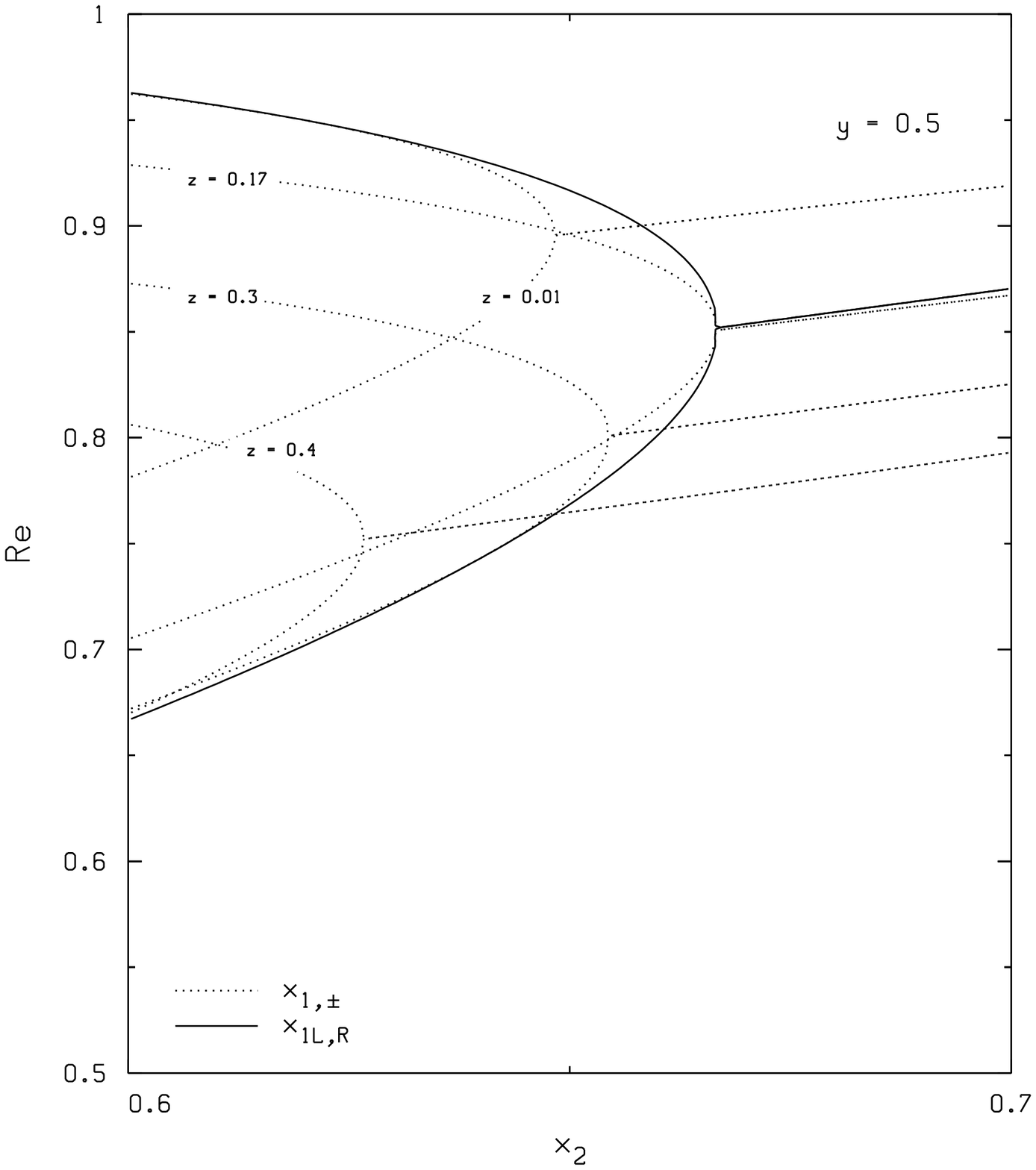,height=12cm,angle=0}}
\caption[]{Behavior, at the normal threshold $s = (m_1+m_3+m_5)^2$ of branch
points and apparent singularities for $S^{221}$ as a function of $x_2$.}
\label{irs5}
\efi
%--
Since there is no ultraviolet divergency we may set $\ep = 0$, obtaining
%--
\bq
S^{221} = - \frac{1}{s}\,\intfxy{x_1}{x_2}\,(1-x_2)\,
\intfx{y}\int_{\scriptstyle X(1-y)}^{\scriptstyle
y+X(1-y)}\,dz\,X_{\bba}^{-1}.
\eq
%--
Our first step will be to use the following relation:
%--
\bq
\chi_{\bba}^{-1} = -\,\frac{4}{b_{\bba}}\,\Bigl\{ 1 -
\Bigl[ y\,\partial_y + \frac{1}{2}\,(z+Z)\partial_z\Bigr]\,
\ln\chi_{\bba}\Bigr\}.
\eq
%--
We obtain
%--
\bq
S^{221} = \frac{4}{b_{\bba}^{\ssR}}
\intfxy{x_1}{x_2}\,(1-x_2)\,\intfx{y}
\int_{\scriptstyle X(1-y)}^{\scriptstyle y+X(1-y)}\,dz\,
\Bigl\{ 1 - \Bigl[ y\,\partial_y + \frac{1}{2}\,(z+Z)\partial_z\Bigr]\,
\ln\chi_{\bba}\Bigr\}.
\eq
%--
where $b_{\bba}^{\ssR} = x_2(1-x_2)\Lambda$. Three new quantities are
needed:
%--
\bqa
\chi_{\bba}^1 &=& \chi_{\bba}(z=y+X(1-y)) = (1-X)^2 y^2 - \Bigl[
1 - \mu^2_5 + \mu^2_{xx} -2\,X (1-X)\Bigr] y + \mu^2_{xx} + X^2,  \nl
\chi_{\bba}^2 &=& \chi_{\bba}(z=X(1-y)) = X^2 y^2 +
\Bigl[ \mu^2_4 - \mu^2_{xx} - 2\,X^2\Bigr] y + \mu^2_{xx} + X^2,
\nl
\chi_{\bba}^3 &=& \chi_{\bba}(y=1,z=y) = y^2 -\mu^2_{45}\,y + \mu^2_4.
\label{babresidues}
\eqa
%--
Therefore, using
%--
\bq
\intfx{y}\int_{\scriptstyle X(1-y)}^{\scriptstyle y+X(1-y)}\,dz =
\int_{0}^{\scriptstyle X}\,dz\,
\int_{\scriptstyle 1-\frac{z}{X}}^{\scriptstyle 1}\,dy +
\int_{\scriptstyle X}^{\scriptstyle 1}\,dz\,
\int_{\scriptstyle \frac{z-X}{1-X}}^{\scriptstyle 1}\,dy,
\eq
%--
we obtain
%--
\bqa
S^{221} &=& \frac{4}{s\,b_{\bba}^{\ssR}}\intfxy{x_1}{x_2}\,
\Bigl\{ \frac{3}{2}\,\intfx{y}
\int_{\scriptstyle X(1-y)}^{\scriptstyle y+X(1-y)}\,dz\,\ln\chi_{\bba}  - 
\intfx{y}\,\ln \chi^3_{\bba} \nl
{}&-& \frac{1}{2}\,\intfx{y}\,\Bigl[ (X - y + X y + Z)\,\ln\chi^1_{\bba} -
(X + X y + Z)\,\ln\chi^2_{\bba}\Bigr] + \frac{1}{2}\Bigr\}
\eqa
%--
Next we change variable, $z \to z'$, so that $0 \le z' \le 1$ and write
all $\chi$'s as polynomials in $x_1$ with coefficients that are functions
of $x_2,y$ and z. In particular we introduce
%--
\bq
\chi_{\bba}^1 = \frac{X_{\bba}^1}{x_2(1-x_2)^2},  \qquad
X_{\bba}^1 = A^1_{\xi} x^2_1 + B^1_{\xi} x_1 + C^1_{\xi}.
\eq
%--
we find
%--
\bqa
A^1_{\xi} &=& Y\,\Bigl[ 1 - x_2 + Y x_2\Bigr],  \quad
B^1_{\xi} = -\,Y\,\Bigl[ (1-x_2) \mu^2_{12} + 2\,Y x_2\Bigr],
\nl
C^1_{\xi} &=& Y (1-x_2)\,( - 2\,x^2_2 y + x_2 \mu^2_{32} + x^2_2 + \mu^2_1)
+ Y^2 x^3_2 + x_2 (1-x_2)^2 (\mu^2_5-Y)\,y
\eqa
%--
\bq
\chi_{\bba}^2 = \frac{X_{\bba}^2}{x_2(1-x_2)^2},  \qquad
X_{\bba}^2 = A^2_{\xi} x^2_1 + B^2_{\xi} x_1 + C^2_{\xi}.
\eq
%--
\bqa
A^2_{\xi} &=& Y\,( 1 - x_2 + Y x_2),
\quad
B^2_{\xi} = - Y\,\Bigl[ (1-x_2) ( \mu^2_{12} + 2 x_2) + 2 Y x^2_2\Bigr],
\nl
C^2_{\xi} &=& Y (1-x_2) ( \mu^2_{32} x_2 + x^2_2 + \mu^2_1) + Y^2 x^3_2 +
x_2 (1-x_2)^2 \mu^2_4 y.
\eqa
%--
Finally we introduce
%--
\bqa
\chi_{\bba}^0 &=& \frac{X_{\bba}^0}{x_2(1-x_2)^2},  \qquad
X_{\bba}^0 = A^0_{\xi} x^2_1 + B^0_{\xi} x_1 + C^0_{\xi},
\eqa
%--
where
%--
\bqa
A^0_{\xi} &=& (1-y)\,(1- x_2),  \quad
B^0_{\xi} = -(1-y)\,(1-x_2)\,(\mu^2_{54} x_2 +\mu^2_{12}),  \nl
C^0_{\xi} &=& x_2 (1-x_2)^2\,(z^2 - \mu^2_{45} z + \mu^2_4 y) -
(1-y)(1-x_2)\,\Bigl[(\mu^2_4 - \mu^2_5) x^2_2 - \mu^2_{32} x_2 -
\mu^2_1\Bigr].
\eqa
%--
Here we used
%--
\bq
Y = 1 - y.
\eq
%--
At this point we must realize that the situation for $S^{221}$ is different
from all previous cases. Indeed $S^{221}$ and all its sub-diagrams are
convergent and, for this reason, only logarithms appear after the first
iteration and not logarithms squared.
Therefore, the strategy for distorting the $x_1$ integration contour will be
the following.
Let us introduce variables $\Delta_i = x_1 - x_{1i}$, where $x_{1i}$ is any
root of the argument of a logarithm, and split the logarithms according to
%--
\bq
\ln\lpar A^n_{\xi} x^2_1 + B^n_{\xi} x_1 + C^n_{\xi} - i\,\delta\rpar =
\ln A^n_{\xi} + \eta(A^n_{\xi},\Delta_{\ssL}\Delta_{\ssR}) +
\eta(\Delta_{\ssL},\Delta_{\ssR}) + \sum_{i=\ssL,\ssR}\,\ln(x_1-x_{1i}),
\eq
%--
where $\eta$ is the usual Veltman's function,
$\eta(a,b) = \ln(ab) - \ln a - \ln b$,
and logarithms are defined with a cut along the negative real axis.
The final expression for the diagram will be the sum of terms proportional
to $\eta$-functions and terms of the form $\ln(x_1-x_{1i})/\Lambda_{\ssR}$.
The former do not create a problem in distorting, when needed, the
integration
contour.
For the latter, however, if one of the roots of $\lambda_{\ssR}$, or both,
is
real and internal to the interval $[0,1]$ then the distortion for the
corresponding term must be examined with care. If the imaginary part of
$x_{1i}$ is positive(negative) then we move the contour into the
negative(positive) imaginary half-plane, so that the cut of the logarithm
will never be crossed. If $x_{1i}$ is real and negative no problem will arise
and we can distort in any of the two ways. If its is real and positive then
the distortion is fixed according to the $i\,\delta$ prescription: for
instance
the roots of $a x^2_1 + b x_1 + c - i\delta$ corresponding to $b^2-4\,ac >
0$ are such that
%--
\bq
\Imb x_{1,\{\ssL,\ssR\}} = \mp \,{\rm sign}(a)\,\delta.
\eq
%--
The reason why we cannot apply this argument to the other cases discussed
before is linked to the presence of $\ln^2$-terms. After the splitting of
the logarithm we will encounter terms proportional to $\ln(x_1-x_{1\ssL})
\ln(x_1-x_{1\ssR})$. When the roots are complex conjugated we will have
cuts both in the positive and in the negative imaginary half-planes and
the possibility of distorting the contour ceases when they approach the
real axis.
%--
\subsection{Evaluation of $S^{221}$: method II\label{esbabmtwo}}
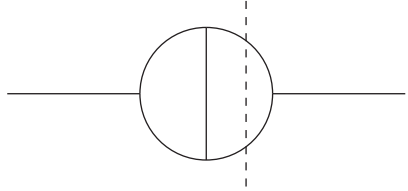
\begin{figure}[th]
\vspace{0.5cm}
\[
  \vcenter{\hbox{
  \begin{picture}(150,0)(0,0)
  \Line(0,0)(50,0)
  \CArc(75,0)(25,0,90)
  \CArc(75,0)(25,-180,0)
  \CArc(75,0)(25,90,180)
  \Line(75,25)(75,-25)
  \Line(100,0)(150,0)
  \DashLine(90,35)(90,-35){3.}
  \end{picture}}}
\]
\vspace{0.5cm}
\caption[]{One of the two-particle cuts of $S^{221}$.}
\label{twopcut}
\end{figure}

%--
An alternative evaluation of $S^{221}$ follows when we change variable in
\eqn{defs5}, $y= 1-y'$, so that
%--
\bqa
{}&{}&
\intfx{y}\int_{\scriptstyle X(1-y)}^{\scriptstyle y+X(1-y)}\,dz\,
(1-y)^{\ep/2}\,(az^2+bz+cy+d)^{-1-\ep} =
\nl
{}&{}&
\intfx{y}\int_{\scriptstyle Xy}^{\scriptstyle 1-(1-X)y}\,dz\,
y^{\ep/2}\,(az^2+bz-cy+d')^{-1-\ep}.
\eqa
%--
This method has the advantage that
$d'= c + d = m^2_4$,
with a corresponding change of $b_{\bba} \to b'_{\bba}$, where
$b'_{\bba} = \mu^4_{45} - 4\,\mu^2_4$,
is $x_{1,2}$-independent. In this way the $x_1 - x_2$ integration is free
of numerical instabilities. The only point in the external parameter space
where a failure will occur is for $s = (m_4 \pm m_5)^2$, representing normal 
and pseudo thresholds associated to the two-particle cut of \fig{twopcut}.
They are non-leading Landau singularities. The fact that $b'_{\bba}$ is
$x$ independent means that, with this method, we can perform the
$x$-integration without having to distort the integration contour. As
observed the method fails at the non-leading Landau singularities 
corresponding to the one-loop complementary sub-diagram of $S^{221}$
(in this case there are two sub-diagrams with the same number of lines).
When we are in this kinematical configuration and $s \ne (m_1 \pm m_2)^2$
the strategy will be to interchange the order of the operations,
$q_1 \leftrightarrow q_2$. When $(m_1 \pm m_2)^2 = (m_4 \pm m_5)^2$ the
method fails. Using this approach, we set $\ep = 0$ and get
%--
\bqa
S^{221} &=& -\,\frac{1}{s}\,\intfxy{x_1}{x_2}\,\intfx{y}\,
\int_{Xy}^{1+(X-1)y}\,dz  \nl
{}&\times& \Bigl[x_2(1-x_2)\Bigr]^{-1}\,\chi^{-1}_{\bba}(x_1,x_2,1-y,z-Xy),
\eqa
%--
\bqa
\chi_{\bba}(x_1,x_2,1-y,z-Xy) &=&
z^2 - \mu^2_{45}\,z + (\mu^2_{xx} - \mu^2_4 + X\,\mu^2_{45})\,y + \mu^2_4.
\eqa
%--
Successively we use the identity
%--
\bq
\chi_{\bba}^{-1-\ep} = -\,\frac{4}{\lambda(1,\mu^2_4,\mu^2_5)}\,\Bigl[
1 + \frac{y}{\ep}\,\partial_y + \frac{z - \mu^2_{45}/2}{2\,\ep}\,
\partial_z\Bigr]\,\chi^{-\ep}_{\bba}.
\eq
%--
The derivation continues, as usual, with an integration by parts, where
we use
%--
\bq
\intfx{y}\,\int_{Xy}^{1+(X-1)y}\,dz = \int_0^X\,dz\,\int_0^{z/X}\,dy +
\int_X^1\,dz\,\int_0^{(z-1)/(X-1)}\,dy.
\eq
%--
Next we define
%--
\bq
\mu^2_{xx}= \frac{\nu^2_{xx}}{x_2(1-x_2)},
\eq
%--
and introduce four quadratic forms:
%--
\bqa
\chi^1_{\bba} &=& x_2 (x_1-x_2) (1-x_2) y \mu^2_{45}
       + x_2 (1-x_2) (1-x_1)  y  ( \mu^2_{45} - 2 )  \nl
{}&+& x_2 (1-x_2)^2   ( 1 - y \mu^2_4 + \mu^2_4 - \mu^2_{45} )
       + x_2 (1-x_1)^2  y^2
       + (1-x_2)  y \nu^2_{xx},
\nl
\chi^2_{\bba} &=& x_2 (x_1-x_2)^2 y^2
       + x_2 (1-x_2)^2 \mu^2_4   (  1 - y )
       + (1-x_2) y \nu^2_{xx},  \nl
\chi^3_{\bba} &=& x_2 (x_1-x_2) (1-x_2) y \mu^2_{45}
       + x_2 (1-x_2) (1-x_1) z ( \mu^2_{45} - 2 )  \nl
{}&+& x_2 (1-x_2)^2   ( 1 - y \mu^2_4 + \mu^2_4 - \mu^2_{45} )
       + x_2 (1-x_1)^2 z^2
       + (1-x_2) y \nu^2_{xx},  \nl
\chi^4_{\bba} &=& x_2 (x_1-x_2) (1-x_2) \mu^2_{45} ( y - z )
       + x_2 (x_1-x_2)^2 z^2  \nl
{}&+& x_2 (1-x_2)^2 \mu^2_4  ( 1 - y )
       + (1-x_2)  y \nu^2_{xx}.
\eqa
%--
Note that, in the limit $x_2 \to 1$, we obtain
%--
\bq
\chi^1_{\bba} = \chi^2_{\bba} = (1 - x_1)^2\,y^2, \qquad
\chi^3_{\bba} = \chi^4_{\bba} = (1 - x_1)^2\,z^2.
\eq
%--
Similarly, in the limit $x_2 \to 0$, we obtain
$\chi^i_{\bba} = \nu^2_{xx}\,y$.
The result for $S^{221}$ is as follows:
%--
\bqa
S^{221} &=& -\,\frac{4}{s\,\lambda(1,\mu^2_4,\mu^2_5)}\,
\intfxy{x_1}{x_2}\,\frac{1}{x_2}\,J_{\bba}(x_1,x_2,y,z),
\eqa
%--
where the function $J_{\bba}$ is
%--
\bq
J_{\bba} = \int_0^1\,dy\,\int_y^1\,dz\,J^2_{\bba} +
\intfx{y}\,J^1_{\bba},
\eq
%--
and
%--
\bqa
J^1_{\bba} &=& \frac{1}{2}\,\frac{y(1-x_1)}{(1-x_2)^2}\,\Bigl[
\ln\chi^1_{\bba} - \ln\chi^2_{\bba}\Bigr]  \nl
{}&+& \frac{1}{1-x_2}\,\Bigl[ \frac{1}{2}\,(1 - \frac{1}{2}\,\mu^2_{45})\,
\ln\chi^1_{\bba} + \frac{1}{2}\,(y + \frac{1}{2}\,\mu^2_{45})\,
\ln\chi^2_{\bba}\Bigr],
\nl
J^2_{\bba} &=& \frac{3}{2}\,\frac{1-x_1}{(1-x_2)^2}\,\Bigl[
\ln\chi^4_{\bba} - \ln\chi^3_{\bba}\Bigr] - \,\frac{1}{1-x_2}\,
\Bigl[ \frac{3}{2}\,\ln\chi^4_{\bba} + 1\Bigr].
\eqa
%--
It is straightforward to show that in the limit $x_2 = 0$
%--
\bq
J_{\bba}(x_1,0,y,z) = 0,
\eq
%--
so that we obtain
%--
\bqa
S^{221} &=& -\,\frac{4}{s\,\lambda(1,\mu^2_4,\mu^2_5)}\,
\intfxy{x_1}{x_2}\,\lpar\frac{I^{\bba}(x_1,x_2)}{x_2}\rpar_+.
\eqa
%--
The integrand $I^{\bba}_+$ is defined as follows:
%--
\bq
I^{\bba}_+ = \int_0^1\,dy\,\int_y^1\,dz\,I^2_{\bba} +
\intfx{y}\,I^1_{\bba},
\eq
%--
with auxiliary functions $I^{(1,2)}_{\bba}$ given by
%--
\bqa
I^2_{\bba} &=& -\,\frac{x_2}{1-x_2} - \frac{3}{2}\,(1-x_1)\,\Bigl[
\frac{1}{(1-x_2)^2}\,\ln\chi^3_{\bba}(x_1,x_2,y,z) -
\ln\chi^3_{\bba}(x_1,0,y,z)\Bigr]  \nl
{}&+& \frac{3}{2}\,(1-x_1)\,\Bigl[
\frac{1}{(1-x_2)^2}\,\ln\chi^4_{\bba}(x_1,x_2,y,z) -
\ln\chi^4_{\bba}(x_1,0,y,z)\Bigr]  \nl
{}&-& \frac{3}{2}\,\Bigl[ \frac{1}{1-x_2}\,\ln\chi^4_{\bba}(x_1,x_2,y,z) -
\ln\chi^4_{\bba}(x_1,0,y,z)\Bigr],
\nl
I^1_{\bba} &=& \frac{1}{2}\,y (1-x_1)\,\Bigl[ \frac{1}{(1-x_2)^2}\,
\ln\chi^1_{\bba}(x_1,x_2,y) - \ln\chi^1_{\bba}(x_1,0,y)\Bigr]  \nl
{}&-& \frac{1}{2}\,y (1-x_1)\,\Bigl[ \frac{1}{(1-x_2)^2}\,
\ln\chi^2_{\bba}(x_1,x_2,y) - \ln\chi^2_{\bba}(x_1,0,y)\Bigr]  \nl
{}&+& \frac{1}{2}\,(1 - \frac{1}{2}\,\mu^2_{45})\,\Bigl[
\frac{1}{1-x_2}\,\ln\chi^1_{\bba}(x_1,x_2,y) -
\ln\chi^1_{\bba}(x_1,0,y)\Bigr],
\nl
{}&+& \frac{1}{2}\,(y + \frac{1}{2}\,\mu^2_{45})\,\Bigl[
\frac{1}{1-x_2}\,\ln\chi^2_{\bba}(x_1,x_2,y) -
\ln\chi^2_{\bba}(x_1,0,y)\Bigr].
\eqa
%--
%--
\subsubsection{The case $s = (m_4 \pm m_5)^2$}
%--
As already pointed out the derivation fails in this case because $\mu^2_{45}
=
\pm 2\,\mu_4$. The alternative solution is quite similar to the one adopted
for other topologies. If $\mu^2_{45} = -\,2\,\mu_4$ we can change variable
$z' =z + \mu_4$ so that
%--
\bq
\chi_{\bba} \to z^2 + (\mu^2_{xx} - \mu^2_4 - 2\,X\mu_4)\,y,
\eq
%--
with $z_- \le z \le z_+$ and
%--
\bq
z_- = X\,y + \mu_4, \qquad z_+ = 1 + (X-1)\,y + \mu_4,
\eq
%--
where $z_{\pm} \ge 0$. We change again variable $z^2 = t$ and use
%--
\bq
(t+a\,y)^{-1} = \partial_t\,\ln(t+a\,y),
\eq
%--
and integrate by parts obtaining
%--
\bqa
\frac{1}{2}\,\intfx{y}\,\int_{z^2_-}^{z^2_+}\,dt\,t^{-1/2}\,
\partial_t\,\ln(t+a\,y) &=&
\frac{1}{2}\,\intfx{y}\,\Bigl[ \frac{1}{z_+}\,\ln(z^2_++a\,y)  \nl
{}&-& \frac{1}{z_-}\,\ln(z^2_-+a\,y) + \frac{1}{2}\,
\int_{z^2_-}^{z^2_+}\,dt\,t^{-3/2}\,\ln(t+a\,y) \Bigl].
\eqa
%--
If $\mu^2_{45} = 2\,\mu_4$ this derivation is of little use because the new
limits of integration can be negative. However, in both cases we can write
%--
\bq
\mu^2_4 = \frac{1}{4}\,\Bigl[ \mu^4_{45} - \lambda(1,\mu^2_4,\mu^2_5)\Bigr],
\eq
%--
introduce $z_m = \mu^2_{45}/2$ and transform $z' = z - z_m$. With
$z_- = X\,y - z_m$, and $z_+ = 1 + (X-1)\,y - z_m$ we obtain
%--
\bq
\intfx{y}\,\int_{z_-}^{z_+}\,dz\,(z^2 + a\,y - \frac{1}{4}\,\lambda)^{-1},
\eq
%--
which gives
%--
\bqa
\frac{1}{2}\,\intfx{y}\,R^{-1}\,\Bigl[
\ln\frac{z_+-R}{z_--R} - \ln\frac{z_++R}{z_-+R}\Bigr],
\eqa
%--
where we have introduced
%--
\bq
R^2 = \frac{1}{4}\,\lambda - a\,y, \quad
a = \mu^2_{xx} - \mu^2_4 - 2\,X\mu_2,
\eq
%--
showing that no singularity will appear for $\lambda \equiv
\lambda(1,\mu^2_4,\mu^2_5) = 0$.
%--
\subsection{Derivative of $S^{221}$ and infrared poles\label{dsbabip}}
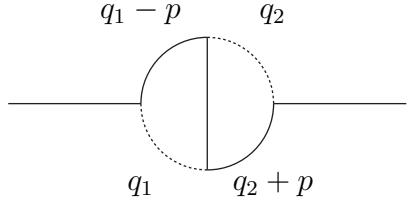
\begin{figure}[th]
\vspace{0.5cm}
\[
  \vcenter{\hbox{
  \begin{picture}(150,0)(0,0)
  \Line(0,0)(50,0)
  \DashCArc(75,0)(25,0,90){1.}
  \CArc(75,0)(25,90,180)
  \DashCArc(75,0)(25,180,270){1.}
  \CArc(75,0)(25,270,360)
  \Line(75,25)(75,-25)
  \Line(100,0)(150,0)
  \Text(50,30)[cb]{$q_1 - p$}
  \Text(100,30)[cb]{$q_2$}
  \Text(50,-35)[cb]{$q_1$}
  \Text(100,-35)[cb]{$q_2+p$}
  \end{picture}}}
\]
\vspace{0.5cm}
\caption[]{A special case of $S^{221}$ topology. Solid lines denote a
massive
particle, dashed lines a massless one.}\label{special}
\end{figure}

%--
In this section we study another infrared divergent object, the on-shell
derivative of the $S^{221}$ integral corresponding to the configuration a)
where
%--
\bq
\mbox{a)} \quad m_2 = m_4 = 0, m_1 = m_3 = m_5 = m \quad \mbox{and}\quad
p^2 = - m^2.
\eq
%--
It is important to recall that a necessary condition for the presence of
infrared divergences is that the Landau equations are fulfilled.
In the case of $S^{221}$ we see that the system of Landau equations,
\eqn{lsystem} is satisfied for this configuration of parameters if
$\alpha_3 = - \alpha_1$ and $\alpha_5 = \alpha_1$.
$S^{221}$ itself is not infrared divergent but the on-shell derivative is.
Let us consider the diagram of \fig{special} that corresponds to
configuration a). Solid lines correspond to the particle of mass $m$, while 
dashed lines give the massless particle. After taking the derivative with 
respect to $p^2$ we
will have two contributions where the propagators $(q_1-p)^2+m^2$ and
$(q_2+p)^2+m^2$ will appear with power $-2$ respectively. To study
qualitatively the infrared behavior we neglect $q^2_1, q^2_2$ and
$\spro{q_1}{q_2}$ compared with $\spro{p}{q_1}$ and $\spro{p}{q_2}$ in
propagator denominators, the so-called eikonal approximation. We further
define a superficial degree of infrared divergence for an arbitrary scalar
two-loop diagram
%--
\bq
\mu_{\ssI\ssR}(G) = 2\,n - 1 -2\,n_m - n_{\ssM},
\eq
%--
where $n_m$ is the number of massless lines and $n_{\ssM}$ the number of
massive lines. For the diagram of \fig{special} (after $p^2$-differentiation
and for on-shell $p$) we obtain $\mu_{\ssI\ssR}(S^{221}_p)= -1$. This shows
the presence of an infrared pole singularity.
%--
%--
We have
%--
\bqa
S^{221}_p &=& \frac{\partial}{\partial p^2}\,S^{221} = -\,
\lpar\frac{\tHss}{\pi}\rpar^{\ep}\,\egam{2+\ep}\,
\intfxy{x_1}{x_2}\,\Bigl[x_2(1-x_2)\Bigr]^{-1-\ep/2}  \nl
{}&\times& \intfx{y}\,\int_{Xy}^{1+(X-1)y}\,dz\,y^{\ep/2}\,
\Bigl[ z(z-1) - \rho\,y\Bigr]\,U^{-2-\ep}_{\bba},
\label{anotherd}
\eqa
%--
where the quadratic form $U_{\bba}$ is defined by
%--
\bq
U_{\bba} = -p^2\,z^2 + (p^2+m^2)\,(z - X\,y) + m^2_{xx}\,y,
\eq
%--
\bq
X = \frac{x_1-x_2}{1-x_2}, \qquad \rho = \frac{1-x_1}{x_2}\,X,
\eq
%--
and $m^2_{xx}$ is given in \eqn{defmxx}. Next we define the on-shell limit
of
$U_{\bba}$,
%--
\bq
U_{\bba}(p^2 = - m^2) = m^2\,\chi_{\bban{\rm os}},
\quad
\chi_{\bban{\rm os}} = z^2 + \eta\,y, \quad \eta =
\frac{(1-x_1+x_2)^2}{x_2(1-x_2)}.
\label{defeta}
\eq
%--
Inserting this result into \eqn{anotherd} we obtain the on-shell limit
of the derivative, expressed as
%--
\bqa
S^{221;\rm os}_p &=& -\,\lpar\frac{\tHss}{\pi\,m^2}\rpar^{\ep}\,
\egam{2+\ep}\,\frac{1}{m^4}\intfxy{x_1}{x_2}\,
\Bigl[x_2(1-x_2)\Bigr]^{-1-\ep/2}\,{\cal S}_{\bba},
\nl
{\cal S}_{\bba} &=& \intfx{y}\,\int_{Xy}^{1+(X-1)y}\,dz\,y^{\ep/2}\,
\Bigl[ z(z-1) - \rho\,y\Bigr]\,(z^2 + \eta\,y)^{-2-\ep}.
\label{starting}
\eqa
%--
In the evaluation we must keep $\ep \ne 0$ and one cannot expand around
$\ep =0$ after the $y-z$ integration because of the presence of overlapping
infrared divergences; here we face a novel feature and the procedure is made
even more complicated. Our solution is derived by adapting the general
algorithm of~~\cite{Binoth:2000ps}. Before explaining in detail how the
residues of the infrared poles and the finite part can be cast into a form
which allows for a reliable numerical integration we notice that the
$z$-integration can be performed in \eqn{starting} after using the identity:
%--
\bq
\Bigl[ 1 - \frac{1}{\mu} ( y \partial_y + \frac{z}{2} \partial_z)\Bigr]\,
 ( z^2 + \eta y)^{\mu} = 0.
\eq
%--
We define the following integrals:
%--
\bq
I_{ij} = \intfx{y}\,\int_{Xy}^{1+(X-1)y}\,dz\,y^{\ep/2}\,z^i y^j\,
          ( z^2 + \eta y)^{-2-\ep},
\eq
%--
and obtain
%--
\bqa
I_{20} &=& \frac{1}{1-\ep}\,\intfx{y}\,\Bigl[ y^{\ep/2}\,Q^{-2-\ep}\,
( 1 + \xi y - \xi^2 y^2 - \xi^3 y^3 ) + y^{1-\ep/2}\,L^{-2-\ep}\,
( 1 + \xi )^3\Bigr],
\nl
I_{10} &=& -\,\frac{1}{\ep}\,\intfx{y}\,\Bigl[
y^{\ep/2}\,Q^{-2-\ep}\,(1 - \xi^2 y^2) + y^{-\ep/2}\,L^{-2-\ep}\,(1 +
\xi)^2\Bigr],
\nl
I_{01} &=& \frac{1}{1-\ep}\,\intfx{y}\,\Bigl\{ y^{1+\ep/2}\,
Q^{-2-\ep}\,( 1 - \xi y) + y^{-\ep/2}\,L^{-2-\ep}\,( 1 + \xi)\Bigr],
\eqa
%--
with $\xi = X - 1$ and where
%--
\bq
Q = ( 1 + \xi y )^2 + \eta y, \qquad L = \eta + X^2 y.
\eq
%--
The relevant combination of integrals is
%--
\bq
{\cal I} = I_{20} - I_{10} - \rho\,I_{01}.
\eq
%--
Collecting the various terms we obtain
%--
\bqa
{\cal S}_{\bba} &=& \intfx{y}\,\Bigl\{
\frac{1}{\ep}\,\Bigl[ y^{\ep/2}\,( 1 - \xi^2 y^2 )\,Q^{-2-\ep} +
y^{-\ep/2}\,X^2\,L^{-2-\ep} \Bigr] + \frac{1}{1-\ep}  \nl
{}&\times& \Bigl[ y^{\ep/2}\,( 1 + (\xi-\rho)y - (\xi-\rho)\,
\xi y^2 - \xi^3 y^3)\,Q^{-2-\ep} + y^{-\ep/2}\,( X^2\,y - \rho )\,X\,L^{-2-\ep}
\Bigr]\Bigr\}.
\label{restart}
\eqa
%--
From its definition, \eqn{defeta}, we see that $\eta = 0$ for $x_1 = 1$ and
$x_2 = 0$. While
the quadratic form in $y$ contains a constant term, the linear form
$\eta + X^2 y$ vanishes if the parameters $\{x'_1 = 1-x_1, x_2, y\}$ are set
to zero. This is the origin of the overlapping divergence that we must
disentangle. First of all we change variable in order to factorize the
divergent terms: $x_1 = 1 - x'_1$ and,successively, $x_2 = x_1 x'_2$,
so that divergences will arise from the behavior of the integrands at the
point $x'_1 = 0$. After the transformation we get
%--
\bq
\xi = -\,\frac{x_1}{1-x_1 x_2}, \qquad \eta = x_1\,
\frac{(1+x_2)^2}{x_2 (1-x_1 x_2)}.
\eq
%--
When we exchange the $x_1 - x_2$ integrations the limits become
$0 \le x_2 \le +\,\infty$ and $0 \le x_1 \le 1/(1+x_2)$. We change again
variable, $x_1 = x'_1/(1+x_2)$, so that $0 \le x'_1 \le 1$ and
%--
\bq
\xi = -\,\frac{x_1}{1+x_2-x_1 x_2}, \qquad
\eta = x_1\,\frac{(1+x_2)^2}{x_2 (1+x_2-x_1 x_2)}.
\eq
%--
\subsubsection{Defoliation of IR divergent integrals}
%--
At this point we introduce a simple example to explain the method. Consider
%--
\bq
I = \intfxx{x}{y}\,Q^{-2-\ep}, \qquad Q = x(1+x) + y^2.
\eq
%--
A minimal set of zeros is a minimal set of parameter values that satisfy
$Q = 0$. In our simple case we have $\{x=y=0\}$. A defoliation is the
procedure that allows for a factorization of the divergent part (in the limit 
$\ep \to 0$) through different steps. In our case,
%--
\bq
I = \Bigl\{ \intfxy{x}{y} + \intfxy{y}{x} \Bigr\}\, Q^{-2-\ep} =
\intfxx{x}{y}\,\Bigl[ x^{-1-\ep}\,Q_{\{x=0\}}^{-2-\ep} +
y^{-1-\ep}\,Q_{\{y=0\}}^{-2-\ep} \Bigr\},
\eq
%--
where the defoliated Q's are:
%--
\bq
Q_{\{x=0\}} = 1 + x + x\,y^2, \qquad Q_{\{y=0\}} = x\,(1 + x\,y) + y.
\eq
%--
Since the first form never vanishes over the whole parametric space,
including the border, the procedure stops here for the first integral, while 
for the second one has to iterate until the condition is met.
To give an example, if we start with $Q = x(1-x)+y^2$ the
first defoliation gives $x(1-xy)+y$ but also $1-x+xy^2$ which has a set of
zeros at $\{x=1,y=0\}$; then one has to transform, $x = 1-x'$, and iterate.

Note that the algorithm is perfectly defined for all integrands that may
vanish only at the boundaries of the parametric space.
$S^{221}_p$ is the first example of a strategy which differs considerably
from the minimal BT approach. In the latter the main attempt is toward
{\em raising} powers in the integrand while here the powers are untouched
and the whole idea is about factorizing the singular behavior into simple
factors that can be integrated analytically, leaving non singular terms to
be integrated numerically.

Other cases, where the Feynman integrand is not semi-positive defined
(presence of thresholds) are more difficult to analyze. However this never
happens for two-point functions and their derivatives. Infrared
singularities are finally extracted by using Taylor expansion, typically
%--
\bq
\intfxx{x}{y}\,x^{-1-\ep}\,f(x,y,\ep) = -\,\frac{1}{\ep}\,\intfx{y}\,
f(0,y,\ep) + \intfxx{x}{y}\,\Bigl( \frac{f(x,y,\ep)}{x^{1+\ep}}\Bigr)_+.
\eq
%--
Note that only now one can expand around $\ep = 0$. Various branches in the
procedure are not uniquely defined and one is led to the search of the path
which brings the minimal number of iterations.
%--
\subsubsection{$S^{221}_p$ method I}
%--
After these preliminar considerations we go back to \eqn{restart}.
After all transformations we rewrite $Q$ and $L$ obtaining
%--
\bqa
Q &=&
(1+x_2-x_1x_2)^2\,x_2 + (1+x_2^2)\,(1+x_2-x_1x_2)\,x_1\,y + x_1^2\,x_2\,y^2,
\\
L &=&
(1+x_2)^2\,(1+x_2-x_1x_2)\,x_1 + (1+x_2)^2\,(1-x_1)^2\,x_2\,y.
\eqa
%--
The integration range is from $0$ to $1$ for $x_1$ and $y$ and from
$0$ to $\infty$ for $x_2$. It is easily seen that $Q$ and $L$ may vanish
only for zero values of the parameters. In order to extract the divergences we
use the relation:
%--
\bqa
\intfxx{x}{y}\,f(x,y) &=&
\intfxy{x}{y}\,f(x,y) + \intfxy{y}{x}\,f(x,y)  \\
&=& \intfxx{x}{y}\,\big( x\,f(x,xy) + y\,f(xy,y) \big)
\eqa
%--
Since our functions $f=Q,L$ are polynomials to a power $-2-\ep$,
we have $f(x,xy)=x^{-2-\ep}g(x,y)$, where $g$ has a zero at $x=0$
of lower order.
We split the integration over $x_2$ in two parts,
the first in the interval $[0,1]$ and the second in $[1,\infty]$. In the
former we iterate the procedure until we reach polynomials that do not
vanish inside the parametric space. In the latter we perform the procedure
only for $x_1$ and $y$.

In the limit $x_2 \to \infty$ some of the integrals may lead to a
divergency because of the presence of divergent factors, other than in $Q$
and $L$. In that case we perform the transformation $x_2 \to 1/x_2$
and iterate the procedure for $x_2 \in [0,1]$.
At the end we obtain integrands where the divergences are contained 
only in the following factors:
%--
\bq
x_1^{-1-\ep}, \quad x_1^{-1-2\ep},\quad x_2^{-1-\ep}, \quad \mbox{or}\quad
y^{-1-2\ep}.
\eq
%--
The final answer is obtained by using the following relation:
%--
\bq
\intfx{y}\,y^{-1-2\ep}\,f(y) = \intfx{y}\,y^{-2\ep}\,\Bigl(
\frac{f(y)}{y} \Bigr)_+ - \frac{1}{2\ep}\,f(0).
\eq
%--
In this way, after the expansion around $\ep = 0$ we obtain
%--
\bq
\intfxy{x_1}{x_2}\,
\Bigl[x_2(1-x_2)\Bigr]^{-1-\ep/2}\,{\cal S}_{\bba} =
\frac{1}{\ep^2}\,{\cal S}_{\bba}^2 +
\frac{1}{\ep}\,{\cal S}_{\bba}^1 +
{\cal S}_{\bba}^0
\eq
%--
where ${\cal S}_{\bba}^2$, ${\cal S}_{\bba}^1$ and ${\cal S}_{\bba}^0$ are
expressed in terms of smooth integrands. Furthermore, ${\cal S}_{\bba}^2$ is
the sum of integrals which can be calculated analytically:
%--
\bqa
{\cal S}_{\bba}^2 &=&
- \frac{1}{2}\,\int_0^{\infty} dx_2 \intfx{y}\,
  \frac{x_2}{1+x_2}(1+x_2+x_2y)^{-2}  \nl
{}&-& \frac{1}{2}\,\int_1^{\infty} dx_2 \intfx{x_1}\,
  \frac{x_2}{1+x_2}(x_1+x_2+x_1x_2)^{-2}  \nl
&-&
 \frac{1}{2}\,\intfxx{x_1}{x_2}\,
  \frac{1}{1+x_2}(1+x_2+x_1x_2)^{-2}
- \frac{1}{2}\,\intfx{x_1}\,(1+x_1)^{-2}  \nl
&-&
 \frac{1}{2}\,\intfxx{x_1}{x_2}\,
  \frac{x_2}{1+x_1x_2}(1+x_2+x_1x_2)^{-2}
- \frac{1}{2}\,\intfx{x_2}\,(1+x_2)^{-2}  \nl
&=&
  \frac{1}{2}\,(\ln 2 - 1) - \frac{1}{2}\,\big( \ln 2 - \ln \frac{3}{2}
\big)
+ \frac{1}{2}\,\big( \ln 3 - 2 \ln 2 - \frac{1}{2} \big)  \nl
{}&+& \frac{1}{2}\,\big( 3 \ln 2 - 2 \ln 3 - \frac{1}{2} \big) = -1.
\eqa
%--
For the single IR pole we obtain a relatively long expression that will be
reported in Appendix D.
The expression of ${\cal S}_{\bba}^0$ is too long and of little interest
from
the analytical point of view. Therefore it will not be reported here.
%--
\subsubsection{$S^{221}_p$ method II}
%--
To check the correctness of method I we have also used a different strategy.
In \eqn{starting} after changing variables as described after \eqn{restart}
we exchange the order of integration, so that $x_2 \in [0,\infty]$ and
$x_1 \in [0,1/(1+x_2)]$. Two other changes of variables will bring all
integrals in the interval $[0,1]$: they are $x_1 = x'_1/(1+x_2)$ and
$x_2 = 1/x'_2-1$. After rationalization we end up with two polynomials to
the power $-2-\ep$. They are
%--
\bqa
Q^t &=&  y x_1 (1 - x_1 x_2)\, \Bigl[ x_2^2 + ( 1 - x_2)^2 \Bigr] +
         y^2 x_1^2  x_2 ( 1 - x_2)^2 + ( 1 - x_1 x_2)^2 x2,
\nl
L^t &=&  y x_1^2 x_2 (1-x_2)^2
       + y x_2 (1-x_1 x_2)^2
       + x_1 (1-x_1 x_2)\,\Bigl[ 2 x_2 (1-x_2) (1-y) + x_2^2 + (1-x_2)^2
\Bigr].
\nl
\eqa
%--
We have written the two polynomials in a form that makes clear their
semi-positive definiteness over the cube $ 0 \le \{x_1,x_2,y\} \le 1$.
For more than two variables, $x_1 \dots x_n$, defoliation is based on
inserting the following identity,
%--
\bq
1 = \sum_{l=1}^n\,\prod_{j\ne l=1}^n\,\theta(x_l \ge x_j),
\eq
%--
in the integral. Several defoliations are needed and we summarize the
results
in the following list. First $Q^t$,
%--
\bqa
\{x_1=x_2=0\} \quad &:&  \quad Q^t \to Q^t_1 + Q^t_2,
\nl
\{x_1=x_2=1\} \quad &:&  \quad Q^t_1 \to Q^t_{11} + Q^t_{12},
\nl
\{y=x_2=0\} \quad &:&  \quad Q^t_{11} \to Q^t_{111} + Q^t_{112},
\nl
\{y=x_1=0\} \quad &:&  \quad Q^t_{12} \to Q^t_{121} + Q^t_{122},
\nl
\{y=x_2=0\} \quad &:&  \quad Q^t_{2} \to Q^t_{21} + Q^t_{22},
\nl
\{y=x_1=x_2=1\} \quad &:&  \quad Q^t_{21} \to Q^t_{211} + Q^t_{212} +
Q^t_{213},
\nl
\{x_1=x_2=1\} \quad &:&  \quad Q^t_{22} \to Q^t_{221} + Q^t_{222},
\nl
\{y=x_2=0\} \quad &:&  \quad Q^t_{221} \to Q^t_{2211} + Q^t_{2212},
\nl
\{y=x_1=0\} \quad &:&  \quad Q^t_{222} \to Q^t_{2221} + Q^t_{2222},
\eqa
%--
and then $L^t$,
%--
\bqa
\{y=x_1=0\} \quad &:&  \quad L^t \to L^t_1 + L^t_2,
\nl
\{x_1=x_2=0\} \quad &:&  \quad L^t_1 \to L^t_{11} + L^t_{12},
\nl
\{y=x_1=x_2=1\} \quad &:&  \quad L^t_{11} \to L^t_{111} + L^t_{112} +
L^t_{113},
\nl
\{y=x_1=x_2=1\} \quad &:&  \quad L^t_{12} \to L^t_{121} + L^t_{122} +
L^t_{123},
\nl
\{x_1=x_2=1\} \quad &:&  \quad L^t_2 \to L^t_{21} + L^t_{22}.
\nl
\eqa
%--
In the above procedure it is understood that the appropriate change of
variables is performed to bring the zeros at the origin at each step.
At the end of any chain all polynomials are strictly positive defined in the
unit cube. One final step is needed to disentangle integrals of the
following
form:
%--
\bq
J(a,b) = \intfxx{x}{y}\,(1-x)^{-1-a\ep}\,(1-x y)^{-1-b\ep}\,f(x,y),
\eq
%--
where $f(x,y)$ is analytic in the integration domain. We obtain
%--
\bqa
J(a,b) &=& \intfxx{x}{y}\,\Bigl[ x^{-1-(a+b)\ep}\,(1+y-x y)^{-1-b\ep}\,
f(1-x,1-xy)  \nl
{}&+& x^{-1-a\ep} y^{-1-(a+b)\ep}\,(1+x-x y)^{-1-b\ep}\,f(1-xy,1-y).
\eqa
%--
Finally all divergent integrals factorize, an example being as follows:
%--
\bq
K(a,b) = \intfxx{x}{y}\,x^{-1-a\ep}\,y^{-1-b\ep}\,F(x,y),
\eq
%--
\bqa
K(a,b) &=& \frac{1}{ab\ep^2}\,F(0,0) - \frac{1}{a\ep}\intfx{y}\,\Bigl(
\frac{F(0,y)}{y^{1+b\ep}}\Bigr)_+ - \frac{1}{b\ep}\,\intfx{x}\,\Bigl(
\frac{F(x,0)}{x^{1+a\ep}}\Bigr)_+  \nl
{}&+& \intfxx{x}{y}\,\Bigl(\frac{F(x,y)}{x^{1+a\ep}y^{1+b\ep}}\Bigr)_{++},
\eqa
%--
where we have introduced a $'+'$ and a $'++'$ distribution
%--
\bq
\intfx{x}\,\Bigl(\frac{g(x)}{x^{1+a\ep}}\Bigr)_+ =
\intfx{x}\,\frac{g(x)-g(0)}{x}\,\Bigl[ 1 - a\ep\,\ln x + \ord{\ep^2}\Bigr],
\eq
%--
\bqa
\intfxx{x}{y}\,\Bigl(\frac{h(x,y)}{x^{1+a\ep}y^{1+b\ep}}\Bigr)_{++} &=&
\intfxx{x}{y}\,\frac{h(x,y)-h(0,y)-h(x,0)+h(0,0)}{x y}  \nl
{}&\times& \Bigl[ 1 - \ep\,(a \ln x + b \ln y ) + \ord{\ep^2}\Bigr].
\eqa
%--
Also in this case the residue of the double IR pole can be computed
analytically and gives $- 1$. The explicit expressions for the residue of
the single IR pole, as well as for the finite part will not be reported.
%--
\subsubsection{Another configuration for $S^{221}_p$}
%--
There is another configuration where \eqn{lsystem} is satisfied,
%--
\bq
\mbox{b)} \quad m_1 = m_2 = m_4 = m_5 = m, m_3 = 0 \quad \mbox{and}\quad
s = 0.
\eq
%--
Actually this configuration gives a nice example of a necessary condition
which, however, is not sufficient. Let us consider the corresponding
derivative,
%--
\bqa
S^{221|b}_p &=& \frac{\partial}{\partial p^2}\,S^{221|b} = -\,
\lpar\frac{\tHss}{\pi}\rpar^{\ep}\,\egam{2+\ep}\,
\intfxy{x_1}{x_2}\,\Bigl[x_2(1-x_2)\Bigr]^{-1-\ep/2}  \nl
{}&\times& \intfx{y}\,\int_{Xy}^{1+(X-1)y}\,dz\,y^{\ep/2}\,
\Bigl[ z(z-1) - \rho\,y\Bigr]\,U^{-2-\ep}_{\bbab},
\label{anotherdp}
\eqa
%--
where the quadratic form $U_{\bbab}$ is defined by
%--
\bq
U_{\bbab} = - p^2 z^2 + p^2 z + (m^2_{xx} - m^2 - X p^2) y + m^2,
\eq
%--
and $m^2_{xx}$ is given in \eqn{defmxx}. Next we define the $p^2 \to 0$
limit of $U_{\bbab}$,
%--
\bq
U_{\bbab}(p^2 = 0) = \frac{m^2}{x_2}\,\chi_{\bbabn{\rm os}}, \quad
\chi_{\bbabn{\rm os}} = ( 1 - x_2 )\,y + x_2.
\eq
%--
Although the point $y = x_2 = 0$ is the candidate for generating an infrared
divergence we can prove, by direct calculation, that the result is infrared
finite. When we set $\ep = 0$ it is easy to see that the $y,z$ integrations
are trivial and
%--
\bq
{\cal S}{\bbab} = \intfxy{x_1}{x_2}\,{\cal F}_{\bbab},
\quad
{\cal F}_{\bbab} = \sum_{n=2}^{6}\,\frac{A_n(x_1)}{(1-x_2)^n} +
\sum_{n=3}^{7}\,\frac{B_n(x_1)}{(1-x_2)^n}\,\ln x_2,
\eq
%--
where the various coefficients are:
%--
\bqa
A_6 &=& 3 - 6 x_1 + 3 x_1^2,
\quad
A_5 = - \frac{17}{2} + 14 x_1 - \frac{11}{2} x_1^2,
\quad
A_4 = 7 - \frac{13}{2} x_1 + \frac{1}{2} x_1^2,
\nl
A_3 &=& - 2 + \frac{1}{2} x_1,
\quad
A_2 = \frac{1}{3},
\quad
B_7 = 3 - 6 x_1 + 3 x_1^2,
\nl
B_6 &=& - 10 + 17 x_1 - 7 x_1^2,
\quad
B_5 = 11 - 13 x_1 + 3 x_1^2
\quad
B_4 = - 5 + 3 x_1,
\quad
B_3 = 1.
\eqa
%--
The non trivial part of the $x_2$-integration is performed recursively by
using the following relation:
%--
\bqa
J_{-n} &=& \int_{0}^{x_1}\,dx_2\,\frac{\ln x_2}{(1-x_2)^n},
\qquad J_{-2} = \ln(1-x_1) - \frac{x_1}{x_1-1}\,\ln x_1.
\nl
n\,J_{-(n+1)} &=& (n-1)\,J_{-n} + \frac{1}{n-1} -
\frac{1}{n-1}\,\frac{1}{(1-x_1)^{n-1}} + \frac{x_1\,\ln x_1}{(1-x_1)^n},
\eqa
%--
Finally the $x_1$-integration uses the following results~\footnote{For the
first equation we disagree with the result of p. 20, 
ref.~\cite{Devoto:1984tc}}:
%--
\bqa
\intfx{x}  x^{-4}\,\ln(1-x) &=& - \intfx{x} (x^{-3}+\frac{1}{2}
x^{-2} + \frac{1}{3} x^{-1} ) - \frac{11}{18},  \nl
\intfx{x} x^{-3}\,\ln (1-x) &=& - \intfx{x} ( x^{-2} + \frac{1}{2}
x^{-1} ) - \frac{3}{4},  \nl
\intfx{x} x^{-2}\,\ln(1-x) &=& - \intfx{x} x^{-1} - 1,  \nl
\intfx{x} x^{-1}\,\ln(1-x) &=&  - \zeta(2),
\label{infint}
\eqa
%--
\bq
\intfx{x} x^n_1 \ln x =  - \frac{1}{(n+1)^2},  \quad
\intfx{x} x^n_1 \ln(1-x))
= -\frac{1}{n+1}\,\sum_{l=1}^{n+1}\,\frac{1}{l}.
\eq
%--
The final result is
%--
\bq
{\cal S}{\bbab} = -\,\frac{13}{72},
\eq
%--
proving that the b-configuration is infrared finite.
%--
\subsection{Tensor integrals of the $S^{221}$ family\label{itsbabf}}
%--
The scalar $S^{221}$ integral has been computed at $n = 4$ and the absence
of double-logarithms has shown some advantage for the integration in the
complex plane. On the contrary, tensor integrals of the same family are 
ultraviolet divergent and this feature seems lost. However, this problem is
encompassed by the procedure of {\em scalarization}. Scalarization can never
be complete, in contrast with the one-loop case~\cite{Passarino:1979jh}, and 
we will stop the procedure whenever an irreducible numerator is encountered.
A tensor integral will be irreducible when the numerator cannot be expressed
as a linear combination of those propagators which are present at that 
particular moment of the reduction. In a sense we scalarize the obvious
and explicitly compute the rest.

Let us define the following function:
%--
\bq
\pi^4\,S^{2-i/2,1,2-j/2}_5(\mu_1\dots \mu_i\,|\,\nu_1\dots \nu_j)
= \mu^{2\ep}\,\intmomsii{n}{q_1}{q_2}  \nl
\frac{q^{\mu_1}_1\dots q^{\mu_i}_1\,q^{\nu_1}_2\dots q^{\nu_j}_2}
{(q^2_1+m^2_1)\dots ((q_2+p)^2+m^2_5)},
\eq
%--
where we assume $i+j \le 3$ and where $\alpha$ and $\beta$ have been
changed in order to account for the effective degree of convergence:
for $i+j \ge 2$ the function is overall divergent, for $i \ge 2$ the
$\alpha\gamma$ sub-diagram is divergent, etc.
Note that we always indicate the total number of internal lines.

Let us consider some example,
%--
\bq
S^{1/2,1,2}_5(p,p,p\,|\,0),
\eq
%--
and perform the reduction. Since there are {\em enough} propagators, the
reduction is particularly simple, 
%--
\bqa
\spro{q_1}{p} &=& \frac{1}{2}\,\Bigl\{
\Bigl[(q_1+p)^2+m^2_2\Bigr]-\Bigl[q_1^2+m^2_1\Bigr]+m^2_1-m^2_2-p^2\Bigr\},
\nl
\spro{q_2}{p} &=& \frac{1}{2}\,\Bigl\{
\Bigr[(q_2+p)^2+m^2_5\Bigr]-\Bigl[q_2^2+m^2_4\Bigr]+m^2_4-m^2_5-p^2\Bigr\}.
\nl
\spro{q_1}{q_2} &=& -\,\frac{1}{2}\,\Bigl\{\Bigl[(q_1-q_2)^2+m^2_3\Bigr]-
\Bigl[q^2_1+m^2_1\Bigr]-\Bigl[q^2_2+m^2_4\Bigr]-m^2_3+m^2_1+m^2_4\Bigr\},
\eqa
%--
etc. and give the following result:
%--
\bqa
8\,S^{1/2,1,2}_5(p,p,p\,|\,0) &=&
(m^2_1 - m^2_2 - p^2)^3\,S^{221}(0\,|\,0)  \nl
{}&+& (m^2_1 - m^2_2 - p^2)^2\,
S^{121}_4(0\,|\,0\,|p,m_1,m_3,m_4,m_5)  \nl
{}&+& \Bigl[ 4\,p^2\,(m^2_1 - m^2_2) -
 7\,p^4 - (m^2_1-m^2_2)^2\Bigr]\,
S^{121}_4(0\,|\,0\,|\,-p,m_2,m_3,m_5,m_4)  \nl
{}&+& 2\,(m^2_1 - m^2_2 - p^2)\,
S^{1/2,2,1}_4(p\,|\,0\,|\,p,m_1,m_3,m_4,m_5)  \nl
{}&+& 2\,(5\,p^2 - m^2_1 + m^2_2)\,
S^{1/2,2,1}_4(p\,|\,0\,|\,-p,m_2,m_3,m_5,m_4)  \nl
{}&+& 4\,\Bigl[S^{021}_4(p,p\,|\,0\,|\,p,m_1,m_3,m_4,m_5)
- S^{021}_4(p,p\,|\,0\,|\,-p,m_2,m_3,m_5,m_4)\Bigr]
\nl
\eqa
%--
Another interesting example of scalarization is represented by
%--
\bqa
8\,S^{1,1,3/2}_5(p,p\,|\,p) &=&
- p^4_{12}\,p^2_{45}\,S^{221}_5(0\,|\,0) +
p^2_{12}\,p^2_{45}\,S^{121}_4(0\,|\,0\,|\,p,m_1,m_3,m_4,m_5)  \nl
{}&-& p^2_{45}\,(p^2_{12} +
2\,p^2)\,S^{121}_4(0\,|\,0\,|\,-p,m_2,m_3,m_5,m_4)
\nl
{}&+& p^2_{12}\,\Bigl[ S^{121}_4(0\,|\,0\,|\,p,m_4,m_3,m_1,m_2) -
S^{121}_5(0\,|\,0\,|\,-p,m_5,m_3,m_2,m_1)\Bigr]  \nl
{}&+& 2\,p^2_{45}\,\Bigl[ S^{1/2,2,1}_4(p\,|\,0\,|\,-p,m_2,m_3,m_5,m_4) -
S^{1/2,2,1}_4(p\,|\,0\,|\,p,m_1,m_3,m_4,m_5)\Bigr]  \nl
{}&+& ( p^2 + m^2_2 + p^2_{12})\,S^{111}_3(0\,|\,0\,|\,-p,m_2,m_3,m_4) +
p^2_{12}\,\Bigl[ S^{111}_3(0\,|\,0\,|\,p,m_1,m_3,m_5)  \nl
{}&-& S^{111}_3(0\,|\,0\,|\,0,m_1,m_3,m_4)\Bigr]
+ (p^2 - m^2_2)\,S^{111}_3(0\,|\,0\,|\,-p,m_2,m_3,m_4)  \nl
{}&-& (p^2_{12} + 2\,p^2)\,S^{111}_3(0\,|\,0\,|\,0,m_2,m_3,m_5)
- 2\,\Bigl[ S^{1/2,1,1}_3(p\,|\,0\,|\,p,m_1,m_3,m_5)  \nl
{}&+& S^{1/2,1,1}_3(p\,|\,0\,|\,-p,m_2,m_3,m_4)\Bigr]
\nl
\eqa
%--
where we have introduced the following combinations:
$p^2_{ij}= p^2 - m^2_i + m^2_j$. 
%--
Other relevant examples can be obtained by permutation of masses and all of
them indicate that we need at most the scalar $S_5$-function plus
a combination of tensor integrals with a number of internal legs which
is $\le 4$.
%--
\subsection{A realistic example\label{are}}
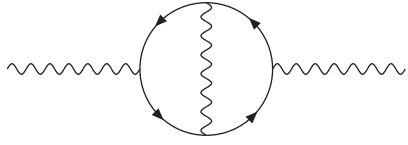
\begin{figure}[th]
\vspace{0.5cm}
\[
  \vcenter{\hbox{
  \begin{picture}(150,0)(0,0)
  \Photon(0,0)(50,0){2}{7}
  \ArrowArc(75,0)(25,0,90)
  \ArrowArc(75,0)(25,90,180)
  \ArrowArc(75,0)(25,180,270)
  \ArrowArc(75,0)(25,270,360)
  \Photon(75,25)(75,-25){2}{7}
  \Photon(100,0)(150,0){2}{7}
  \end{picture}}}
\]
\vspace{0.5cm}
\caption[]{A contribution to the photon self-energy in QED}\label{QED}
\end{figure}

\begin{figure}[th]
\vspace{0.5cm}
\[
  \vcenter{\hbox{
  \begin{picture}(150,0)(0,0)
  \Photon(0,0)(50,0){2}{7}
  \ArrowArc(75,0)(25,0,180)
  \ArrowArc(75,0)(25,180,360)
  \Photon(100,0)(150,0){2}{7}
  \Text(50,-5)[cb]{{\Large$\times$}}
  \end{picture}}}
\qquad
  \vcenter{\hbox{
  \begin{picture}(150,0)(0,0)
  \Photon(0,0)(50,0){2}{7}
  \ArrowArc(75,0)(25,0,180)
  \ArrowArc(75,0)(25,180,360)
  \Photon(100,0)(150,0){2}{7}
  \Text(100,-5)[cb]{{\Large$\times$}}
  \end{picture}}}
\]
\vspace{0.5cm}
\caption[]{The one-loop subtraction diagrams, containing a one-loop
counter-term (represented by a $\times$), associated to the two-loop
diagram of \fig{QED}.}
\label{subQED}
\end{figure}
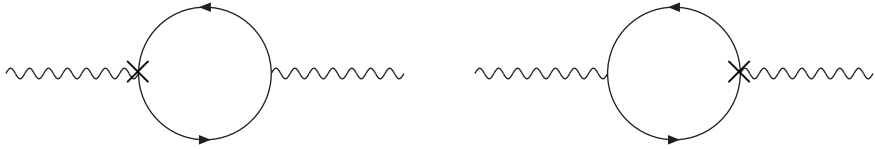

%--
To discuss some realistic example of tensorial reduction in the $S_5$-family
let us consider one of the two-loop diagrams that in QED contribute to the
photon self-energy, see \fig{QED}.
%--
%--
We derive the following expression for the vacuum polarization:
%--
\bq
\Pi^{(2)}_{\mu\nu} = \pi^4\,e^4\,\Pi^{(2)}_0\,(\drii{\mu}{\nu} -
\frac{p_{\mu}p_{\nu}}{p^2}), \qquad  \Pi^{(2)}_0 = \frac{1}{3-\ep}\,S.
\eq
%--
For the function $S$ we simplify the numerator and obtain
%--
\bqa
S &=& 4\,\Bigl[ (\ep-2)\,p^4 + 8\,m^4 + 2\,\ep p^2 m^2\Bigr]\,
S^{221}(p^2;m,m,0,m,m)  \nl
{}&+& 2\,\Bigl[ 8\,m^2 - (8 - 2\,\ep - \ep^2)\,p^2\Bigr]\,
B^2_0(p^2;m,m) + 8\,(2 - \ep)\,p^2\,B^2_1(p^2;m,m) \nl
{}&-& 16\,\Bigl[2\,m^2 + (\ep - 2)\,p^2\Bigr]\,S^{121}_4(p^2;m,0,m,m)
+ 4\,\ep\,(2 - \ep)\,S^{111}(p^2;m,0,m)  \nl
{}&+& 8\,(\ep - 2)\,S^{111}(0;m,0,m) + 8\,(2 - \ep)\,A_0(m)\,B_0(p^2;m,m).
\eqa
%--
Note that $S$ is ultraviolet divergent and must be associated with the
corresponding subtraction diagrams of \fig{subQED}.
%--
%--
\section{Numerical Results\label{numres}}
%--
The whole set of algorithms presented in the previous sections has been
translated into a FORM code~\cite{Vermaseren:2000nd} and the output has been
used to create a FORTRAN code.

In this section we will present numerical results for two-loop two-point
functions (self-energies), based on the results developed in the previous
sections. In the following we present a detailed comparison among different
formulations for the same topology, different integration routines and,
for few special cases, between our numerical results and the available
analytical calculations.
Typically we have used three different sub-routines for numerical
integration: D01EAF, a multi-dimensional adaptive quadrature over 
hyper-rectangle, multiple integrands; D01GBF, a multi-dimensional quadrature 
over hyper-rectangle, Monte Carlo method; D01GDF, a multi-dimensional 
quadrature, general product region, number-theoretic method.
%--
\subsection{Numerical results for $S^{121}$}
%--
To check our results for this topology we consider two special cases,
namely a) $m_1 = m, m_i = 0$ for $i \ge 2$ and b) $m_2 = m_4 = m, m_1 =
m_3 = 0$ which correspond to Eq. (38) and to Eq.(45) of
ref.~\cite{Bauberger:1995by}.
Results and comparisons for configuration b) are shown in \tabn{tablea}.
The first entry is based on method II, \eqn{sabamii}, and the
NAG~\cite{naglib} subroutine D01EAF; the second entry refers to the same 
\eqn{sabamii} but uses the NAG subroutine D01GBF. The latter usually returns 
a smaller, underestimated,integration error. The former has a larger error, 
typically $0.06\%$ for the real part and $1.7\%(0.9\%,0.5\%)$ for the 
imaginary part at $81\,\GeV(83\,\GeV,85\,GeV)$. Relative deviations with the 
analytical results are $\le 0.006\%$ for the real part and 
$0.09\%(0.05\%,0.04\%)$ for the imaginary part ($0.07\%,0.02\%,0.01\%$ for 
D01GBF). As expected the largest errors occur around threshold.

For configuration a) we report results and comparisons in \tabn{tableb}.
%--
\begin{table}[ht]\centering
\begin{tabular}{|l|l|l|l|l|}
\hline
 & & & &\\
$\sqrt{s}\,$[GeV]  &  $\Reb\,S^{121}$  & $\Reb\,S^{121}$ (anal.)  &
                      $\Imb\,S^{121}$  & $\Imb\,S^{121}$ (anal.)  \\
 & & & & \\
\hline
 & & & & \\
75 & 54.27(3)   & 54.2624 & 0.00 & 0.00  \\
   & 54.2652(1) &         &      &       \\
 & & & & \\
\hline
 & & & & \\
77 & 53.11(3)   & 53.1076 & 0.00 & 0.00  \\
   & 53.1227(4) &         &      &       \\
 & & & & \\
\hline
 & & & & \\
79 & 51.63(3)   & 51.6333 & 0.00 & 0.00  \\
   & 51.6348(2) &         &      &       \\
 & & & & \\
\hline
 & & & & \\
80 & 50.59(3)   & 50.5868 & 0.00 & 0.00  \\
   & 50.6040(2) &         &      &       \\
 & & & & \\
\hline
 & & & & \\
81 & 49.35(3)   & 49.3415 & -0.58(1)   & -0.585457 \\
   & 49.3791(4) &         & -0.5850(5) &           \\
 & & & & \\
\hline
 & & & & \\
83 & 47.68(3)   & 47.6763 & -2.18(2)   & -2.18366  \\
   & 47.6781(4) &         & -2.183(5)  &           \\
 & & & & \\
\hline
 & & & & \\
85 & 46.53(3)   & 46.5277 & -3.88(2)   & -3.87974  \\
   & 46.5088(4) &         & -3.880(5)  &           \\
 & & & & \\
\hline
\hline
\end{tabular}
\vspace*{3mm}
\caption[]{The topology $S^{11}$ for $m_1 = m_3 = 80\,$GeV and $m_2 =
m_4 = 0$ as a function of $\sqrt{s}$. First entry is from \eqn{sabamii}
and NAG subroutine D01EAF and the analytical result of
ref.~\cite{Bauberger:1995by}. Second entry is from \eqn{sabamii}
and NAG subroutine D01BGF. The UV-pole is $\ep = 1$ and the unit of mass
is $\tHs = 1\,$GeV.}
\label{tablea}
\end{table}
\normalsize
%--
%--
\begin{table}[ht]\centering
\begin{tabular}{|l|l|l|l|l|}
\hline
 & & & &\\
$\sqrt{s}\,$[GeV]  &  $\Reb\,S^{121}$  & $\Reb\,S^{121}$ (anal.)  &
                      $\Imb\,S^{121}$  & $\Imb\,S^{121}$ (anal.)  \\
 & & & & \\
\hline
 & & & & \\
75 & 40.34(3)   & 40.3502 & -49.77(3)   & -49.7720  \\
   & 40.3501(4) &         & -49.7756(3) &           \\
 & & & & \\
\hline
 & & & & \\
77 & 41.16(3)   & 41.1696 & -49.94(3)   & -49.9374  \\
   & 41.1952(5) &         & -49.9387(5) &           \\
 & & & & \\
\hline
 & & & & \\
79 & 41.96(3)   & 41.9693 & -50.10(3)   & -50.0985  \\
   & 41.9512(4) &         & -50.1045(3) &           \\
 & & & & \\
\hline
 & & & & \\
80 & 42.36(3)   & 42.3621 & -50.18(3)   & -50.1775  \\
   & 42.3447(4) &         & -50.1756(3) &           \\
 & & & & \\
\hline
 & & & & \\
81 & 42.74(3)   & 42.7502 & -50.26(3)   & -50.2556  \\
   & 42.7302(4) &         & -50.2691(3) &           \\
 & & & & \\
\hline
 & & & & \\
83 & 43.50(3)   & 43.5127 & -50.41(3)   & -50.4090  \\
   & 43.5399(4) &         & -50.4033(4) &           \\
 & & & & \\
\hline
 & & & & \\
85 & 44.25(4)   & 44.2576 & -50.57(3)   & -50.5593  \\
   & 44.2784(4) &         & -50.5566(2) &           \\
 & & & & \\
\hline
\hline
\end{tabular}
\vspace*{3mm}
\caption[]{The topology $S^{11}$ for $m_1 = 80\,$GeV and $m_2 =
m_3 = m_4 = 0$ as a function of $\sqrt{s}$. First entry is from
\eqn{sabamii}
and NAG subroutine D01EAF and the analytical result of
ref.~\cite{Bauberger:1995by}. Second entry is from \eqn{sabamii}
and NAG subroutine D01BGF. The UV-pole is $\ep = 1$ and the unit of mass
is $\tHs = 1\,$GeV.}
\label{tableb}
\end{table}
\normalsize
%--
In Appendix E we give the analytical evaluation of the imaginary part of
$S^{121}$ which receives a contribution from a two-particle cut and a
three-particle cut. Comparisons with the numerical result are shown in
\tabns{tableca}{tablecb}. The relative deviation with respect to the 
analytical result is $\le 0.04\%$ over a wide range of energies, below and 
above the three-particle cut.
%--
\begin{table}[ht]\centering
\begin{tabular}{|l|l|l|l|}
\hline
 & & &\\
$\sqrt{s}\,$[GeV]  &  $\Imb\,S^{121}$ &  $\Imb\,S^{121}$
  & $\Imb\,S^{121}$ (anal.)  \\
 & & & \\
\hline
 & & & \\
above two-p cut & & & \\
 & & & \\
\hline
\hline
 & & & \\
180 & 15.90(3) & 15.8909(5) & 15.8957  \\
 & & & \\
\hline
 & & & \\
190 & 23.26(3) & 23.2562(5) & 23.2592  \\
 & & & \\
\hline
 & & & \\
200 & 28.41(4) & 28.4148(4) & 28.4059  \\
 & & & \\
\hline
 & & & \\
210 & 32.39(4) & 32.3947(5) & 32.3834  \\
 & & & \\
\hline
 & & & \\
220 & 35.62(4) & 35.6124(4) & 35.6137  \\
 & & & \\
\hline
 & & & \\
230 & 38.30(4) & 38.2994(4) & 38.2953  \\
 & & & \\
\hline
 & & & \\
240 & 40.59(4) & 40.5769(5) & 40.5929  \\
 & & & \\
\hline
 & & & \\
250 & 42.57(4) & 42.5749(5) & 42.5613  \\
 & & & \\
\hline
\hline
\end{tabular}
\vspace*{3mm}
\caption[]{The imaginary part of $S^{121}$ for $m_1 = m_3 = 80\,$GeV and
$m_2 =  m_4 = 91\,$GeV as a function of $\sqrt{s}$. First entry is from
\eqn{sabamii} and NAG subroutine D01EAF. Second entry is from
the same equation and NAG subroutine D01BGF and third one is the analytical
result of ref.~\cite{Bauberger:1995by} as reported in Appendix E.}
\label{tableca}
\end{table}
\normalsize
%--
\begin{table}[ht]\centering
\begin{tabular}{|l|l|l|l|}
\hline
 & & &\\
$\sqrt{s}\,$[GeV]  &  $\Imb\,S^{121}$ &  $\Imb\,S^{121}$
  & $\Imb\,S^{121}$ (anal.)  \\
 & & & \\
\hline
 & & & \\
above three-p cut & & & \\
 & & & \\
\hline
\hline
 & & & \\
260 & 44.30(5) & 44.3069(5) & 44.2915  \\
 & & & \\
\hline
 & & & \\
270 & 45.84(4) & 45.8391(5) & 45.8339  \\
 & & & \\
\hline
 & & & \\
280 & 47.24(4) & 47.2457(5) & 47.2316  \\
 & & & \\
\hline
\hline
\end{tabular}
\vspace*{3mm}
\caption[]{The imaginary part of $S^{121}$ for $m_1 = m_3 = 80\,$GeV and
$m_2 =  m_4 = 91\,$GeV as a function of $\sqrt{s}$. First entry is from
\eqn{sabamii} and NAG subroutine D01EAF. Second entry is from
the same equation and NAG subroutine D01BGF and third one is the analytical
result of ref.~\cite{Bauberger:1995by} as reported in Appendix E.}
\label{tablecb}
\end{table}
\normalsize
%--
In \tabn{tabled} we scan the threshold regions, $\sqrt{s} = m_3 + m_4$ and
$\sqrt{s} = m_1 + m_2 + m_4$. Even at one MeV above threshold we are able
to reproduce the imaginary part with a relative precision of less than $1\%$

\begin{table}[ht]\centering
\begin{tabular}{|l|l|l|l|l|l|}
\hline
 & & & & &\\
$\sqrt{s}\,$[GeV]  &  $\Imb\,S^{121}$ (I) &  $\Imb\,S^{121} (II)$
  & $\Imb\,S^{121}$ (anal.) & $|$I/anal - 1$|$ & $|$II/anal - 1$|$ \\
& & & & & \\
\hline
& & & & & \\
around the two-p cut & && & & \\
& & & & & \\
\hline
\hline
& & & & & \\
171.1 & 1.26(1) & 1.25423(5) & 1.24533 & $0.87\pc$ & $0.71\pc$ \\
& & & & & \\
\hline
& & & & & \\
171.2 & 1.89(1) & 1.87835(5) & 1.87846 & $0.13\pc$ & $0.01\pc$ \\
& & & & & \\
\hline
& & & & & \\
171.3 & 2.38(1) & 2.37516(5) & 2.37086 & $0.28\pc$ & $0.18\pc$ \\
& & & & & \\
\hline
& & & & & \\
171.4 & 2.80(2) & 2.80222(5) & 2.79280 & $0.43\pc$ & $0.34\pc$ \\
& & & & & \\
\hline
\hline
& & & & & \\
around the three-p cut & & & & & \\
& & & & & \\
\hline
\hline
 & & & & & \\
262.1 & 44.63(4) & 44.6222(5) & 44.6337 & $< 0.01\pc$ & $0.03\pc$ \\
& & & & & \\
\hline
& & & & & \\
262.2 & 44.65(4) & 44.6446(3) & 44.6392 & $0.01\pc$ & $0.03\pc$ \\
& & & & & \\
\hline
& & & & & \\
262.3 & 44.67(4) & 44.6617(4) & 44.6589 & $0.02\pc$ & $0.01\pc$ \\
& & & & & \\
\hline
& & & & & \\
262.4 & 44.68(4) & 44.6983(5) & 44.6718 & $0.02\pc$ & $0.06\pc$ \\
& & & & & \\
\hline
\hline
\end{tabular}
\vspace*{3mm}
\caption[]{A scan of the imaginary part of $S^{121}$ for $m_1 = m_3 =
80\,$GeV
and $m_2 =  m_4 = 91\,$GeV around the two- and three-particle cuts. First
entry is from \eqn{sabamii} and NAG subroutine D01EAF. Second entry is from
\eqn{sabamii} and NAG subroutine D01BGF and third one is the analytical
result of ref.~\cite{Bauberger:1995by} as reported in Appendix E.}
\label{tabled}
\end{table}
\normalsize
%--
\subsection{Numerical results for $S^{131}$}
%--
The interesting case for $S^{131}$ is represented by $m_3 = m_5$. No
analytical
results are usually shown in the literature since all authors write
%--
\bq
S^{131}(p^2;m_1,m_2,m_3,m_4,m_3) = -\,\frac{\partial}{\partial\,m^2_3}\,
S^{121}(p^2;m_1,m_2,m_3,m_4).
\eq
%--
We have done the same and compared internally $S^{131}$ with the numerical
derivative of $S^{121}$. However, numerical differentiation of functions
that are already the result of numerical integration enjoy poor reputation, a
fact that we confirm. Unless a extremely accurate but time consuming evaluation
of the multi-dimensional integral is performed we end up with large
uncertainties in the derivative. Fortunately we can still check our results,
at least for the imaginary part where an analytical approach is available.
The results are shown in \tabn{tablee} and the larger error in the
derivative, for larger energies, is due to the three-particle cut.
%--
\begin{table}[ht]\centering
\begin{tabular}{|l|l|l|}
\hline
 &  &\\
$\sqrt{s}\,$[GeV]  &  $\Imb\,S^{131}$ &
$-\,\frac{\partial}{\partial\,m^2_3}\,\Imb\,S^{121}$  \\
 & & \\
\hline
\hline
 & & \\
7  & -1.551(2)   & -1.551(3)   \\
 & & \\
\hline
 & & \\
12 & -0.3047(3)  & -0.3047(4)  \\
 & & \\
\hline
 & & \\
17 & -0.1494(1)  & -0.1494(11) \\
 & & \\
\hline
 & & \\
22 & -0.09166(9) & -0.09183(282) \\
 & & \\
\hline
\hline
\end{tabular}
\vspace*{3mm}
\caption[]{A comparison between $S^{131}$ and the numerical derivative of
$S^{121}$ for $m_3 = m_5 = 1\,$GeV and $m_1 = m_2 =  m_4 = 5\,$GeV as a
function of $\sqrt{s}$. The UV-pole is $\ep = 1$ and the unit of mass
is $\tHs = 1\,$GeV.}
\label{tablee}
\end{table}
\normalsize
%--
The behavior of $S^{131}$ for $m_5 = m_3$ and around the normal threshold
$\sqrt{s} = m_3 + m_4$ is shown in \fig{sacath} and also reported in
\tabn{tablef} where we see that numerical precision is worst on the high
energy side of the threshold, where the imaginary part in non-zero. However,
it is enough to go $100\,$MeV away from threshold to be better that $0.1\%$
($0.3\%$ at $50\,$MeV).
%--
\begin{table}[ht]\centering
\begin{tabular}{|l|l|l|l|}
\hline
 & & &\\
$\sqrt{s}\,$[GeV]  &  $\Reb\,S^{131}$ & $\Imb\,S^{131}$ &
$-\,\frac{\partial}{\partial\,m^2_3}\,\Imb\,S^{121}$  \\
& & & \\
\hline
\hline
& & & \\
5.80 & -2.46581(1) & 0.00 & 0.00 \\
& & & \\
\hline
& & & \\
5.85 & -2.76299(1) & 0.00 & 0.00 \\
& & & \\
\hline
& & & \\
5.90 & -3.20714(1) & 0.00 & 0.00 \\
& & & \\
\hline
& & & \\
5.95 & -3.99512(2) & 0.00 & 0.00 \\
& & & \\
\hline
& & & \\
6.05 & -3.21(1)   & -5.28(1)   & -5.2794(5) \\
& & & \\
\hline
& & & \\
6.10 & -1.820(2)  & -4.381(4)  & -4.3735(8) \\
& & & \\
\hline
& & & \\
6.15 & -1.202(2)  & -3.834(1)  & -3.8354(3) \\
& & & \\
\hline
& & & \\
6.20 & -0.8387(5) & -3.458(1)  & -3.4588(5) \\
& & & \\
\hline
& & & \\
6.25 & -0.5954(5) & -3.1725(7) & -3.1719(4) \\
& & & \\
\hline
\hline
\end{tabular}
\vspace*{3mm}
\caption[]{$S^{131}$ and the numerical derivative of
$S^{121}$ for $m_3 = m_5 = 1\,$GeV and $m_1 = m_2 = m_4 = 5\,$GeV around
the normal threshold $\sqrt{s} = 6\,$GeV. The UV-pole is $\ep = 1$ and
the unit of mass is $\tHs = 1\,$GeV.}
\label{tablef}
\end{table}
\normalsize
%--
\bfi
\centerline{
\epsfig{file=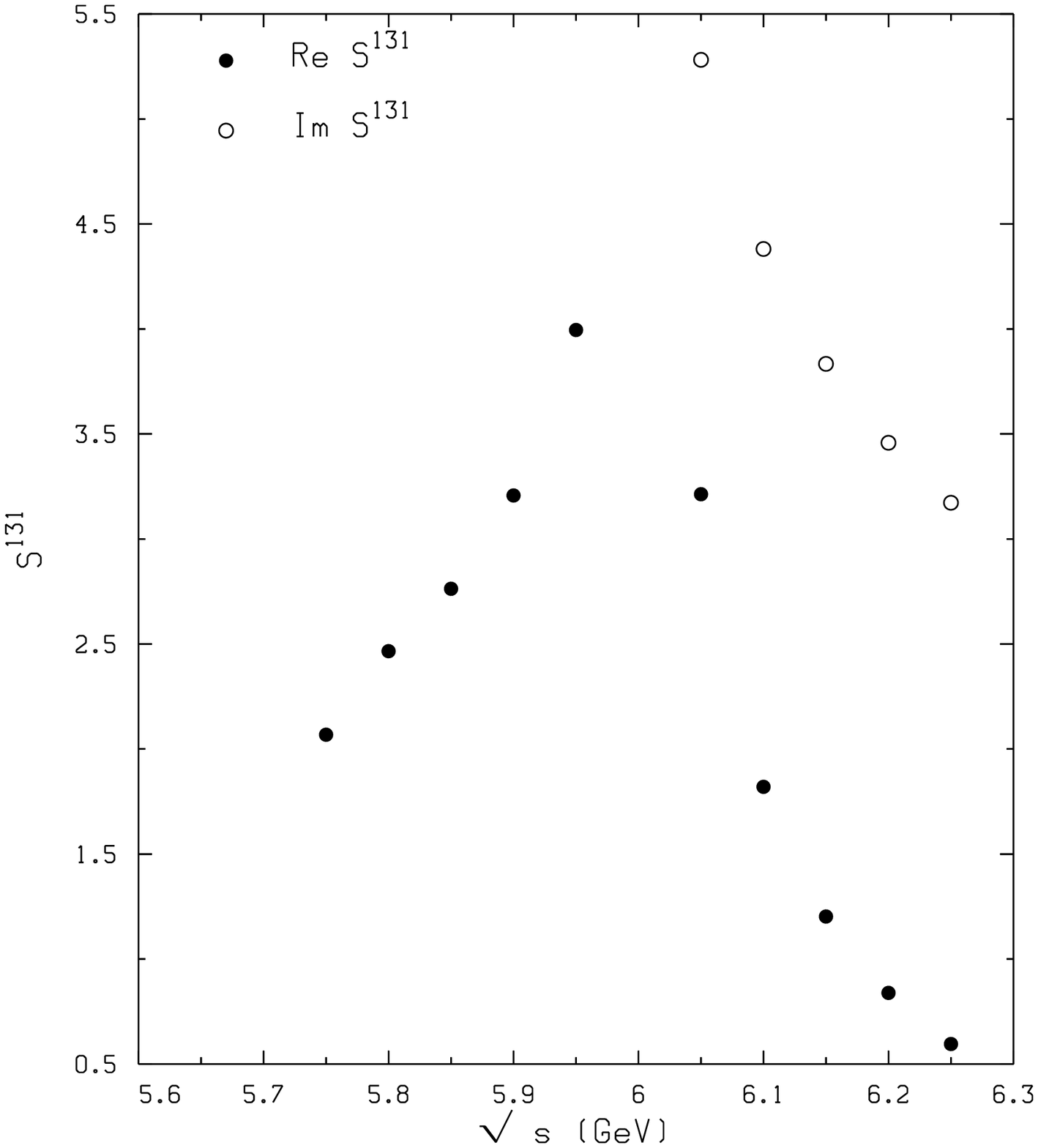,height=15cm,angle=0}}
\caption[]{The behavior of $S^{131}$, \fig{tops5}, around the normal
threshold $s = (m_3+m_4)^2$.}
\label{sacath}
\efi
%--
\subsection{Numerical results for $S^{221}$}
%--
For $S^{221}$ we have compared our results with those of
ref.~\cite{Bauberger:1995hx} where the authors have introduced a
one-dimensional integral representation of $S^{221}$. In particular we refer
to Tab. 1 of ref.~\cite{Bauberger:1995hx}. Comparisons are shown in
\tabn{tableg} where we have selected their numerical results with the
smaller error. In this case we use the NAG subroutine D01GDF which calculates 
an approximation to the integral using the Korobov-Conroy number theoretic
method with a Monte-Carlo error estimate; our two entries in \tabn{tableg}
refer to the choice of $10193\,\cdot\,101$ and $22807\,\cdot\,151$ points
respectively. Note that with their choice of masses, $m_n = \sqrt{n}\,$GeV,
the normal thresholds are at $2.414\,$GeV, $4.236\,$GeV (two-particle cuts)
and $4.968\,$GeV, $5.146\,$GeV (three-particle cuts) respectively.
Only the last energy is above all normal thresholds.
%--
\begin{table}[ht]\centering
\begin{tabular}{|l|l|l|l|l|}
\hline
& & &  &\\
$s\,[\GeV^2]$ & $\Reb\,S^{221}$ &
$\Reb\,S^{221}$ref.~\cite{Bauberger:1995hx} &
$\Imb\,S^{221}$ & $\Imb\,S^{221}$ref.~\cite{Bauberger:1995hx} \\
& & & & \\
\hline
\hline
& & & & \\
$\sst 0.1$ & $\sst -0.287205 \pm 0.20\,\times\,10^{-4}$ &
      $\sst -0.287238 \pm 0.25\,\times\,10^{-5}$ &
      $\sst 0.00$ & $\sst 0.00$ \\
    & $\sst -0.287237 \pm 0.35\,\times\,10^{-5}$ & & & \\
& & & & \\
\hline
& & & & \\
$\sst 0.5$ & $\sst -0.294578 \pm 0.15\,\times\,10^{-4}$ &
      $\sst -0.294952 \pm 0.25\,\times\,10^{-5}$ &
      $\sst 0.00$ & $\sst 0.00$ \\
    & $\sst -0.294590 \pm  0.23   \,\times\,10^{-5}$ & & & \\
& & & & \\
\hline
& & & & \\
$\sst 1.0$ & $\sst -0.304543  \pm 0.21\,\times\,10^{-4}$ &
      $\sst -0.304521  \pm 0.25\,\times\,10^{-5}$ &
      $\sst 0.00$ & $\sst 0.00$ \\
    & $\sst -0.304522  \pm 0.22\,\times\,10^{-5}$ & & & \\
& & & & \\
\hline
& & & & \\
$\sst 5.0$ & $\sst -0.452499 \pm 0.17\,\times\,10^{-4}$ &
      $\sst -0.452520 \pm 0.28\,\times\,10^{-5}$ &
      $\sst 0.00$ & $\sst 0.00$ \\
    & $\sst -0.452518 \pm  0.31\,\times\,10^{-4}$ & & & \\
& & & & \\
\hline
& & & & \\
$\sst 10.0$ & $\sst -0.488036 \pm 0.59\,\times\,10^{-3}$ &
       $\sst -0.488153 \pm 0.23\,\times\,10^{-5}$ &
       $\sst -0.353197 \pm  0.69\,\times\,10^{-3}$ &
       $\sst -0.353217 \pm 0.23\,\times\,10^{-5}$ \\
     & $\sst -0.488224 \pm 0.18\,\times\,10^{-3}$ & &
       $\sst -0.353082 \pm 0.18 \,\times\,10^{-3}$ &  \\
& & & & \\
\hline
& & & & \\
$\sst 50.0$ & $\sst 0.173897 \pm  0.49\,\times\,10^{-4}$ &
       $\sst 0.173901 \pm  0.22\,\times\,10^{-5}$ &
       $\sst -0.118089 \pm 0.40\,\times\,10^{-4}$ &
       $\sst -0.118080 \pm 0.22\,\times\,10^{-5}$ \\
     & $\sst 0.173900 \pm  0.18\,\times\,10^{-4}$ & &
       $\sst -0.118082 \pm 0.20\,\times\,10^{-4}$ &  \\
& & & & \\
\hline
\hline
\end{tabular}
\vspace*{3mm}
\caption[]{$S^{221}$ compared with the numerical result of Tab. 1 in
ref.~\cite{Bauberger:1995hx}. Here $m^2_n = n\,\GeV^2$, for $n=1,\dots,5$.
The second entry refers to a run with a much larger number of points in
sub-routine D01GDF.}
\label{tableg}
\end{table}
\normalsize
%--
There are two configurations where $S^{221}$ can be computed using the
appropriate expansion: let variables $x,y$ be defined as in \eqn{defxy},
we can give the $|y| \ll 1$ and the $|x| \ll 1$ expansion of
$S^{221}(s;m,0,0,0,0)$, as discussed in Appendix F. Results are shown in
\tabn{tableh}. By using \eqn{expat} we can compare the low-$s$ behavior of
$S^{221}(s;m,m,m,0,0)$ which is shown in \tabn{tablei}.
%--
\begin{table}[ht]\centering
\begin{tabular}{|l|l|l|l|l|}
\hline
& & &  &\\
$s\,[\GeV^2]$ & $\Reb\,S^{221}$ & $\Reb\,S^{221}\,$ expansion&
$\Imb\,S^{221}$ & $\Imb\,S^{221}\,$ expansion \\
& & & & \\
\hline
\hline
& & & & \\
0.1  & 0.3740(6)  & 0.373988 & 0.27957(9) & 0.279970  \\
& & & & \\
\hline
& & & & \\
1.0  & 0.19599(4) & 0.195930 & 0.20847(6) & 0.208394  \\
& & & & \\
\hline
& & & & \\
10.0 & 0.07144(3) & 0.071467 & 0.14067(3) & 0.140692  \\
& & & & \\
\hline
\hline
& & & & \\
5.0  & -1.61744(9)  & -1.61746  & 0.4909(2)   & 0.490807 \\
& & & & \\
\hline
& & & & \\
10.0 & -0.79270(2)  & -0.79270  & 0.14061(5)  & 0.140692 \\
& & & & \\
\hline
& & & & \\
50.0 & -0.150366(2) & -0.150362 & 0.007492(3) & 0.007492 \\
& & & & \\
\hline
\hline
\end{tabular}
\vspace*{3mm}
\caption[]{$S^{221}$ compared with the expansions described in Appendix F
for $m_1 = 100\,$GeV($m_1 = 1\,$GeV) upper(lower) part of the table. The
sum over $n$ in \eqn{expao} is restricted to $n \le 20$.}
\label{tableh}
\end{table}
\normalsize
%--
%--
\begin{table}[ht]\centering
\begin{tabular}{|l|l|l|l|l|}
\hline
& & &  &\\
$s\,[\GeV^2]$ & $\Reb\,S^{221}$ & $\Reb\,S^{221}\,$ expansion&
$\Imb\,S^{221}$ & $\Imb\,S^{221}\,$ expansion \\
& & & & \\
\hline
\hline
& & & & \\
0.1  & 0.0496(2)   & 0.049542  & 0.01566(3)  & 0.015709  \\
& & & & \\
\hline
& & & & \\
0.5  & 0.04154(5)  & 0.041506  & 0.015722(8) & 0.0157145 \\
& & & & \\
\hline
& & & & \\
1.0  & 0.03805(3)  & 0.038051  & 0.015717(9) & 0.0157211 \\
& & & & \\
\hline
& & & & \\
5.0  & 0.030085(9) & 0.0300729 & 0.015776(4) & 0.0157739 \\
& & & & \\
\hline
& & & & \\
10.0 & 0.026682(9) & 0.0266732 & 0.015836(6) & 0.0158406  \\
& & & & \\
\hline
\hline
\end{tabular}
\vspace*{3mm}
\caption[]{$S^{221}$ compared with the expansions described in Appendix F
for $m_1 = m_2 = m_3 = 100\,$GeV and $m_4 = m_5 = 0$. The
sum over $n$ in \eqn{expat} is restricted to $n \le 20$.}
\label{tablei}
\end{table}
\normalsize
%--
\section{Conclusions}
%--
For one-loop diagrams the analytical way has been fully successful,
resulting
in compact expressions containing at most di-logarithms~%\cite{Bern:1994kr}.
For multi-loop diagrams, special cases have been equal successful, see
for instance ref.~\cite{Barbieri:1972as} and ref.~\cite{Devoto:1984tc},
resulting in expressions containing generalized Nielsen
poly-logarithms~\cite{polyl}.
However, this approach can hardly been extended beyond one-loop for general
massive diagrams. For instance, the work of ref.~\cite{Bauberger:1995by}
and~\cite{Bauberger:1995hx} already shows that for two-loop integrals which
contain a massive three-particle cut, the results in general cannot be
expressed in terms of generalized poly-logarithms.
Sometimes special cases of multi-loop diagrams can be expressed with the
help of generalized hypergeometric functions or of other higher
transcendental
functions, but in the end this is equivalent to some series representation
that always has a restricted region of convergence. In other words, from a
practical point of view, they can hardly be extended above the closest
normal or pseudo-threshold.

This paper represents the second in a series devoted to numerical evaluation
of multi-loop Feynman diagrams with the declared intent of covering all
regions in the external parameters. In the first
paper~\cite{Passarino:2001wv}
we have introduced the minimal Bernstein-Tkachov algorithm and discussed the
simplest two-loop topology, $S^{111}$ or sunset diagram.

In this paper we extended the previous work to cover all two-loop
two-point functions, i.e. the self-energies. Each topology has been
written with an integral representation in the Feynman parametric space
which is particularly suited for numerical integration. For two external legs 
it turns out that distortion of the integration hyper-contour can be avoided.

Already dealing with the sunset topology we discovered this remarkable
property that has been generalized now to all two-loop two-point
functions but that remains true for all two-loop three-point planar
topologies~\cite{preparation}. For all these diagrams the matrix $H$ of
\eqn{defHKL} is singular. As a consequence, a change of variables is always
possible in the quadratic form $V$ so that the $B$-coefficient of
\eqn{functr}
becomes independent on Feynman parameters, at least if we use only one
iteration of the algorithm.
In this way we are able to be closer to the original idea of the BT-method.
However, as already stressed in the original paper~\cite{Tkachov:1997wh},
this $B$ will vanish at some threshold. We have analyzed all Landau
singularities for the two-loop self-energies and have to show that this
occurs
at some non-leading Landau singularity of the diagram. Here some additional
analytical work is needed before starting the numerical evaluation.

A second motivation for this work was to start a comprehensive analysis
of infrared divergent multi-loop diagrams, from the point of view of their
numerical evaluation.
Clearly the approach described in this paper is primarily intended for
evaluation of massive multi-loop diagrams. However, QED and QCD will be part
of any realistic calculation and they usually lead to infrared
singularities.
Therefore, although the massless world is most efficiently treated in QED
and QCD within the analytical approach, any multi-loop calculation in the
standard model or beyond, being plagued by infrared divergences, requires to 
find a way of dealing with them in a purely numerical approach.

For all two-loop two-point functions the relevant infrared objects are the
on-shell derivative which are the building block of wave-function
renormalization. For some of them the infrared divergent
configurations are simple enough to allow for BT treatment with a consequent
and straightforward analytical evaluation. The remaining ones, in
particular when there is the presence of overlapping infrared
configurations, require a novel approach.
Our solution is derived by adapting the general algorithm of
ref.~\cite{Binoth:2000ps}: the residues of the infrared poles and the finite
part of a multi-loop diagram can be cast into a form which allows for a
reliable numerical integration.

Work is in progress to extend the numerical evaluation to two-loop vertices,
including non-infrared on-shell derivatives of self-energies,
where the FORM codes for all topologies, including the non-planar one, have
been developed~\cite{preparation}.
%--
\Acknowledgments
The contribution of Andrea Ferroglia to the general project of 
algebraic-numerical evaluation of Feynman diagrams and to several discussions 
concerning this paper is gratefully acknowledged.
%--
\section{Appendix A}
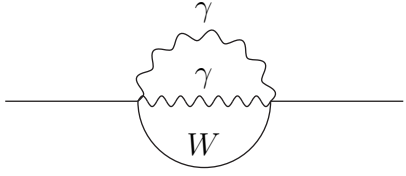
\begin{figure}[th]
\vskip 5pt
\[
  \vcenter{\hbox{
  \begin{picture}(150,90)(0,0)
  \Line(0,50)(50,50)
  \CArc(75,50)(25,-180,0)
  \PhotonArc(75,50)(25,0,180){2}{7}
  \Line(100,50)(150,50)
  \Photon(50,50)(100,50){2}{7}
  \Text(75,80)[cb]{$\gamma$}
  \Text(75,55)[cb]{$\gamma$}
  \Text(75,30)[cb]{$\wb$}
  \end{picture}}}
\]
\vspace{-0.5cm}
\caption[]{An $S^{111}$-like contribution to the $\wb$ self-energy.}
\label{saaa}
\end{figure}

%--
Consider the topology $S^{111}$ with $m_1 = m_2 = 0$ and $m_3 = m$ of
\fig{saaa}. Let us
consider the derivative
%--
%--
\bq
\pi^4\,\frac{\partial}{\partial\,p^2}\,S^{111} =
-\,\frac{\mu^{2\ep}}{p^2}\,\intmomii{n}{q_1}{q_2}\,\frac{\spro{p}{(q_2+p)}}
{q^2_1\,(q_1-q_2)^2\,((q_2+p)^2+m^2)^2}.
\eq
%--
We combine the first two propagators with a Feynman parameter $x$, perform
the
$q_1$ integration and combine the two resulting propagators with a new
parameter $y$. After the $q_2$-integration we put $p^2 = -m^2$ and obtain
%--
\bq
S^{111}_{p}\mid_{\rm os} = \Bigl(\frac{\tHss}{\pi\,m^2}\Bigr)^{\ep}\,
\egam{\ep}\,B(1-\frac{\ep}{2},1-\frac{\ep}{2})\,B(2-2\,\ep,1+\frac{\ep}{2}),
\eq
%--
where $B$ is the Euler beta-function. The the on-shell derivative
contains a single pole of ultraviolet origin and no infrared pole.
%--
\section{Appendix B}
%--
For completeness we report in this Appendix a list of master
integrals~\cite{Devoto:1984tc} that are needed in the evaluation of
$S^{121}_p$, see \eqn{sabap}. Given $X$ we introduce an auxiliary variable 
$x$ defined as $X = x-1$ and obtain 
%--
\bqa
{}&{}& \intfx{y}\,y^n \ln y = -\,\frac{1}{(n+1)^2},
\nl
{}&{}& \intfx{y}\,y \ln (1+Xy) = \frac{1}{2\,X^2}\,\Bigl[ (X^2-1) \ln x -
\frac{1}{2}\,X^2 + X\Bigr],
\nl
{}&{}& \intfx{y}\,y \ln (1+\frac{y}{X}) = \frac{1}{2}\,X^2 \Bigl[
(\frac{1}{X^2}-1) (\ln x-\ln X) - \frac{1}{2\,X^2} +
\frac{1}{X}\Bigr],
\nl
{}&{}& \intfx{y}\,\ln (1+\frac{y}{X}) =  x  \ln (1+\frac{1}{X})-1,
\nl
{}&{}& \intfx{y}\,\ln (1+Xy) = \frac{x}{X} \ln x-1, \quad
\intfx{y}\,(y+\frac{1}{X})^{-1} = \ln x,
\nl
{}&{}& \intfx{y} y \ln^2 y = \frac{1}{4}, \quad
\intfx{y} \ln^2 y = 2,
\nl
{}&{}& \intfx{y} y \ln^2(1+ X y) = \frac{1}{X^2}\,\Bigl[
x (\frac{1}{2}x - 1)\,\ln^2 x + x (2 - \frac{1}{2} x)\,\ln x + 
\frac{1}{4}\,x^2 - 2 x + \frac{7}{4}\Bigr],
\nl
{}&{}& \intfx{y} \ln^2(1+X y) = \frac{1}{X}\,\Bigl[
 x  \ln^2 x  - 2  x  (\ln x  - 1) - 2\Bigr],
\nl
{}&{}& \intfx{y} (y+\frac{1}{X})^{-1}\,\ln(1 + X y) = X \intfx{y} 
(1+X y)^{-1}\,\ln(1+X y),
\nl
{}&{}& \intfx{y} (1+X y)^{-1}\,\ln(1+X y) = \frac{1}{2\,X} \ln^2  x ,
\nl
{}&{}& \intfx{y} (y+\frac{1}{X})^{-1}\,\ln y  = \li{2}{-\,\frac{1}{X}}, \quad
\intfx{y} y^{-1}\,\ln(1+\frac{y}{X}) = -\,\li{2}{-\,\frac{1}{X}},
\nl
{}&{}& \intfx{y} y^{-1}\,\ln(1+X y) = -\,\li{2}{-X},
\nl
{}&{}& \intfx{y} y \ln y \ln(1+\frac{y}{X}) = \frac{1}{4} X^2 (1-\frac{1}{X^2})
\Bigl[\ln x -\ln X\Bigr] - \frac{1}{2} X^2\,\li{2}{-\,\frac{1}{X}} +
\frac{1}{4} -  \frac{3}{4} X,
\nl
{}&{}& \intfx{y} \ln y \ln(1+\frac{y}{X}) = X \Bigl\{
- (1+\frac{1}{X}) \Bigl[ \ln x  - \ln X\Bigr] + \li{2}{-\,\frac{1}{X}})
\Bigr\} + 2,
\nl
{}&{}& \intfx{y} y \ln y \ln(1+ X y) = \frac{1}{4\,X^2} (1-X^2) \ln x -
\frac{1}{2\,X^2} \li{2}{-\,X} + \frac{1}{4} - \frac{3}{4\,X},
\nl
{}&{}& \intfx{y} \ln y \ln(1+X y) = \frac{1}{X} \Bigl[ - x  \ln x  +
\li{2}{-\,X} + 2,
\nl
{}&{}& \intfx{y} \ln(1+\frac{y}{X}) =  x  \Bigl[\ln x  - \ln X\Bigr] - 1,
\nl
{}&{}& \intfx{y} (y+X)^{-1} = \ln x -\ln X, \quad
\intfx{y} y^n= \frac{1}{n+1}.
\eqa
%--
\section{Appendix C}
%--
In this appendix we list all the integrals that are relevant to compute
the on-shell behavior of $S^{131}_p$, see \eqn{sacap}. We will introduce
the following terminology:
%--
\bq
{\cal I}(\mu,\nu,a) = \intfx{x}\,\Bigl[x(1-x)\Bigr]^{-\ep/2}\,
{\cal J}(\mu,\nu,a),
\quad
{\cal J}(\mu,\nu,a) = \intfx{y}\,y^{\mu}\,(1+a\,y)^{\nu}.
\eq
%--
We obtain the following results:
%--
\bqa
{\cal J}(\frac{\ep}{2}-1,-1-\ep,X) &=&  \frac{2}{\ep}\;
\hyper{1+\ep}{\frac{\ep}{2}}{1+\frac{\ep}{2}}{-X}
\nl
{}&=& \frac{2}{\ep}\,x^{1+\ep}\;\hyper{1+\ep}{1}{1+\frac{\ep}{2}}{1-x}.
\eqa
%--
The corresponding $x$ integral gives
%--
\bqa
{\cal I}(\frac{\ep}{2}-1,-1-\ep,X) &=&
\frac{2}{\ep}\,{{\egam{1+\ep/2}}\over {\egam{1+\ep}}}\,\sum_{n=0}^{\infty}
{{\egam{n+1+\ep}}\over {\egam{n+1+\ep/2}}}\,\intfx{x}\,
x^{1+\ep/2}(1-x)^{n-\ep/2}
\nl
{}&=& \frac{2}{\ep}\,{{\egam{1+\ep/2}\egam{2+\ep/2}}\over
{\egam{1+\ep}}}\,\sum_{n=0}^{\infty}\,
{{\egam{n+1+\ep}\egam{n+1-\ep/2}}\over
{\egam{n+1+\ep/2}\egam{n+3}}}
\nl
{}&=& \frac{2}{\ep} + 1 + \ord{\ep}.
\eqa
%--
A second class of integrals,
%--
\bqa
\intfx{y}\,y^{\ep/2}\,(1+X\,y)^{-1-\ep} &=& {\cal
J}(\frac{\ep}{2},-1-\ep,X),
\nl
\intfx{y}\,y^{-\ep/2}\,(y+X)^{-1-\ep} &=&
X^{-1-\ep}\,{\cal J}(-\,\frac{\ep}{2},-1-\ep,\frac{1}{X}),
\eqa
%--
are finite in the limit $\ep \to 0$. We obtain
%--
\bq
\intfxx{x}{y}\,\frac{1}{1+X\,y} = \zeta(2) - 1, \qquad
\intfxx{x}{y}\,\frac{1}{y+X} = 1.
\eq
%--
A third class of integrals can be reduced to the previous ones:
%--
\bqa
{}&{}&
\intfxx{x}{y}\,\Bigl[x(1-x)]^{-\ep/2}\,y^{\ep/2-1}\,(1+X\,y)^{-\ep} =
\nl
{}&{}&
\intfxx{x}{y}\,\Bigl[x(1-x)]^{-\ep/2}\,\Bigl\{
\frac{2}{\ep}\,(1+X)^{-\ep} + 2\,\ln(1+X) + \ep\,\Bigl[
\li{2}{-X} - \ln^2(1+X)\Bigr]\Bigr\}
\nl
{}&=& \intfx{x}\,\Bigl\{ \frac{2}{\ep}\,x^{\ep/2}(1-x)^{-\ep/2} -
2\,\lnx + \ep\,\Bigl[ \li{2}{\frac{x-1}{x}} + \ln (1-x)\,\ln x\Bigr]
\Bigr\}
\nl
{}&=& \frac{2}{\ep}\,\egam{1+\frac{\ep}{2}}\egam{1-\frac{\ep}{2}} +
2 + \ep\,{\cal C} = \frac{2}{\ep} + 2 +
\Bigl[{\cal C} + \frac{1}{2}\,\psi'(1)\Bigr]\,\ep,
\eqa
%--
where we used the relation
%--
\bq
\egam{1 \pm \frac{\ep}{2}} = 1  \mp \frac{1}{2}\,\gamma \ep + \frac{1}{8}\,
\Bigl[ \gamma^2 + \psi'(1)\Bigr]\,\ep^2 + \ord{\ep^3}.
\eq
%--
The constant is
%--
\bq
{\cal C} =
\intfx{x}\,\Bigl[ \li{2}{x} - \frac{1}{2}\,\ln^2 x +
2\ln x\,\ln(1-x) - \zeta(2)\Bigr],
\eq
%--
where we used
%--
\bq
\li{2}{\frac{x-1}{x}} = \li{2}{x} + \ln x\,\ln(1-x) - \frac{1}{2}\,\ln^2 x -
\zeta(2).
\eq
%--
After integration we obtain
%--
\bq
{\cal C}= 2 - 2\,\zeta(2),
\eq
%--
where
%--
\bq
\intfx{x}\,\Bigl\{\li{2}{x}\,;\,\ln x\,\ln(1-x)\,;\,\ln^2 x\Bigr\} =
\Bigl\{\zeta(2) - 1\,;\,2 - \zeta(2)\,;\,2\Bigr\}.
\eq
%--
Similarly
%--
\bq
\intfxx{x}{y}\,\Bigl[x(1-x)]^{-\ep/2}\,y^{-\ep/2-1}\,(y+X)^{-\ep} =
-\,\frac{2}{\ep} - 2 - \Bigl[ 2 + \frac{5}{2} \zeta(2) \Bigr] \,\ep +
\ord{\ep^2}.
\eq
%--
\section{Appendix D}
%--
In this Appendix we report the explicit expression for the residue of the
single infrared pole ${\cal S}_{\bba}^1$ defined in \eqn{babresidues}.
%--
\bqa
{\cal S}_{\bba}^1
&=&
\intfx{x_2}\intfxx{x_1}{y}\,\Bigl\{\,x_2\,(1+x_1)\,
\big[ (1+x_1)^2 - (1-x_1x_2)^2x_2^2y^2 \big]\,A_{011}^{-2}  \nl
&+&
x_2^2\,y\,(1+x_1)\,
\big[ (1+x_1)^2 - (1-x_1x_2y)^2y^2 \big]\,A_{012}^{-2}  \nl
&+&
x_1\,(1+x_2)\,
\big[ (1+x_2)^2 - (1-x_1)^2x_1^2x_2^2y^2 \big]\,A_{021}^{-2}  \nl
&+&
x_1^2\,y\,(1+x_2)\,
\big[ (1+x_2)^2 - (1-x_1y)^2x_2^2y^2 \big]\,A_{022}^{-2}  \nl
&+&
(1+x_2-x_1x_2^2)\,
\big[ (1+x_2-x_1x_2^2)^2 - x_1^2x_2^2y^2 \big]\,A_{1}^{-2}  \nl
&+&
(1+x_1x_2-x_1^2x_2)\,
\big[ (1+x_1x_2-x_1^2x_2)^2 - x_1^2x_2^2y^2 \big]\,A_{21}^{-2}  \nl
&+&
x_2\,(1+x_1x_2y-x_1^2x_2y)\,
\big[ (1+x_1x_2y-x_1^2x_2y)^2 - x_1^2y^2 \big]\,A_{22}^{-2} \, + \, 1\Bigr\}
  \nl
&+&
\int_0^{\infty}dx_2\,\intfxx{x_1}{y}\,\frac{x_2}{1+x_2}\,
\Big\{ \Big[ \frac{(1-x_1)^3 B_{1}^{-2}}{x_1} \Big]_+\,+\,
\frac{(1-x_1)^2}{1+x_2}\,B_{1}^{-2} \Big\}  \nl
&+&
\frac{1}{2}\,\int_0^{\infty}dx_2\intfx{y}\,\frac{1}{1+x_2}\,
B_{1}^{-2}(x_1=0)\,\big[ x_2 L_{b1}(x_1=0) + 1 \big]  \nl
&+&
\int_1^{\infty}dx_2\intfxx{x_1}{y}\,\frac{x_2}{1+x_2}\,
\Big\{ \Big[ \frac{(1-x_1y)^3 B_{2}^{-2}}{y} \Big]_+\,+\,
\frac{(1-x_1y)^2}{1+x_2}\,B_{2}^{-2} \Big\} \nl
&+&
\frac{1}{2}\,\int_1^{\infty}dx_2\intfx{x_1}\,\frac{1}{1+x_2}\,
B_{2}^{-2}(y=0)\,\big[ x_2 L_{b2}(y=0) + 1 \big] \nl
&+&
\intfx{x_2}\intfxx{x_1}{y}\,\Bigl\{ \frac{1}{1+x_2}\,
\Big[ \frac{(1-x_1x_2y)^3 B_{21}^{-2}}{y} \Big]_+  \nl
&+&
\frac{x_1x_2}{(1+x_2)^2}\,
(1-x_1x_2y)^2\,B_{21}^{-2} +
\frac{x_2}{1+x_1x_2}\,
\Big[ \frac{(1-x_1y)^3 B_{22}^{-2}}{y} \Big]_+  \nl
&+&
\frac{x_1x_2}{(1+x_1x_2)^2}\,
(1-x_1y)^2\,B_{22}^{-2}\Bigr\}  \nl
&+&
\frac{1}{2}\,\intfxx{x_2}{x_1}\,\Bigl\{
\Big[ \frac{B_{21}^{-2}(y=0)}{(1+x_2)x_2} \Big]_+  \nl
&+&
\frac{1}{1+x_2}\,
B_{21}^{-2}(y=0)\,\big[ L_{b21}(y=0) + \ln x_2 \big]
+\Big[ \frac{B_{22}^{-2}(y=0)}{(1+x_1x_2)x_2} \Big]_+  \nl
&+&
\frac{x_2}{1+x_1x_2}\,
B_{22}^{-2}(y=0)\,\big[ L_{b22}(y=0) + \ln x_1 \big]\Bigr\}  \nl
&+&
\frac{1}{2}\,\intfx{x_1}\,
B_{21}^{-2}(y=x_2=0)\,\big[ L_{b21}(y=x_2=0) - 1 \big]  \nl
&+&
\frac{1}{2}\,\intfx{x_2}\,
B_{22}^{-2}(y=x_1=0)\,\big[ L_{b22}(y=x_1=0) - 1 \big]
\eqa
%--
where the following functions have been introduced:
%--
\bqa
A_{011} &=&
(1+x_1)^2 + (1+x_2^2)(1+x_1)(1-x_1x_2)y + (1-x_1x_2)^2x_2^2y^2  \nl
A_{012} &=&
(1+x_1)^2x_2 + (1+x_2^2y^2)(1+x_1)(1-x_1x_2y) + (1-x_1x_2y)^2x_2y^2  \nl
A_{021} &=&
(1+x_2)^2 + (1+x_1^2x_2^2)(1+x_2)(1-x_1)y + (1-x_1)^2x_1^2x_2^2y^2  \nl
A_{022} &=&
(1+x_2)^2x_1 + (1+x_1^2x_2^2y^2)(1+x_2)(1-x_1y) + (1-x_1y)^2x_1x_2^2y^2  \nl
A_{1} &=&
(1+x_2-x_1x_2^2)^2 + (1+x_2^2)(1+x_2-x_1x_2^2)x_1y + x_1^2x_2^2y^2  \nl
A_{21} &=&
(1+x_1x_2-x_1^2x_2)^2 + (1+x_1^2x_2^2)(1+x_1x_2-x_1^2x_2)y + x_1^2x_2^2y^2
\nl
A_{22} &=&
(1+x_1x_2y-x_1^2x_2y)^2x_2^2 + (1+x_1^2x_2^2y^2)(1+x_1x_2y-x_1^2x_2y) +
x_1^2x_2y^2  \nl
B_{1} &=&
1 + x_2 - x_1x_2 + (1-x_1)^2x_2y  \nl
B_{2} &=&
(1+x_2-x_1x_2y)x_1 + (1-x_1y)^2x_2  \nl
B_{21} &=&
(1+x_2-x_1x_2^2y)x_1 + (1-x_1x_2y)^2  \nl
B_{22} &=&
(1+x_1x_2-x_1^2x_2y) + (1-x_1y)^2x_2
\eqa
%--
and also
%--
\bqa
L_{a011} &=&
  \ln (A_{011}) - \frac{1}{2} \ln y - \ln (1+x_2)
+ \frac{1}{2} \ln (1-x_1x_2) - \frac{3}{2} \ln (1+x_1)  \nl
L_{a012} &=&
  \ln (A_{012}) - \frac{1}{2} \ln x_2 - \ln (1+x_2y)
+ \frac{1}{2} \ln (1-x_1x_2y) - \frac{3}{2} \ln (1+x_1)  \nl
L_{a021} &=&
  \ln (A_{021}) - \frac{1}{2} \ln y - \ln (1+x_1x_2)
+ \frac{1}{2} \ln (1-x_1) - \frac{3}{2} \ln (1+x_2)  \nl
L_{a022} &=&
  \ln (A_{022}) - \frac{1}{2} \ln x_1 - \ln (1+x_1x_2y)
+ \frac{1}{2} \ln (1-x_1y) - \frac{3}{2} \ln (1+x_2)  \nl
L_{a1} &=&
  \ln (A_{1}) - \frac{1}{2} \ln y + \frac{1}{2} \ln x_1 + \ln x_2
- \ln (1+x_2) - \frac{3}{2} \ln (1+x_2-x_1x_2^2)  \nl
L_{a21} &=&
  \ln (A_{21}) - \frac{1}{2} \ln y + \ln x_1
- \ln (1+x_1x_2) - \frac{3}{2} \ln (1+x_1x_2-x_1^2x_2)  \nl
L_{a22} &=&
  \ln (A_{22}) + \ln x_1 - \frac{1}{2} \ln x_2
- \ln (1+x_1x_2y) - \frac{3}{2} \ln (1+x_1x_2y-x_1^2x_2y)  \nl
L_{b1} &=&
  \ln (B_{1}) + \frac{1}{2} \ln y - \frac{1}{2} \ln x_2
+ \ln (1+x_2) - \frac{3}{2} \ln (1+x_2-x_1x_2)  \nl
L_{b2} &=&
  \ln (B_{2}) + \frac{1}{2} \ln x_1 - \frac{1}{2} \ln x_2
+ \ln (1+x_2) - \frac{3}{2} \ln (1+x_2-x_1x_2y)  \nl
L_{b21} &=&
  \ln (B_{21}) + \frac{1}{2} \ln x_1
+ \ln (1+x_2) - \frac{3}{2} \ln (1+x_2-x_1x_2^2y)  \nl
L_{b22} &=&
  \ln (B_{22}) - \frac{1}{2} \ln x_2
+ \ln (1+x_1x_2) - \frac{3}{2} \ln (1+x_1x_2-x_1^2x_2y).
\eqa
%--
\section{Appendix E\label{appb}}
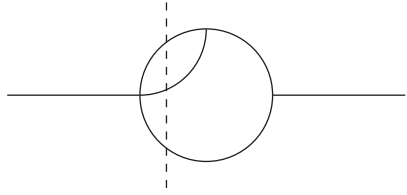
\begin{figure}[th]
\vspace{0.5cm}
\[
  \vcenter{\hbox{
  \begin{picture}(150,0)(0,0)
  \Line(0,0)(50,0)
  \CArc(75,0)(25,0,90)
  \CArc(75,0)(25,-180,0)
  \CArc(75,0)(25,90,180)
  \CArc(50,25)(25,-90,0)
  \Line(100,0)(150,0)
  \DashLine(60,35)(60,-35){3.}
  \end{picture}}}
\]
\vspace{0.5cm}
\caption[]{The three-particle cut of diagram $S^{121}$}
\label{s4threepc}
\end{figure}

%--
In this Appendix we consider all those results that are useful to derive
analytical expressions for the imaginary parts of two-loop two-point
functions. The two-particle cut contribution to the discontinuity of
$S^{121}$, shown in \fig{s4tpc}, is given by
%--
\bq
{\rm disc}_2\,S^{121} = - 2\,i\,\bff{0}{-m^2_3}{m_1}{m_2}\,
\Imb\bff{0}{-s}{m_3}{m_4}.
\eq
%--
Since the one-loop scalar function $B_0$ is
%--
\bq
\bff{0}{\pmoms}{\mone}{\mtwo} = \Ddr - \ln \frac{\mone\mtwo}{\tHss} - R
 + \frac{\mones-\mtwos}{2\pmoms}\ln\frac{\mones}{\mtwos} + 2,
\eq
%--
where
%--
\bq
R = -\frac{\Lambda}{\pmoms}\,
\ln{{\pmoms-\ib\delta+\mones+\mtwos-\Lambda}\over{2\,\mone \mtwo}},
\eq
%--
and where $\Lambda^2 = \lkall{-\pmoms}{\mones}{\mtwos}$,
we must have the imaginary part of $\bff{0}{-s}{m_3}{m_4}$ to
$\ord{\ep}$. It follows
%--
\bqa
\bff{0}{\pmoms}{\mone}{\mtwo} &=& \Ddr - \intfx{x}\,\ln\frac{\chi}{\tHss}
\nl
{}&+& \ep\,\Bigl\{ \frac{1}{4}\,(\gamma + \ln\pi)^2 + \frac{1}{4}\,
\zeta(2) + \frac{1}{2}\,\intfx{x}\,\ln\frac{\chi}{\tHss}\,\Bigl[
\gamma + \ln\pi + \frac{1}{2}\,\ln\frac{\chi}{\tHss}\Bigr]\Bigr\},
\eqa
%--
where, as usual, $\chi= -\pmoms x^2 + (\pmoms + m^2_2 - m^2_1) x + m^2_1$.
Note that, for $s > (m_3+m_4)^2$ and $m^2_3 > (m_1+m_2)^2$ the discontinuity
develops a real part. However, in the same region the discontinuity
coming from the three-particle cut of \fig{s4threepc} also has a real part
and the two cancel so that ${\rm disc}_{2+3}\,S{\aba}$ is purely imaginary.
For completeness we report the result for ${\rm disc}_3\,S^{121}$.
Define first
%--
\bqa
s_{\pm\pm} &=& (\sqrt{s} \pm m_4)^2 - (m_2 \pm m_1)^2,  \qquad
\kappa^2 = \frac{s_{\pm}s_{\mp}}{s_{++}s_{--}}, \nl
x_3 &=& \frac{q_{\mp}}{q_{\pm}}, \qquad
c_1 = \sqrt{s_{++}s_{--}},
\eqa
%--
and also
%--
\bqa
\frac{c_2}{c_1} &=& 4\,m_1m_2\,s_{--},  \qquad
\frac{c_3}{c_1} = 8\,m_1m_2\,(s -m^2_3+m^2_1+m^2_2+m^2_4),  \nl
\frac{c_4}{c_1} &=& -\,8\,m_1m_2\,\frac{(s-m^2_4)^2}{m^2_3},  \qquad
\frac{c_5}{c_1} = 8\,m_1m_2\,\frac{\lambda(s,m^2_3,m^2_4)}{m^2_3},  \nl
\frac{x_4}{x_3} &=& \frac{(m_2-m_1)^2}{(m_2+m_1)^2},  \qquad
\frac{x_5}{x_3} = \frac{m^2_3-(m_2-m_1)^2}{m^2_3-(m_2+m_1)^2}.
\eqa
%--
It follows
%--
\bqa
{\rm disc}_3\,S^{121} &=& \frac{2\,i\pi}{s}\,\Bigl[ c_1\,E(\kappa) +
c_2\,K(\kappa) + c_3\,\Pi(x_3,\kappa) \nl
{}&+& c_4\,\Pi(x_4,\kappa) + c_5\,\Pi(x_5-i\,\delta,\kappa)\Bigr]\,
\theta(s-(m_1+m_2+m_4)^2),
\eqa
%--
where $E,K$ and $\Pi$ are elliptic integrals~\cite{ellip}.
The $i\,\delta$ prescription is needed because in the last complete elliptic
integral of third kind we have, in the region $s > (m_3+m_4)^2$ and
$m^2_3 > (m_1+m_2)^2$, that $x_5 > 1$ and the integral must be understood
as a Cauchy principal value. The corresponding imaginary part cancel
the real part from ${\rm disc}_2$. Adding the two contributions we find
%--
\bq
\Imb S^{121} = -\,\frac{i}{2}\,\Bigl[ {\rm disc_2}\,S^{121} +
{\rm disc}_3\,S^{121}\Bigr].
\eq
%--
%--
For a numerical evaluation of elliptic integrals we use the NAG sub-routines
S21BAF - S21BDF, following the alternative definition by
Carlson~\cite{altdef}.
%--
\section{Appendix F}
%--
In this Appendix we report some special cases of $S^{221}$ for which the
low-(high-)$p^2$ expansion is known~\cite{Bauberger:1995by}. Let us define
the following variables
%--
\bq
x = - \frac{m^2}{s} + i\,\delta, \qquad
y = - \frac{s}{m^2} - i\,\delta.
\label{defxy}
\eq
%--
We obtain the following two expansions:
%--
\bqa
S^{221}(s;m,0,0,0,0) &=& \frac{1}{s}\,\sum_{n=1}^{\infty}\,\Bigl[
\frac{\ln^2 y + 2\,\zeta(2)}{2\,n} - \frac{2}{n^2}\,\ln y + \frac{3}{n^3} +
r_2(n)\Bigr]\,(-y)^n, \quad |y| \ll 1,
\nl
S^{221}(s;m,0,0,0,0) &=& -\,\frac{1}{s}\,\sum_{n=1}^{\infty}\,\Bigl[
\frac{\ln^2 x + 2\,\zeta(2)}{2\,n} - \frac{2}{n^2}\,\ln x + \frac{3}{n^3} +
r_2(n)\Bigr]\,(-x)^n - 6\,\frac{\zeta(3)}{s}, \quad |x| \ll 1,
\nl
\label{expao}
\eqa
%--
where $r_2$ is defined as
%--
\bq
r_2(n) = \frac{1}{n^2}\,\sum_{l=1}^{n-1}\,\frac{n-l}{l^2}.
\eq
%--
Another known expansion is the following:
%--
\bqa
S^{221}(s;m,m,m,0,0) &=& \frac{1}{2\,m^2},\sum_{n=0}^{\infty}\,
\frac{n!\egam{3/2}}{\egam{n+3/2}}\,\lpar\frac{s}{4\,m^2}\rpar^n\,
\Bigl[ \frac{3}{(n+1)^2} - \frac{\ln y}{n+1}\Bigr].
\label{expat}
\eqa

%--
\end{document}